\documentclass[a4paper,fleqn,usenatbib]{mnras}
\hypersetup{draft}

\usepackage{newtxtext,newtxmath}

\usepackage[T1]{fontenc}
\usepackage{ae,aecompl}

\usepackage{graphicx}	
\usepackage{amsmath}	
\usepackage{lineno,hyperref}
\usepackage{pdflscape}
\usepackage{multirow}
\usepackage{wasysym}
\usepackage{longtable}

\usepackage{siunitx}
\usepackage[T1]{fontenc}
\usepackage[utf8]{inputenc}
\usepackage{gensymb}
\usepackage{wasysym}

\DeclareMathOperator{\atantwo}{atan2}

\usepackage[colorinlistoftodos]{todonotes}
\let\Oldtodo\todo
\renewcommand{\todo}[1]{\Oldtodo[inline]{#1}}

\title[Global Meteor Network]{The Global Meteor Network - Methodology and First Results}

\author[D. Vida et al.]{
Denis Vida,$^{1,2}$\thanks{E-mail: dvida@uwo.ca}
Damir \v{S}egon$^{3,4}$,
Peter S. Gural$^{5}$,
Peter G. Brown$^{1,2}$,
Mark J.M. McIntyre$^{6}$,
\newauthor
Tammo Jan Dijkema$^{7}$,
Lovro Pavleti\'{c}$^{8}$,
Patrik Kuki\'{c}$^{9}$,
Michael J. Mazur$^{1}$,
\newauthor
Peter Eschman$^{10}$,
Paul Roggemans$^{11}$,
Aleksandar Merlak$^{12}$
and Dario Zubovi\'{c}$^{4}$
\\
$^{1}$Department of Physics and Astronomy, University of Western Ontario, London, Ontario, N6A 3K7, Canada\\
$^{2}$Institute for Earth and Space Exploration, University of Western Ontario, London, Ontario, N6A 5B8, Canada\\
$^{3}$Astronomical Society Istra Pula, Park Monte Zaro 2, HR-52100 Pula, Croatia\\
$^{4}$Vi\v{s}njan Science and Education Center, Istarska 5, HR-51463 Višnjan, Croatia\\
$^{5}$Gural Software and Analysis LLC, Lovettsville, Virginia, USA\\
$^{6}$Tackley Observatory, 40 Medcroft Road, Tackley, Oxfordshire OX5 3AH, UK\\
$^{7}$ASTRON, Netherlands Institute for Radio Astronomy, Oude Hoogeveensedijk 4, 7991 PD Dwingeloo, The Netherlands\\
$^{8}$Department of Physics, University of Rijeka, HR-51000 Rijeka, Croatia\\
$^{9}$Faculty of Electrical Engineering and Computing, University of Zagreb, HR-10000 Zagreb, Croatia\\
$^{10}$New Mexico Meteor Array, Albuquerque, New Mexico, USA \\
$^{11}$Pijnboomstraat 25, 2800 Mechelen, Belgium \\
$^{12}$Istrastream d.o.o, Hum, Croatia
}

\date{Accepted 2021 July 7. Received 2021 July 6; in original form 2021 January 26.}

\pubyear{2020}

\begin{document}
\label{firstpage}
\pagerange{\pageref{firstpage}--\pageref{lastpage}}
\maketitle

\begin{abstract}
The Global Meteor Network (GMN) utilizes highly sensitive low-cost CMOS video cameras which run open-source meteor detection software on Raspberry Pi computers. Currently, over 450 GMN cameras in 30 countries are deployed.

The main goal of the network is to provide long-term characterization of the radiants, flux, and size distribution of annual meteor showers and outbursts in the optical meteor mass range. The rapid 24-hour publication cycle the orbital data will enhance the public situational awareness of the near-Earth meteoroid environment. The GMN also aims to increase the number of instrumentally observed meteorite falls and the transparency of data reduction methods.

A novel astrometry calibration method is presented which allows decoupling of the camera pointing from the distortion, and is used for frequent pointing calibrations through the night. Using wide-field cameras ($\ang{88}\times\ang{48}$) with a limiting stellar magnitude of $+6.0 \pm 0.5$ at 25 frames per second, over 220,000 precise meteoroid orbits were collected since December 2018 until June 2021. 

The median radiant precision of all computed trajectories is \ang{0.47}, \ang{0.32} for $\sim20\%$ of meteors which were observed from 4+ stations, a precision sufficient to measure physical dispersions of meteor showers. All non-daytime annual established meteor showers were observed during that time, including five outbursts. An analysis of a meteorite-dropping fireball is presented which showed visible wake, fragmentation details, and several discernible fragments. It had spatial trajectory fit errors of only \SI{\sim40}{\metre}, which translated into the estimated radiant and velocity errors of 3 arc minutes and tens of meters per second.
\end{abstract}




\section{Introduction} \label{sec:introduction}

Measuring the activity and variability of meteor showers has been a long term goal of meteor science. Meteor showers provide a direct link in many cases to known parent bodies \citep{vaubaillon2019parent}. Characterization of the absolute flux, variability and particle distribution within streams can therefore provide unique insights into parent body physical properties and evolution \citep[e.g.][]{egal2020halleyids}.

Recording meteor shower activity variations and physical characteristics such as radiants, particle size distributions and orbits is difficult. In particular, flux measurements require persistent, global observations and necessarily cannot be done from a single location. Some of the earliest success in near continuous, global monitoring of meteor shower activity were performed in the 1980s by visual observers distributed across many longitudes during the 1988 Perseid meteor shower \citep{roggemans1989perseid}. 

While visual records of meteor showers have been shown to be valuable, they suffer some drawbacks. Visually, only the strongest showers have reliable activity profiles as background contamination becomes a problem \citep{rendtel2020handbook}. Moreover, visual observations cannot be used to compute individual radiants, orbits or examine ablation characteristics of shower meteors (begin/end height, deceleration, lightcurves) which may be used to infer physical structure of meteoroids \citep{borovicka2019physical}. Optical instruments, however, can measure these quantities.

Historically, optical networks have had a regional focus, often with the goal of recording meteorite producing fireballs \citep{halliday1971photographic}. The lower sensitivity and small number of stations on a global scale have, until recently, prevented such networks from becoming a mainstay for monitoring meteor showers. Since the 1990s, the advent of sensitive and inexpensive digital and video cameras has revolutionized optical meteor astronomy \citep{koten2019meteors}. As a consequence, numerous dedicated regional video networks have been established \citep[e.g.][]{weryk2007southern, sonotaco2009meteor, jenniskens2011cams, molau2013status, vida2014cmn_adapt}. As digital video cameras have become widely used in distributed meteor networks they have also become widely adopted by many amateur and professional observers. This is largely due to the commercial availability of cheap, easy to use and highly sensitive low-light cameras. The most common model of data collection is to have one or more cameras connected to a computer running automated meteor detection - the multi-station data is then sent to a central server where observations are paired and meteor trajectories computed. 

In the last decade, video meteor systems have matured to the point where they have made substantial contributions in a number of areas. This includes the discovery of dozens of new established meteor showers and hundreds of meteor shower candidates \citep{vida2013possible, andreic2014results, jenniskens2016cams, jenniskens2017meteor}, measuring meteor shower flux \citep{molau2014real, ehlert2020measuring}, estimating population indices of meteor showers \citep{molau2015population}, and recovery of meteorites produced by bright fireballs \citep{borovivcka2015instrumentally}. Similar networks have also been used to observe rare meteor shower outbursts and validate the existence of weak meteor showers \citep{holman2012confirmation, segon2014draconids, jenniskens2016camsconfirmation, sato2017detection}. 

While the foregoing networks have largely emerged out of regional efforts, to achieve persistent global sky surveillance both hemispheres and a wide range of longitudes must be instrumented. This ensures that unique short-duration meteor shower outbursts are not missed due to timing (e.g. during the day for some observers) or as a result of poor local weather conditions. For example, no published optical observations of the unexpected 2012 Draconid meteor storm (ZHR $\sim 9000$ at radar sizes) exist because the peak of activity occurred in darkness over central Asia where no meteor cameras were operating at the time \citep{ye2014unexpected}.

A similar challenge exists in trying to record meteorite-producing fireballs where the goal is to provide spatial context in the solar system for recovered samples. Annually, about 10 fireballs drop at least \SI{300}{\gram} of meteorites over an area of 0.5 million square kilometers, the size of a typical current regional network -  e.g. California, France, or Spain \citep{halliday1989flux, bland1996flux}. As fireballs can only be well observed at night and $50\%$ of the planet is covered by clouds at any given time \citep[global average;][]{stubenrauch2013assessment}, only two to three such meteorite dropping fireballs are expected to be optically observed. The number of actual meteorite recoveries will further be reduced due to inhospitable field terrain. 

An initiative aimed at addressing the problem of the small numbers of instrumentally recorded meteorite producing fireballs is the Global Fireball Observatory (GFO) project \citep{devillepoix2020global}. This effort grew out of the regional Desert Fireball Network \citep{howie2017build}. The GFO aims to rapidly expand the rate of meteorite recovery by covering  $2\%$ of the Earth's surface by the early 2020s with modern high resolution DSLR cameras. When complete, the GFO expects to record $\sim50$ meteorite falls a year.

Here we describe the development and architecture of the Global Meteor Network (GMN), an analog to the GFO, but focused on fainter meteors. The broad goal of the GMN is to greatly expand the atmospheric coverage of video networks to a global scale, permitting continuous optical observations of meteors, using a fully open source software and data model. The intent is to make automated, reproducible detections and all associated measurements of individual meteors immediately available for analysis by all in a fully transparent software pipeline. 

Presently, most video meteor networks are either national or regional, with some efforts to merge data sets from different networks \citep[e.g.][]{kornovs2014edmond}. Networks with the largest coverage consist of amateur astronomers or citizen scientists hosting cameras at local observatories or their private residences \citep[e.g.][]{jenniskens2011cams, molau2013status}. Additionally, complete data from these networks are often neither publicly available in near real-time nor are the original data readily available for independent reductions and quality checks. The Cameras for Allsky Meteor Surveillance (CAMS) network is the closest to having a global presence, with stations on all continents except Antarctica \citep{bruzzone2020comparative}. 



Most video networks also store data in mutually incompatible formats, the data calibration procedures are not transparent or well documented and data are reduced using proprietary software. Furthermore, the data are usually published years after being collected. This leads to uncertainty in data quality and makes reproduciblity challenging. Finally, the lags in publication of the data negates its use for real-time near-Earth meteoroid environment situational awareness. High quality data sets that do exist are usually composed of manually calibrated and reduced data \citep[e.g.][]{borovivcka2005survey, campbell2015population}, but manual labour is expensive, severely limits the number of reduced meteors, and introduces subjective biases.

Finally, most existing meteor network  hardware is relatively expensive and specialized from an amateur astronomer perspective. The cost of deploying a single system is currently close to \$1000 USD, including a camera and a computer. This is prohibitive for citizen scientists from low-income countries where sky coverage is particularly desirable. Our goal is to lower the cost of entry into this domain by a factor of 5 to 10 for a single-camera system. 

A major motivator for establishment of the Global Meteor Network (GMN) project is to build a distributed meteor network which is treated as a fully automated decentralized science instrument. In its final mature form, we envision GMN  observing the night sky every night of the year from as many locations around the world as possible using citizen scientists as hosts. To achieve this goal, the installation cost and day-to-day operations must be accessible to an average amateur astronomer. GMN data quality should mimic as closely as possible high-precision manual reductions through implementation of frequent automated astrometric and photometric recalibrations, and by applying rigorous filtering of the final data products. All filtering, calibrations and analysis procedures are based on an open source software architecture and are to be documented. Sufficient Level 0 data products are retained so anyone can fully reprocess detections either with their own software or using manual reduction tools provided with the GMN software suite.

The main science goals of the Global Meteor Network are:
\begin{enumerate}
    \item To provide the public with near real-time awareness of the near-Earth meteoroid environment in the optical meteoroid mass range (mm-sizes and larger) by publishing orbits of meteors within 24 hours of their observation. A yearly average of 1000 orbits/day are set as an initial operational goal, bringing the GMN close to the orbit numbers observed by meteor radars and therefore permitting automated statistical detections of meteor showers \citep{jones2005canadian, Brown2010b}.
    \item To create a continuous record of annual meteor showers and meteor shower outbursts for future research. In particular, the flux, mass indices, and orbits of shower meteors will be computed with each trajectory and the orbit will have an associated covariance matrix.
    \item To provide high-resolution observations of meteorite-producing fireballs with the aim of increasing the number of meteorites with known orbits (only $\sim40$ published orbits up until mid-2021\footnote{List of meteorite orbits: \url{http://www.meteoriteorbits.info/}}) to improve number statistics to constrain meteorite source regions in the Solar System. An additional by-product of GMN high-quality observations is the prospect of recording the formation of fireball wake and details of fragmentation which can be used to estimate the physical properties of large meteoroids \citep{borovivcka2020two, shrbeny2020fireball}.
\end{enumerate}

\subsection{Concept of Operations}

The current model used by most professional video meteor networks is to have a team which applies for funding, builds and distributes meteor camera systems, and handles long-term maintenance of the network, with station operators in passive support roles. In contrast, the Global Meteor Network philosophy is to reduce the time and money spent by scientists on building hardware and on day-to-day operations. GMN provides a blueprint and a list of parts for a meteor camera so anyone can build one. Additionally, building has been outsourced to a private company 
which offers compatible plug and play systems for purchase at low cost. Interested individuals then build and deploy the systems, which are then automatically connected to the main GMN server which collects, processes, correlates and quickly publishes all daily GMN detections.

\begin{figure}
  \includegraphics[width=\linewidth]{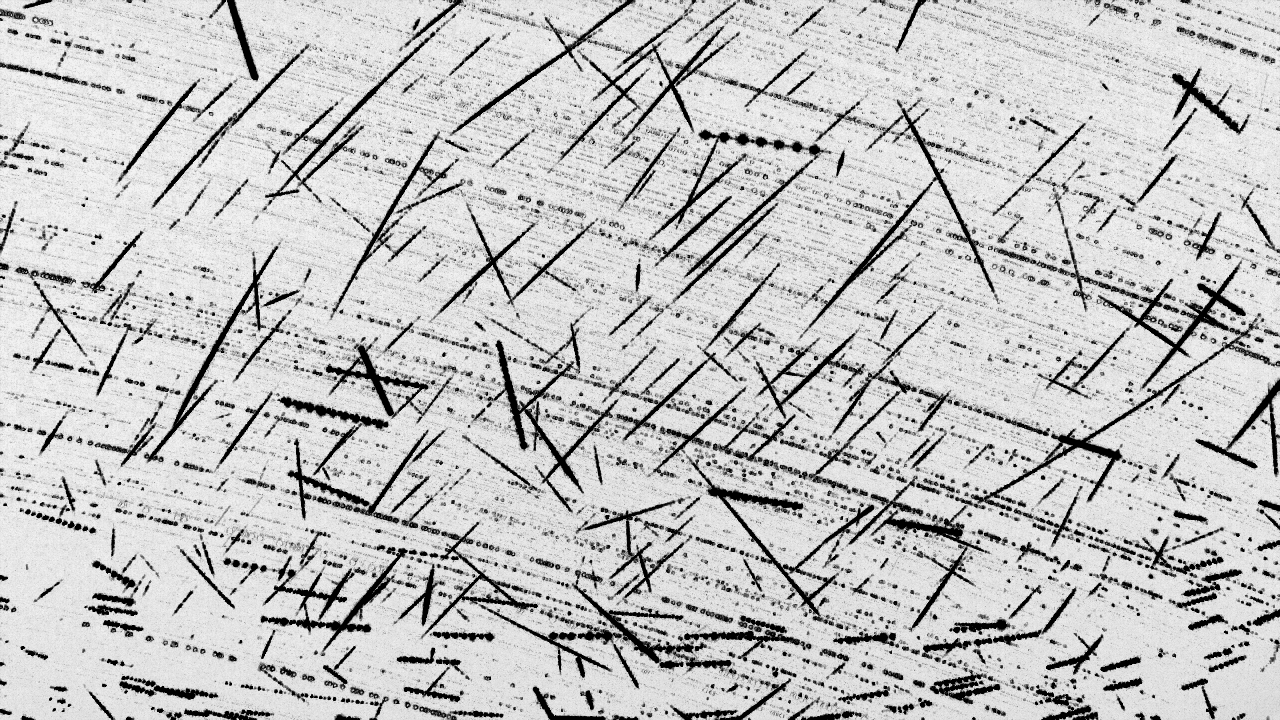}
  \caption{A full night of stacked images showing maximum pixel values during meteor detections. Over 300 meteors are visible in this stack from Oct 9, 2018 during the Draconid outburst as imaged by the HR0010 camera in Croatia (color-inverted). The slanted dashed lines are bright stars which drift through the field over the course of the night relative to the fixed camera. The field of view is $\ang{88} \times \ang{48}$.}
  \label{fig:2018draconids}
\end{figure}

In the most common configuration, the imaging part of a GMN system consists of a low-light Internet Protocol (IP) camera board ($\sim$\$30 USD), a \SI{3.6}{\milli \metre} f/0.95 CCTV lens ($\sim$\$10 USD), and an aluminium housing ($\sim$\$20 USD). The most expensive part of a meteor station, the personal computer, is replaced by a Raspberry Pi (RPi) single-board computer ($\sim$\$35 USD) in the GMN. Together with other accessories, the total cost of parts for one meteor station is around \$200 USD (circa 2021). The stations run Raspberry Pi Meteor Station (RMS) open-source software\footnote{RMS software library: \url{https://github.com/CroatianMeteorNetwork/RMS}} that is simple to configure and requires little to no user attention on a day-to-day basis. This hardware configuration achieves a limiting stellar magnitude of $+6 \pm 0.5^M$ and 10-100 meteor detections on a typical night. This rate may surpass thousands during the peak of the Geminid shower \citep{vida2019overview}. As an example, Figure \ref{fig:2018draconids} shows a co-added image of the 2018 Draconid outburst \citep{egal2018draconid} observed by a GMN camera in Croatia \citep{vida20182018}.

Recognizing that amateur astronomers are an invaluable resource as they can offer their time and enthusiasm to operate and invest a small amount of money towards a meteor camera, the GMN is built around amateur participation. In this way, GMN has outsourced funding, maintenance, and deployment to private parties. The GMN project opened the development of software to contributors from the general public, but still retains control over final software changes, ensuring a stable, common code base. 

As many amateur astronomers tend to install meteor cameras close to their residence, and due to the modular design of the system, maintenance is simple. No component of the system costs more than \$50 USD and most parts can be sourced locally. The network is loosely organized on the global scale and divided into national or regional sub-networks. Some stations consist of lone camera operators who manage several cameras and choose not to closely interact with local networks.

Operationally, once a camera is deployed, the station operator contacts the network coordinator who assigns a unique station code. The code consists of an ISO-2 country code followed by four alphanumeric characters, e.g. \texttt{US001A}. Next, the network coordinator creates an astrometric calibration plate which the operator uploads to their station. After the network coordinator inspects the quality of the data and ensures that the automated recalibration procedure works well (described in detail in Section \ref{subsec:auto_recalib}), the camera is added into the automated trajectory estimation pipeline. Finally, meteor observations from multiple stations are automatically correlated. Associated meteor trajectories and additional metadata are typically available on the Global Meteor Network's website\footnote{Global Meteor Network data: \url{globalmeteornetwork.org/data/}} within 24 hours of when the meteor was observed.

\subsection{Overview of current status}

\begin{figure}
  \includegraphics[width=\linewidth]{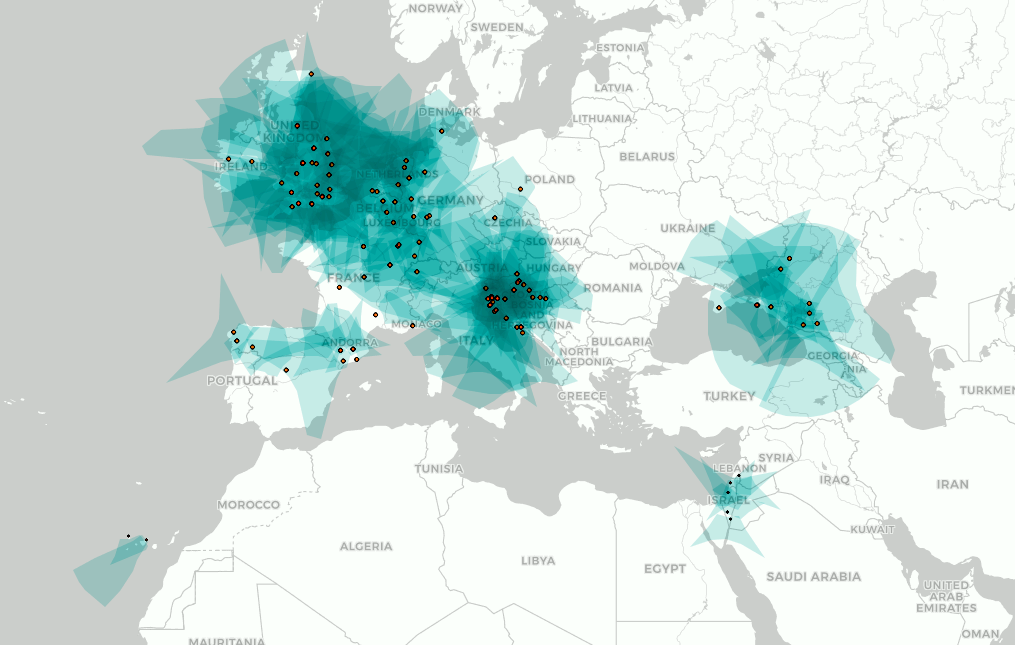}  \hfill
  \includegraphics[width=\linewidth]{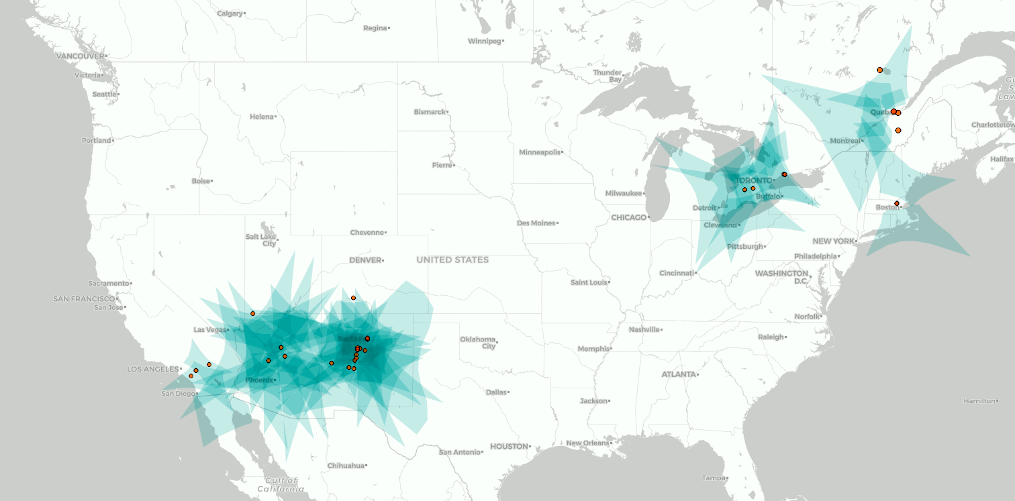}
  \caption{Field of view coverage (light blue) projected to a height of \SI{100}{\kilo \metre} for currently operational GMN cameras (dots) in Europe (top) and North America (bottom). Taken from the meteor map website\protect\footnotemark.}
  \label{fig:gmn_coverage}
\end{figure}

The first GMN systems were deployed in the spring of 2018. Since then, the network has experienced a rapid expansion, with the number of cameras as of mid-2021 in excess of 450 in 30 countries. Figure \ref{fig:gmn_coverage} shows fields of view of currently operational GMN cameras in Europe and North America (a few cameras in Brazil, Australia, and New Zealand are not shown).

\footnotetext{Meteor map website: \url{https://tammojan.github.io/meteormap/}}

The data collected by the GMN in this early period has been used in several studies, demonstrating the versatility of GMN data. \cite{moorhead2020realistic} presented a new method of measuring the radiant dispersion of meteor showers, applying the technique to the Orionids and the Perseids using GMN data. The analysis will be expanded in a future paper. The measured dispersions were tighter than reported by other authors in the past and larger than the internally estimated measurement accuracy based on the Monte Carlo procedure \citep{vida2020estimating}, suggesting that the true, physical radiant dispersion was observed for the first time. 

\cite{VidaPHD} compared high-precision manual radiant measurements of the Orionids done using the Canadian Automated Meteor Observatory's mirror tracking system \citep{weryk2013canadian, vida2020high} with GMN's automated measurements and found good agreement between the two datasets. \cite{roggemans2020hvirginids} used GMN observations of the enhanced activity of the h Virginid meteor shower in 2020 to compute the mean orbit of the shower. In that study, the GMN contributed the most observed trajectories and the GMN mean orbit was in agreement with orbits reported by other networks. Finally, \cite{egal2020halleyids} performed dynamical simulations of the Orionids and the $\eta$-Aquariids, and shown that the radiants observed by the GMN were a good match to their simulations.

We have organized this paper describing the GMN as follows. In Section \ref{sec:methods} we summarize the hardware and software used by the GMN, with a particular focus on calibration methods. We also outline what we have found empirically to be best practices for computing reliable meteor trajectories from video observations, both manually and in an automated way. In Section \ref{sec:results} we present some first preliminary scientific results. In this section we also discuss the data accuracy, the observed orbital and magnitude distribution, and observations of established meteor showers and outbursts that occurred in 2019 and 2020. Finally, we highlight GMN's capability for recording fireballs by presenting an analysis of one recent meteorite-dropping fireball.

\section{Methods} \label{sec:methods}

\subsection{Hardware Evolution}

\cite{zubovic2015advances} were the first to demonstrate that Raspberry Pi 2 single-board computers\footnote{Raspberry Pi foundation: \url{https://www.raspberrypi.org/}} are powerful enough to replace personal computers for video meteor data capture and automated processing. Their prototype meteor camera system had a total cost of only \$150 USD (in parts), an order of magnitude less than any other video meteor systems deployed at the time. As there was no freely available software which would run on Linux operating systems, \cite{vida2016open} developed novel open-source meteor, fireball, and star detection algorithms. 

Initial tests with this system were performed using analog Sony ICX-673 Charge Coupled Device (CCD) cameras and an EasyCap frame grabber \citep{vida2018first} which produced a standard definition ($720 \times 576$) interlaced video at 25 frames per second (FPS). However, this system architecture was potentially weak in terms of long term support and replacement as Sony ceased production of CCD sensors in 2015. Such sensors were ubiquitously used in video meteor work and in 2015 there were no comparable complementary metal–oxide–semiconductor (CMOS) imaging chip alternatives. 

In 2017, \cite{hankey2018} reported first results with a low-cost and highly sensitive Sony IMX290 CMOS sensor embedded in a commercial of the shelf (COTS) IP camera providing high definition ($1920 \times 1080$) video at 25 FPS. These digital cameras send the video stream via an Ethernet connection and support Power-over-Ethernet (PoE), simplifying the installation procedure. This simplifies connecting multiple cameras into a single PoE network switch while running only one cable to each camera. One downside of these new COTS cameras is that they compress the video stream using the H.264 compression standard, which is lossy and requires additional computing to be decompressed introducing additional computational overhead. 

In late 2017, a more powerful RPi 3 became available which \cite{vida2018first} tested using a newer IMX291 based COTS CMOS IP camera paired with a \SI{4}{\milli \metre} f/1.2 lens. This combination produced a field of view (FOV) of around $\ang{90} \times \ang{45}$. They demonstrated that such a setup can achieve a limiting stellar magnitude of around $+5.5^M$ at 25 FPS in light-polluted city conditions. The IMX291 has a full HD ($1920 \times 1080$ px at 25 FPS) 1/2.8" sensor with a pixel pitch of \SI{2.9}{\micro \metre} and a quantum efficiency of $\sim70\%$\footnote{IMX291 datasheet, available from the manufacturer}.  \cite{vida2018first} also showed that good photometric measurements can be performed using these CMOS sensors and that there was a noticeable increase in sensitivity compared to previous CCD-based cameras used for meteor observations. A major advantage of the Raspberry Pi is that it has a powerful Graphical Processing Unit (GPU) which supports hardware H.264 decoding. This reduces CPU overhead, although the resolution needs to be down-sampled to $1280\times720$ px to ensure real-time processing. In addition to the IMX291, cameras based on IMX307 and IMX385 were successfully tested and are in regular use, albeit in smaller numbers. We also note that the RMS software supports any camera that can be opened as a video device in the operating system.

In contrast to analog CCD cameras, which provide interleaved video signals, the IMX291 has a rolling shutter. In analog interleaved video every other row is read out at the same FPS but offset in time by half the FPS rate. This half frame spatial resolution is termed a field. Therefore the fields per second rate is twice the frames per second rate. For an IMX291 rolling shutter the sensor continuously reads out video frames from the top of the frame to the bottom such that the top and the bottom of video frames are not phase synchronised. If meteor positions are measured using such video, \cite{kukic2018correction} have shown that the rolling shutter effect introduces an angular velocity bias of up to several percent for fast meteors observed using moderate field of view systems. They developed a simple time correction which mitigates this effect which is implemented within the GMN software library.

\begin{figure}
  \includegraphics[width=\linewidth]{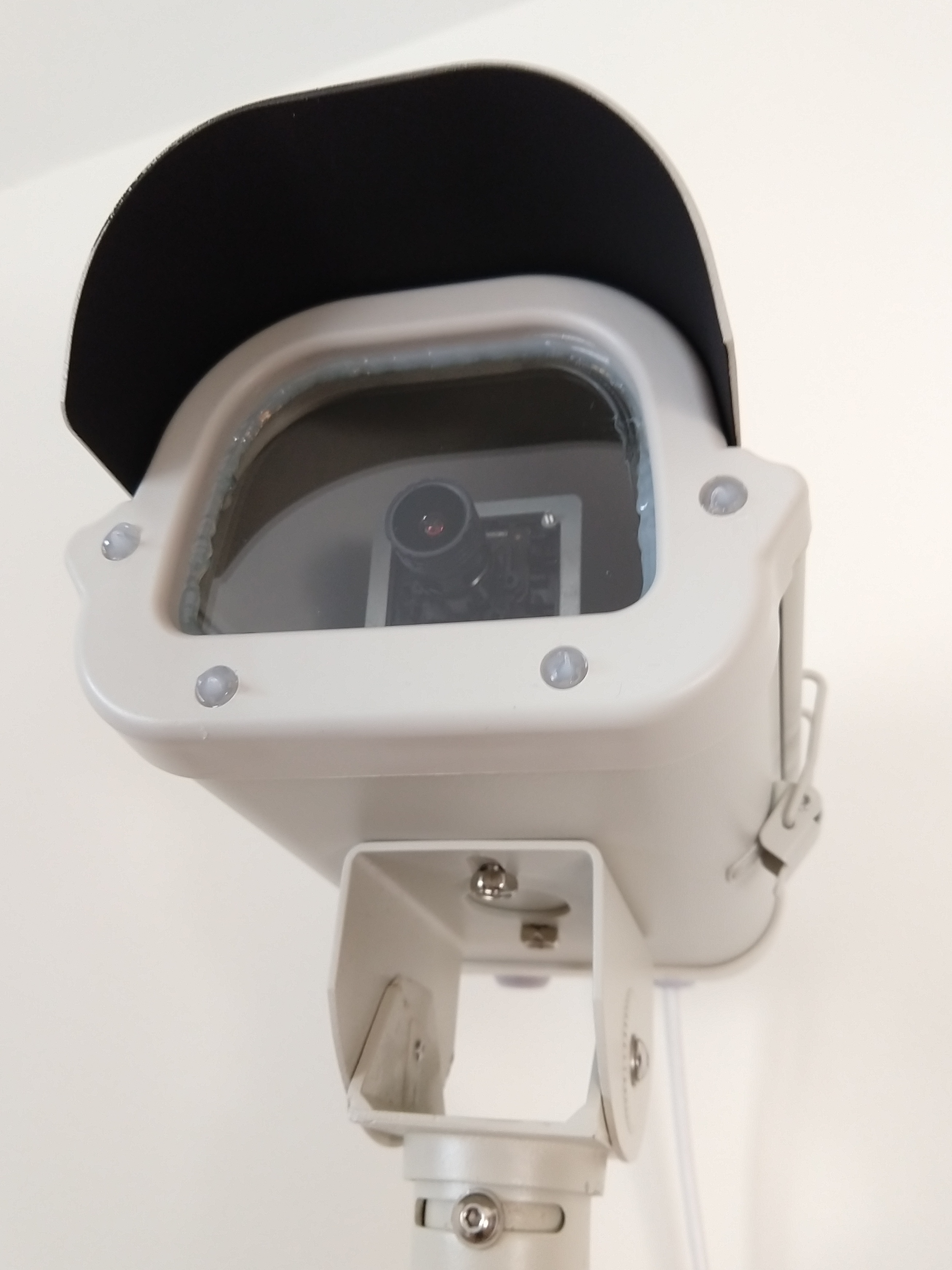} 
  \caption{An assembled GMN camera system. The housing contains the lens and the sensor board (fan and heater optional).}
  \label{fig:camera}
\end{figure}

\begin{figure}
  \includegraphics[width=\linewidth]{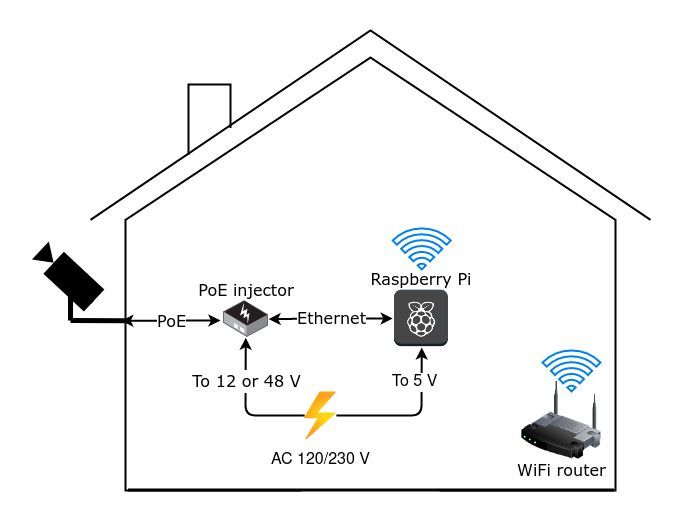}
  \caption{Diagram showing an installation of a GMN system. The Raspberry Pi is not kept inside the camera housing but inside the building to shield it from the elements. The PoE cable may be tens of meters long and carries both the power and the digital video signal. 
  }
  \label{fig:installation}
\end{figure}

Figure \ref{fig:camera} shows a typical Global Meteor Network camera and enclosure (circa mid-2021) while Figure \ref{fig:installation} depicts a common  installation arrangement. As of mid-2021, Raspberry Pi 3 and 4 are the most commonly used boards. The Raspberry Pi is kept inside a building because it would overheat during the summer if kept in the camera housing, and there is a risk of water damage if the housing is not completely waterproofed.

The Raspberry Pis are equipped with an external real-time clock module which keeps the correct time if it reboots. The time is synchronized using the Network Time Protocol (NTP) protocol, which is able to keep the correct time within at least one second. Video frames are first encoded on the camera board, travel as transmission control protocol (TCP) packets via Ethernet, and are finally decoded on the Pi's GPU. As a result, there is a fixed delay from capture to recording of \SI{\sim2.2}{\second} which is taken into account during the trajectory estimation procedure.

\begin{figure*}
  \includegraphics[width=\linewidth]{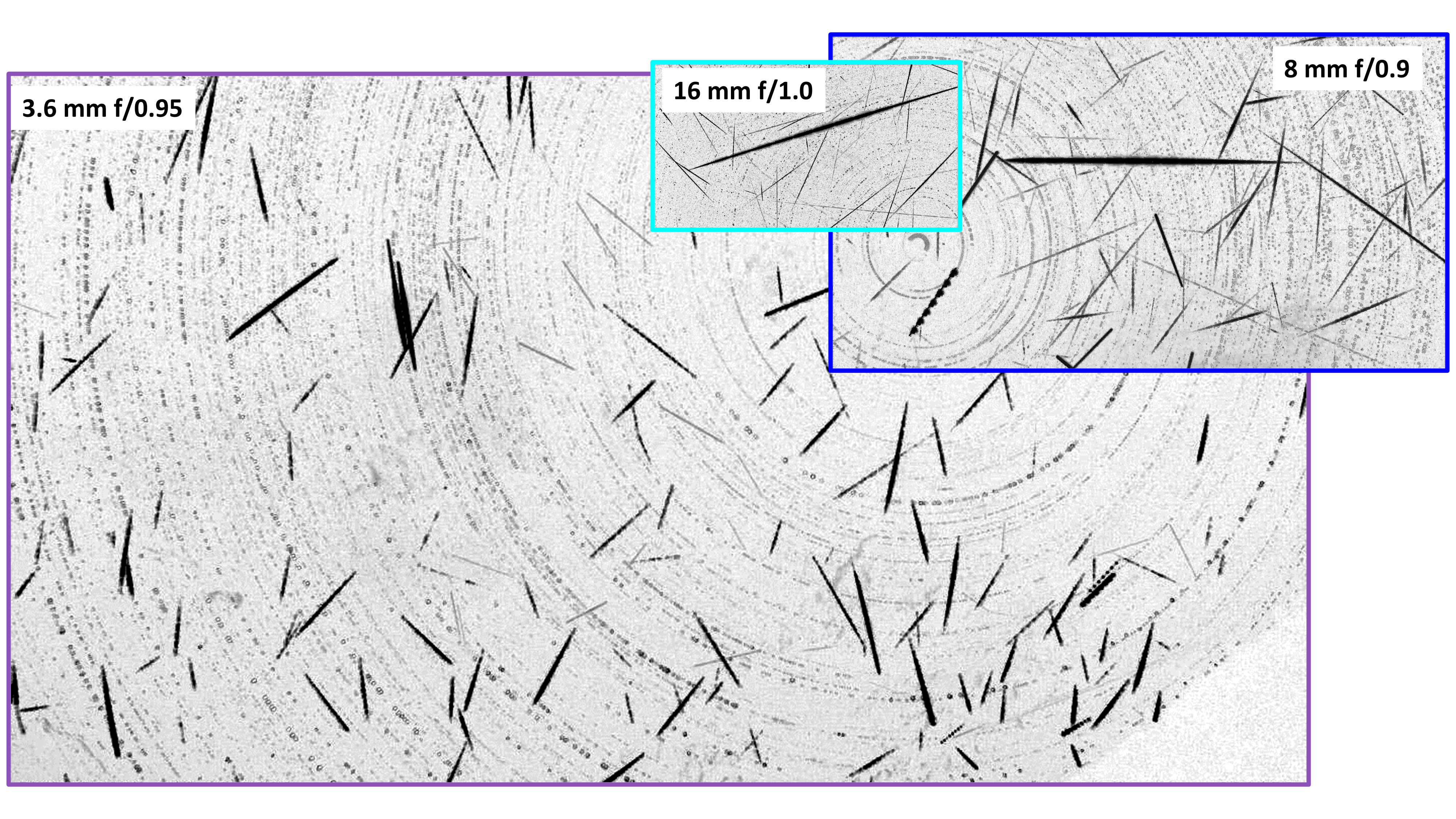}
  \caption{A comparison of the fields of view of lenses used by the GMN (color-inverted). The \SI{16}{\milli \metre} lens provides a 4x better astrometric precision than the \SI{3.6}{\milli \metre} lens, but it only covers $\sim5\%$ of that lens' field of view.}
  \label{fig:fov_comparision}
\end{figure*}

The GMN standard cameras typically use three types of lenses with varying fields of views (see Table \ref{tab:lens_comparison}). A FOV comparison is shown in Figure \ref{fig:fov_comparision}. The \SI{3.6}{\milli \metre} f/0.95 lens affords a field of view of $\ang{88} \times \ang{48}$ and is mostly used in dark locations. The \SI{8}{\milli \metre} f/0.9 lens is used in heavily light polluted areas and has a field of view of $\ang{40} \times \ang{22}$. Finally, the \SI{16}{\milli \metre} f/1.0 lens is useful for measuring very precise meteor trajectories (astrometric precision of $\sim 0.1$ arc minutes) and has a FOV of $\ang{20} \times \ang{11}$.  A dedicated network of twenty three \SI{16}{\milli \metre} systems is deployed in western Croatia.

\begin{table*}
\caption{Comparison of lens fields of view and limiting stellar magnitudes with the IMX291 1/2.9" sensor. An image resolution of $1280\times720$ pixels is assumed. Most of these lenses can be purchased for \$10 - \$20 USD.}
\label{tab:lens_comparison}
\begin{tabular}{lllll}
Focal length & f-number  & FOV                        & Stellar limiting magnitude  & Pixel scale (arcmin/px)\\
\hline\hline 
\SI{3.6}{\milli \metre} & f/0.95 & $\ang{88} \times \ang{48}$ & +6.5 (ideal), +5.5 (city) & 4.1 \\
\SI{6}{\milli \metre} & f/0.95   & $\ang{53} \times \ang{30}$ & +7.0 (ideal), +6.0 (city) & 2.5 \\    
\SI{8}{\milli \metre} & f/0.9    & $\ang{40} \times \ang{22}$ & +7.5 (ideal), +6.5 (city) & 1.9 \\
\SI{16}{\milli \metre} & f/1.0   & $\ang{20} \times \ang{11}$ & +8.0 (ideal), +7.0 (city) & 0.94 \\
\SI{25}{\milli \metre} & f/1.2   & $\ang{13} \times \ang{7}$ & +10.0 & 0.60
\end{tabular}
\end{table*}

\subsection{Data acquisition and meteor detection}

\begin{figure*}
  \includegraphics[width=\linewidth]{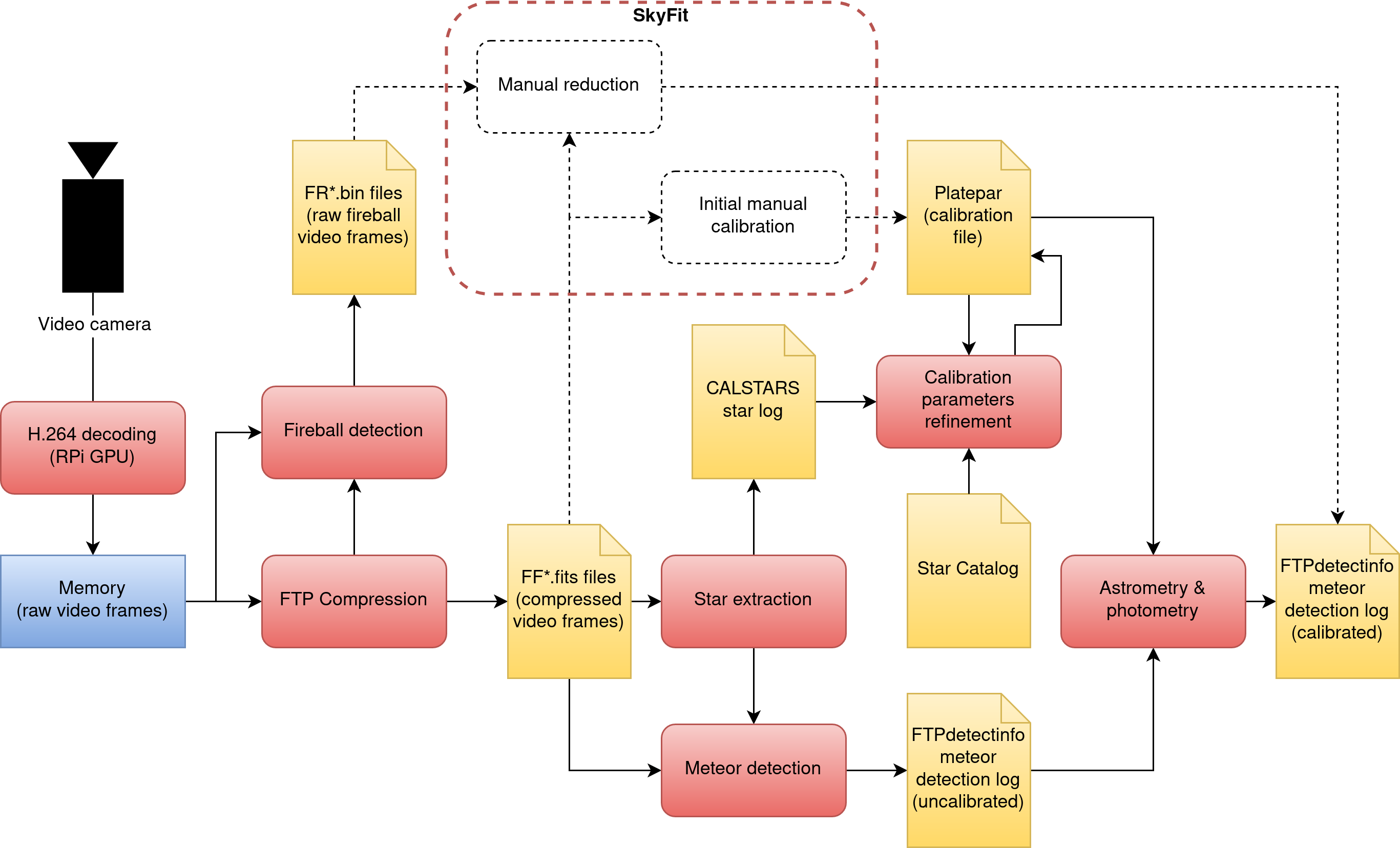}
  \caption{A schematic of the RMS data flow and automated calibration procedure. Solid lines denote automated steps, while dashed lines indicate manual steps. Red rectangles identify algorithms, and yellow rectangles represent data files.}
  \label{fig:rms_pipeline}
\end{figure*}

Figure \ref{fig:rms_pipeline} is a flowchart of the RMS data capture and calibration pipeline. More details can be found in \cite{vida2016open, vida2018first}.

The software automatically schedules data acquisition to begin each night. This starts 30 minutes before and ends 30 minutes after the local astronomical dusk and dawn, respectively. Video frames are compressed using the Four-frame Temporal Pixel (FTP) compression \citep{gural2011california}, and stored to disk. A real-time fireball detector stores raw frames of bright fireballs for posterior use. 

Stars and faint meteors are detected using a separate dedicated detector, and the astrometry and photometry calibration is automatically performed for each potential meteor detection. The end product of this process is a list of meteor detections, with every detection consisting of a set of time, position, and magnitude of the meteor on every video frame. The very first calibration file is created manually and then automatically updated thereafter numerous times every night. Manual reduction of events of interest may be performed if desired. The software automatically manages disk space, and the data is stored using a first-in, first-out approach where the oldest data is deleted to make room for the newest.

The storage memory limitations of the Raspberry Pi are due to its use of micro-SD cards, which are currently limited to between 64 to 256 GB at modest cost points. As a result, full raw video frames ($\sim600$ GB for an 8 hour night) are generally not stored to disk. To address this issue, the RMS software employs the FTP compression method which provides a 64:1 compression ratio. This algorithm takes a block of 256 raw video frames and compresses them into four images which are saved into one Four Frame (FF) file in the Flexible Image Transport System (FITS) format, consisting of the maximum pixel, maximum pixel frame index, average pixel, and standard deviation image, each on an independent per pixel basis.

\begin{figure*}
  \includegraphics[width=\linewidth]{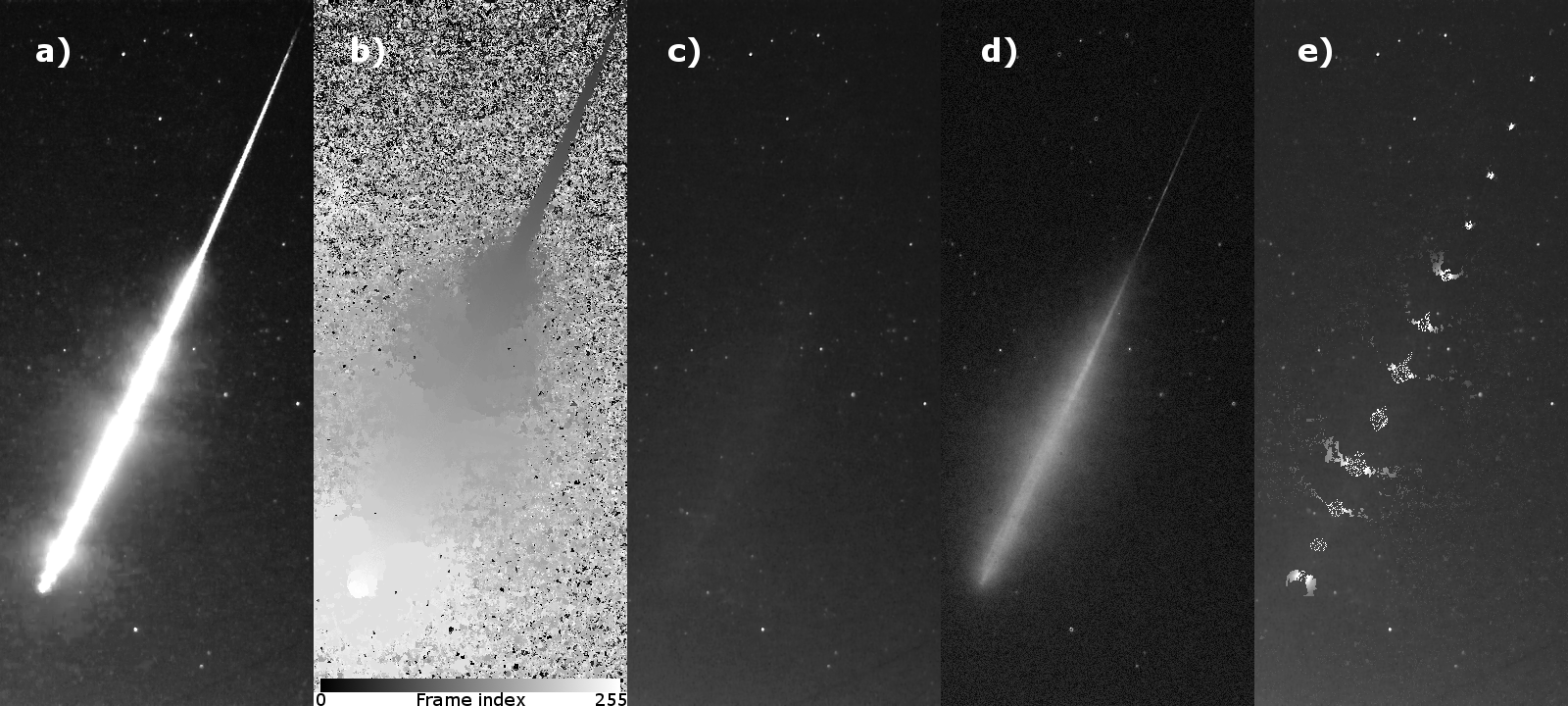}
  \caption{FTP compressed images stored in an FF file of a fireball observed on January 22, 2020 over Southwestern Ontario. a) Maximum pixel image, b) Maximum frame index image, c) Average image showing the background stars, d) Standard deviation image, e) Co-added every tenth reconstructed video frame.}
  \label{fig:ftp_example}
\end{figure*}

In Figure \ref{fig:ftp_example} an example of each of these frames is shown. The maximum pixel image only stores the maximum value of every pixel in the block of 256 frames, as shown in inset a). To preserve the temporal information which allows reconstruction of a close approximation of the original video, the index of the video frame with the brightest pixel value is stored in the maximum frame index image. Note that the number of frames in the block is chosen to correspond to the number of discrete image levels in an 8-bit image ($2^8 = 256$), so the index can be stored as an image level. If multiple frames have the same maximum value, the frame index is chosen with a random number threshold dependent on how many times the value is repeated.

Inset b) in Figure \ref{fig:ftp_example} shows the gradient of the increasing brightness/index with the progression of the fireball towards the bottom of the image. The average and standard deviation images, shown as  insets c) and d), have the top four maximum values removed to mitigate the influence of bright fireballs on the estimate of the background. This works well for fainter meteors, but as can be seen in this case of a bright and slow-moving fireball which lasted over 120 frames, the standard deviation is contaminated by the meteor. 

The average and standard deviation frames are used by the meteor detection algorithm to determine which parts of the image contain a meteor by simply checking if the maximum value is a certain number of standard deviation above the mean. Finally, inset e) shows co-added reconstructed video frames (every tenth) of the fireball using the compressed frames. This reconstruction approach works well for faint meteors and the fainter parts of fireballs, but it creates artifacts for the brighter parts of fireballs simply because it can only store one frame index for one brightest value. This limitation was the reason a dedicated real-time fireball detector was developed \citep{vida2016open} which stores raw frames of fireballs while they are still available in memory and runs prior to the FTP compression. These raw frames can also be later used for manual data reduction (e.g. see Figure \ref{fig:manual_reduction}).

The star and meteor detection algorithms are run in parallel to data acquisition. They asynchronously load the compressed FF files from a queue and may continue to run after the capture ends for the night if processing has not been completed. The data processing typically lasts no more than a few hours after the end of data collection.

The star extractor algorithm detects stars by finding local maxima on the average pixel image and attempts to fit a 2D Gaussian point spread function (PSF) to them. The value of the Gaussian is limited to the saturation point (255 for 8-bit sensors), making it a flat-topped distribution for bright stars that saturate the sensor. In this way, a reliable fit and a rough estimate of the stars unsaturated brightness is obtained. Star candidates which are not round (the smallest ratio between X and Y standard deviations is less than 0.5), or too large ($2\sigma > 5$px), and for which the minimization procedure does not quickly converge are rejected. This method is found to be robust and also provides an estimate of the variation of the PSF across the field of view. Generally, if the \SI{3.6}{\milli \metre} lenses are well mounted and focused, the PSF is constant across the whole field of view with a full width at half maximum (FWHM) of $\sim 3$ px. Given clear skies, 50-200 stars are usually detected on every FF image, depending on the stellar limiting magnitude.

If a minimum of 20 stars are visible, indicating that the skies are at least partially clear, a dedicated faint meteor detector is run on the FF file. This algorithm attempts to detect very faint meteors and it requires more time to run than the fast fireball detector. In summary, segments of 64 frames which are reconstructed from the FF files and thresholded using a user-configurable simple image operation $\textrm{max} > \textrm{avg} + K_1 \sigma + J_1$, where values of $K_1 = 3.5$ and $J_1 = 12$ are most commonly used for IMX291 IP cameras. Next, a set of image morphological operators \citep[clean, bridge, close, thin, clean; see ][]{dougherty2003hands} reduce potential meteors into thin lines while suppressing all background noise. A kernel-based Hough transform \citep{fernandes2008real} is applied to find lines of pixels which may be meteor trails in the image, and line candidates are fed into a 3D line segment finding algorithm (where dimensions are X, Y, and frame number) to check for temporal propagation \citep{vida2016open}.

If the detection covers at least 4 consecutive video frames and has an angular velocity within the range of  \SIrange{2}{51}{\degree \per \second} \citep[expected meteor range; ][]{gural1999rigorous}, centroids and sums of pixel intensity for every frame are computed within a $\sim30$ px wide rectangular segment with a length which varies with the angular velocity such that it encompasses the whole meteor on every frame. Finally, the rolling shutter correction is applied if necessary. For more algorithmic details, see \cite{vida2016open}.

\subsection{Astrometric and photometric calibration}

To produce useful astrometry and photometry for analysis of meteor measurements, transformations of the pixel detections and intensities from the focal plane to celestial coordinates and calibrated intensities on the sky must be performed.

The transformation between image (x,y) coordinates and celestial coordinates ($\alpha$,$\delta$ or equivalent) is performed using an astrometric plate. The GMN employs a simple plane projection approach \citep{smart1977textbook} which uses a novel radial distortion model with odd terms up to the ninth order to construct an astrometric plate. This approach can compensate for lens asymmetry and non-square pixels using a total of only nine coefficients. Alternatively, a third order polynomial in (x,y) together with radial terms can be used. This approach results in an astrometric plate with 12 distortion coefficients per image axis \citep{vida2018first}. A detailed description of the method is given in Appendix \ref{appendix:astrometric_calibration}.

The camera pointing (corresponding to the centre of the optical axis) is defined in equatorial coordinates at a reference time using four affine transform parameters: right ascension, declination, position angle of the Y image axis, and plate scale. If the radial distortion is used, one polynomial describes the forward, and one the reverse mapping. If the polynomial distortion is used, four sets of polynomials describe the distortion - two sets for forward mapping (image to sky) and a second pair of sets for reverse mapping (sky to image), one set for every image axis X and Y.

\begin{figure*}
  \includegraphics[width=\linewidth]{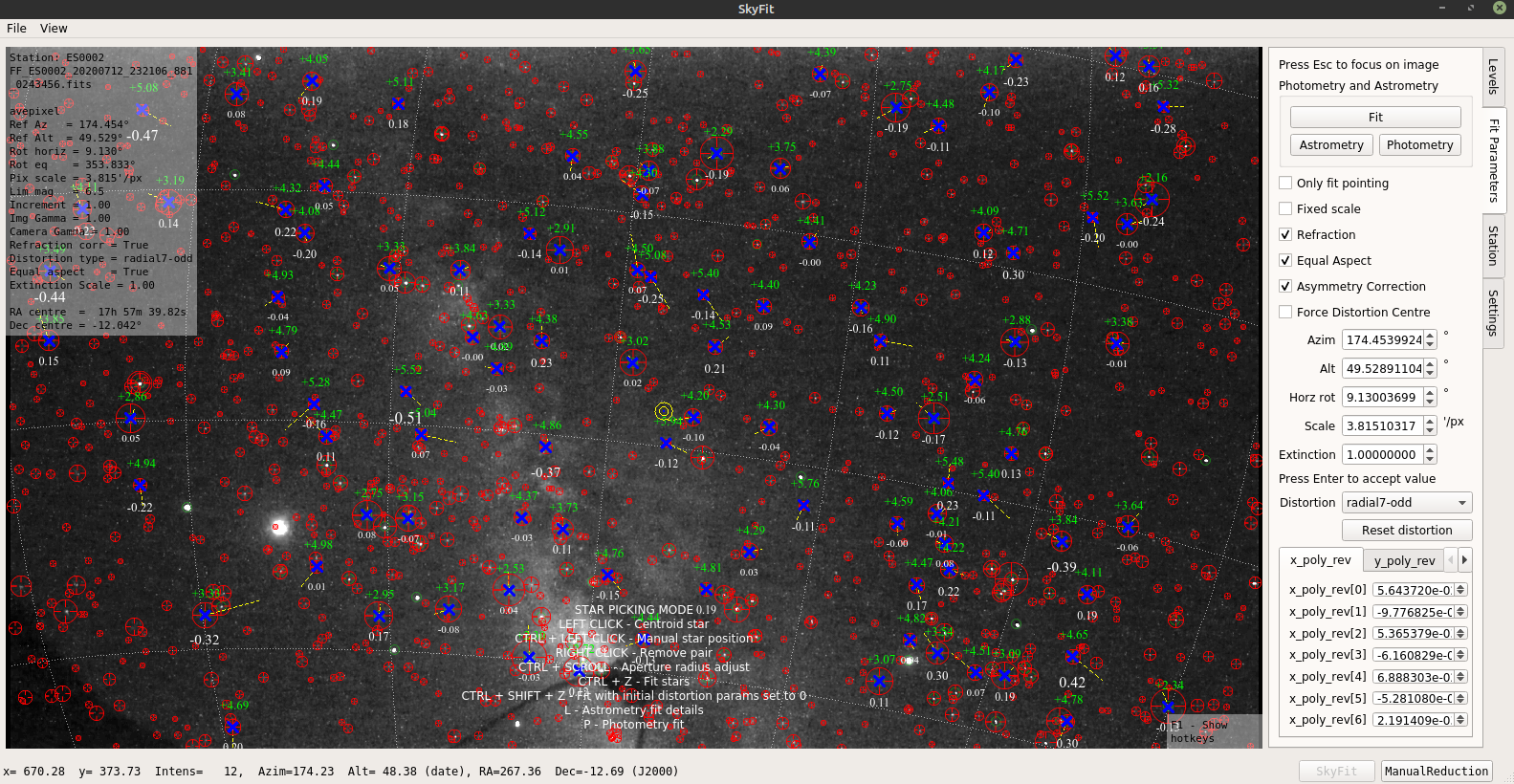}
  \caption{An example of the \texttt{SkyFit} graphical user interface with an overlaid equatorial grid showing a manual astrometric plate produced for a camera at Roque de Los Muchacos, La Palma, Spain (the MAGIC telescope is visible in the bottom of the FOV). The effective stellar limiting magnitude of this 256 frame average is $+6.7^M$. The red markers are catalog stars and their size indicates their magnitude (bigger means brighter). Blue markers indicate paired stars. The yellow lines indicate the absolute value (times 100) and the direction of astrometric fit residuals. The green numbers are catalogue stellar magnitudes while the white numbers below each star are the photometric residual per star for the given fit. The side bar at the right gives the details of the fit, and the status bar at the bottom indicates the coordinates under the mouse cursor (currently pointing just below the middle of the image, as indicated by a yellow annulus).}
  \label{fig:skyfit}
\end{figure*}

When a new camera pointing is made, an initial astrometric calibration is preformed manually using the \texttt{SkyFit} program \citep[based on a previous version from ][]{vida2010croatian} provided in the RMS library. \texttt{SkyFit} allows calibrating and manually reducing RMS format image files, as well as the Global Fireball Observatory \citep{howie2017build} and FRIPON images \citep{colas2020fripon}, or any media format supported by the OpenCV library \citep[e.g. png, jpg, avi, mp4;][]{opencv2000}. Figure \ref{fig:skyfit} shows the \texttt{SkyFit} graphical user interface. If the RMS data is used, the calibration is done on the 256 frame average. Most commonly, the GAIA DR2 star catalog \citep{gaia2018dr2} is used, although others such as the Sky2000 and BSC5 are supported as well (see Appendix \ref{appendix:photometric_calibration}). The resulting meteor measurements are reported in the J2000 epoch, to be consistent with the star catalog.

\subsubsection{IMX291 spectral sensitivity}

\begin{figure}
  \includegraphics[width=\linewidth]{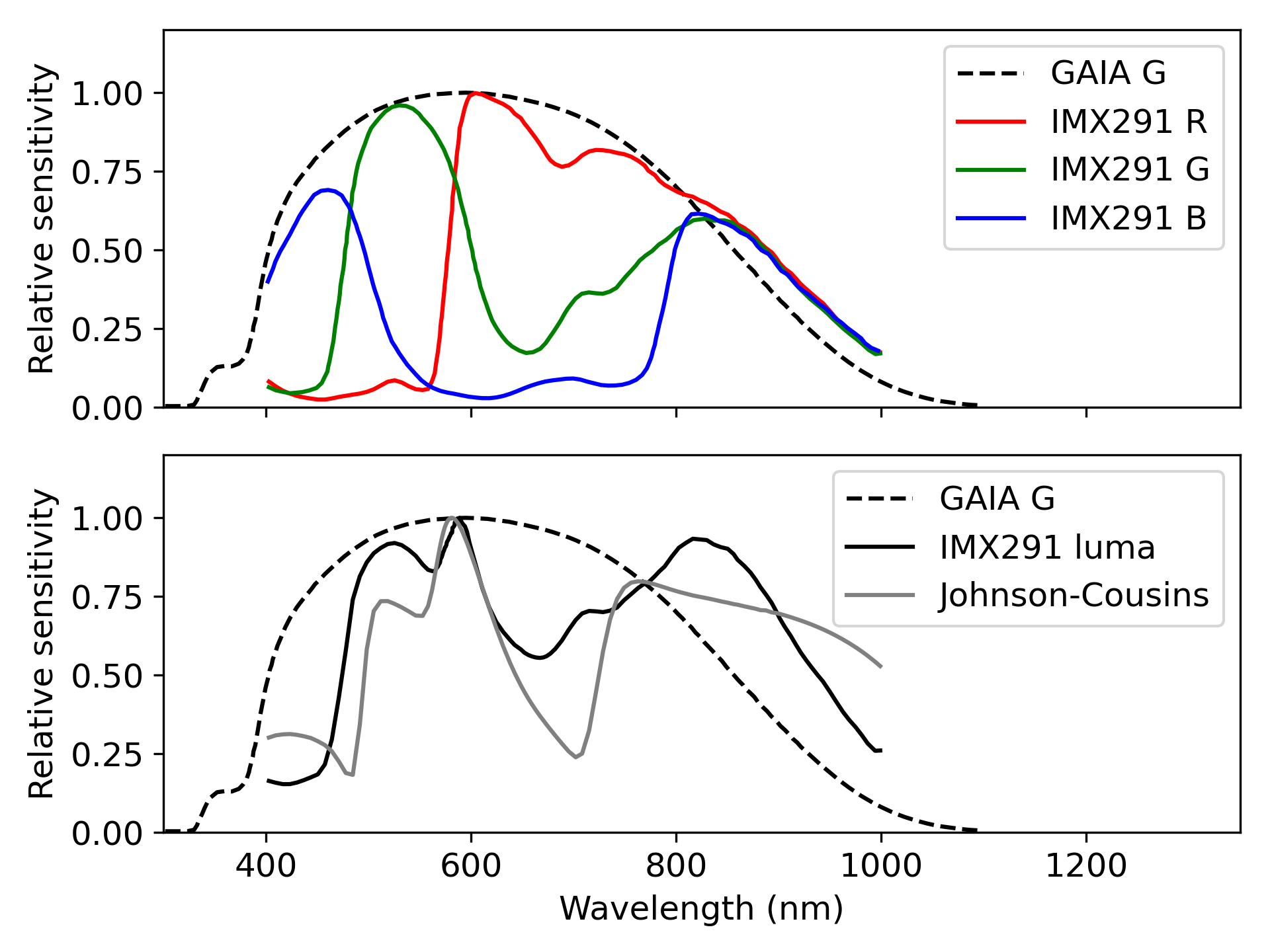}
  \caption{Top: Comparison between the GAIA G spectral passband and individual color filters of the IMX291 sensor. Bottom: IMX291 luma spectral sensitivity (after conversion to black and white) compared to the closest fitted Johnson-Cousins equivalent.}
  \label{fig:spectral_comparison}
\end{figure}

The GAIA G band (not to be confused with the green filter) has a broad spectral sensitivity, with the transmissivity sharply rising at \SI{400}{\nano \metre} and falling off at \SIrange{900}{1000}{\nano \metre} \citep{weiler2018revised}. This range roughly matches the sensitivity of the IMX291 sensor with the IR filter removed (Figure \ref{fig:spectral_comparison}) and we use the G-band as a proxy for the spectral system of the IMX291 chip.

However, the IMX291 sensors have a Bayer RGB filter and as the cameras are operated in the black and white mode, the camera computes the pixel intensity luma value as $Y = 0.299 R + 0.587 G + 0.114 B$ \citep{burger2010principles} (black line in Figure \ref{fig:spectral_comparison}). Assuming no significant influence from the front glass or the lens (which are unknown to us for these low-cost products), we can approximate the spectral sensitivity using a combination of Johnson-Cousins filters \citep{jenniskens2011cams} as $0.15 B + 0.30 V + 0.25 R + 0.30 I$ (gray line in Figure \ref{fig:spectral_comparison}). Nevertheless, we use the GAIA catalog as the spectra of meteors are essentially unknown and not all stars in the Sky2000 catalog have BVRI values. 

Faster meteors radiate $\sim20\%$ more in the near infra-red (\SI{>700}{\nano \metre}) than slower meteors due to stronger atmospheric lines \citep{segon2018infrared}.  Compared to previous Sony CCD sensors \citep[e.g.][]{jenniskens2011cams}, the IMX291 sensors have a second peak of sensitivity at around \SI{850}{\nano \metre}, which may be beneficial for detecting faster meteors relative to earlier CCDs.

\subsubsection{Manual calibration procedure}

To create an initial astrometric plate, the user first manually defines a rough pointing direction and the plate scale. There is also an option to query \texttt{astrometry.net} \citep{lang2010astrometry} for an initial astrometric solution. Then, the user manually pairs image stars to catalog stars. Clicking on a star produces a centroid and a sum of background-subtracted pixel intensities within an annulus of adjustable size. If less than 12 stars are paired, only the four affine parameters are fit, which is useful if a user only wants to readjust the pointing and the distortion is already known. For the initial fit however, at least 40 stars are paired across the whole field of view to ensure a robust estimate of the distortion coefficients. We describe the minimization procedure in detail in Appendix \ref{app:fit_algorithm}. The fit residuals for \SI{3.6}{\milli \metre} lenses and $1280 \times 720$ resolution (plate scale of $\sim 4$ '/px) is usually around 0.2 pixels and 0.8 arcmin when the radial distortion model is used, although this may vary depending on the image quality. Some manual refinement is usually needed to achieve a precise astrometric fit: centroids for stars with high fit residuals are redone with a different spatial aperture, and picks are manually forced to a particular position if the star is not isolated. 

The photometric fit is performed taking system vignetting and atmospheric extinction into account. The procedure is described in detail in Appendix \ref{appendix:photometric_calibration}. The fit is initially performed on the same set of stars selected for astrometry, but some manual refinement is usually needed during which stars with high fit residuals are removed (e.g. stars in clouds or haze), and some new stars are added. An average photometric fit error for the zero-point (intercept) of $\pm 0.15$ magnitudes is usually achieved for the typical GMN setup.

\subsection{Calibration Examples}

\begin{figure*}
  \includegraphics[width=.69\linewidth]{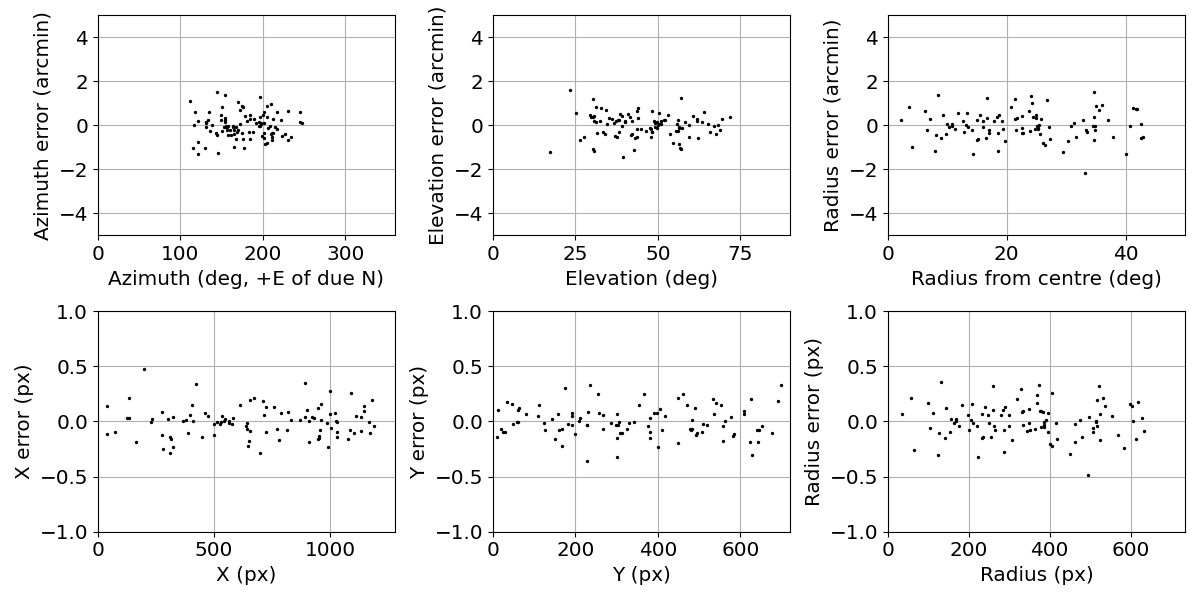}\hfill
  \includegraphics[width=.31\linewidth]{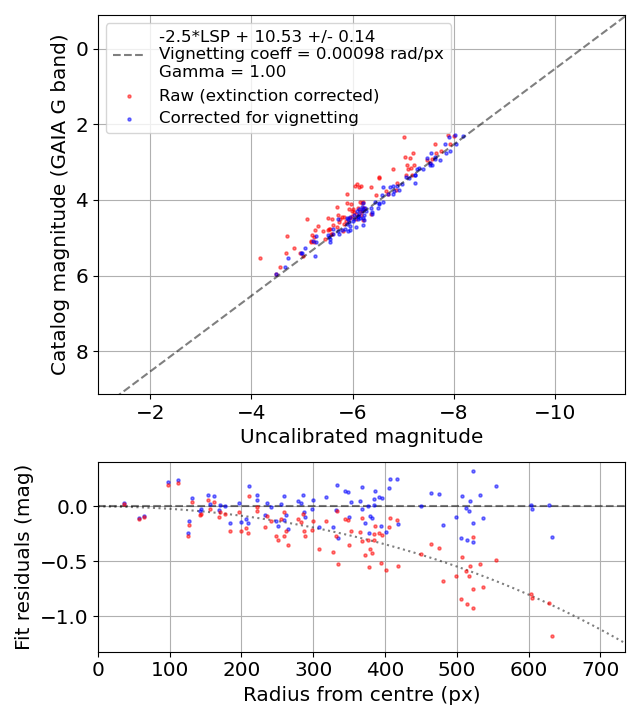} 
  \caption{Calibration details for the GMN station ES0002 with the \SI{3.6}{\milli \metre} lens. Left: Astrometric calibration residuals using the radial distortion model. Right: Photometric calibration residuals. The vignetting coefficient is given in radians per pixel.}
  \label{fig:es0002_calib}
\end{figure*}

For the purpose of illustrating the type of calibration that was used to produce the results presented in this paper, Figure \ref{fig:es0002_calib} shows details of the astrometric and photometric residuals done for a typical GMN station (shown in Figure \ref{fig:skyfit}). The astrometric fit was done using the novel radial distortion model (with odd terms up to the $7^{\textrm{th}}$ order) which was adopted by the GMN in late 2020. The average image and angular fit residuals are 0.19 px and 0.78 arc minutes and show no trends, which is what is expected for a "good" astrometric fit free of systematic errors. Prior to the development of the novel radial method in late 2020, most systems were calibrated using the polynomial method (see Appendix \ref{tab:astrometry_calib_comparison}). Using the same data set as above, the method produced average image and angular fit residuals of 0.19 px and 1.27 arc minutes, respectively. Minor trends in the residuals could have been seen.

Because the sensor is 8-bit, the effective dynamic range for this camera is from magnitude $+7.0^M$ to $+1.0^M$. For objects brighter than $+1.0^M$ the sensor will start to saturate. The photometric error in the zero-point fit is $\pm 0.13$ magnitudes, when vignetting and extinction are taken into account (see Appendix \ref{appendix:photometric_calibration}). We attempted to perform a saturation correction using the numerical simulation method of \cite{kikwaya2010model}, but it appears that the non-linearity of CMOS sensors close to saturation and the H.264 compression prevent the successful application of this approach; thus we do not apply any saturation correction to our data.

\begin{figure*}
  \includegraphics[width=.69\linewidth]{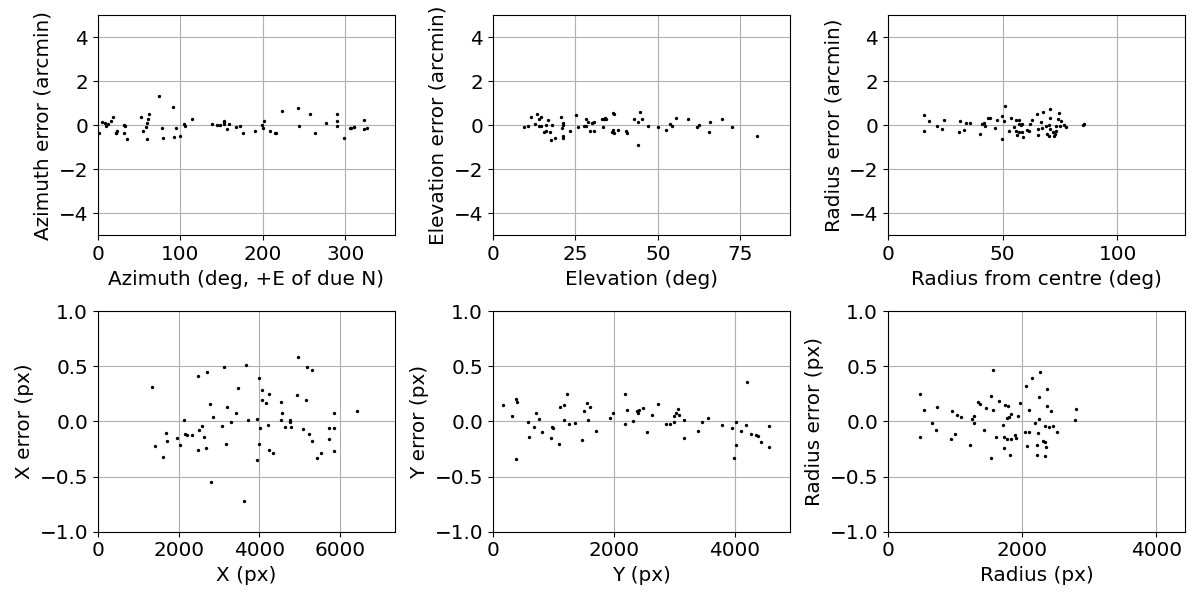}\hfill
  \includegraphics[width=.31\linewidth]{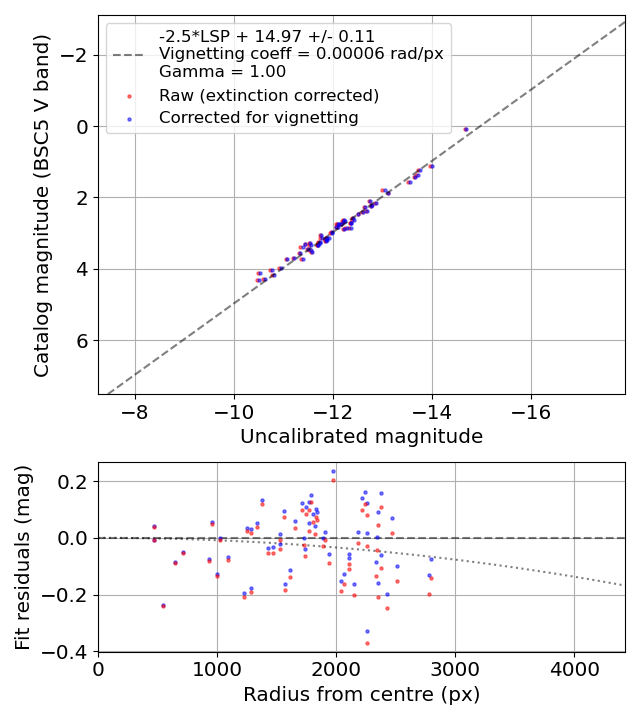} 
  \caption{Calibration details for the all-sky Global Fireball Observatory station at Tavistock, Ontario, Canada, co-located with the Canadian Meteor Orbit Radar and Canadian Automated Meteor Observatory. Left: Astrometric calibration residuals. A radial distortion with odd terms up to the $7^\mathrm{th}$ order was used and the asymmetry correction was applied. Right: Photometric calibration residuals.}
  \label{fig:dfn_calib}
\end{figure*}

\texttt{SkyFit} can also fit astrometric plates using radial distortion models for high-resolution all-sky images, such as those produced by the Global Fireball Observatory system \citep{devillepoix2020global}. Figure \ref{fig:dfn_calib} shows the calibration residuals when \texttt{SkyFit} is applied to one of these cameras. An average angular precision of 0.5 arc minutes can be achieved, limited  mainly by the deformation of the PSF at the edges of the field of view. The photometric fit errors are on the order of $\pm 0.1$ magnitudes. See Appendix \ref{appendix:astrometry_methods_comparison} for a detailed performance comparison between different methods of astrometry calibrations on various types of cameras.

\subsection{Automated recalibration} \label{subsec:auto_recalib}

During early operation of GMN, it was discovered that astrometric calibrations done on images several hours apart were often shifted by several arc minutes. By inspecting the actual and predicted positions of stars, a gradual drift in pointing was noticed. We believe this is caused by thermal expansion of the camera metal housing, mounting bracket and/or the platform or building the camera is attached to. 

To mitigate this effect, we implemented an automated recalibration routine for the camera pointing which is performed on every image with a meteor detection. Because the camera pointing in our astrometric method is independent of the distortion, and we assume that the distortion should remain unchanged (as it depends only on the optics, which is stable), only the affine parameters (reference right ascension, declination, rotation, and scale) are adjusted in this process. Although in theory the scale should remain fixed, in practice we found slight but visible changes which need to be compensated for.

\begin{figure}
  \includegraphics[width=\linewidth]{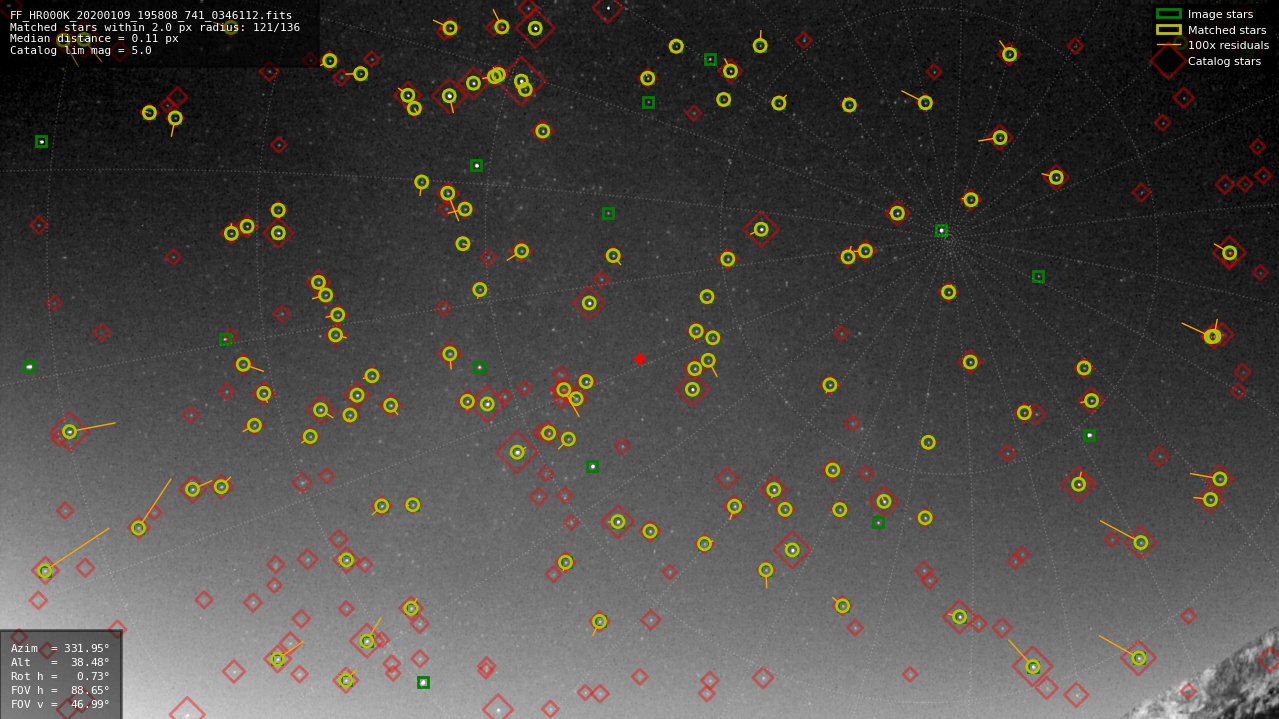}
  \caption{An example of an automated astrometry calibration generated by RMS for an image showing many matched stars in the night of January 9/10, 2020 for the station HR000K (\SI{3.6}{\milli \metre} lens). Yellow circles indicate matched image and catalog stars, with the residuals plotted as yellow lines at $100\times$ their actual value. Green boxes are detected stars and red diamonds are catalog stars that were not matched because either a bright star was missing in the GAIA DR2 catalog, or a fainter star was not picked up by the star extractor.}
  \label{fig:astrometry_calib_report}
\end{figure}

\begin{figure}
  \includegraphics[width=\linewidth]{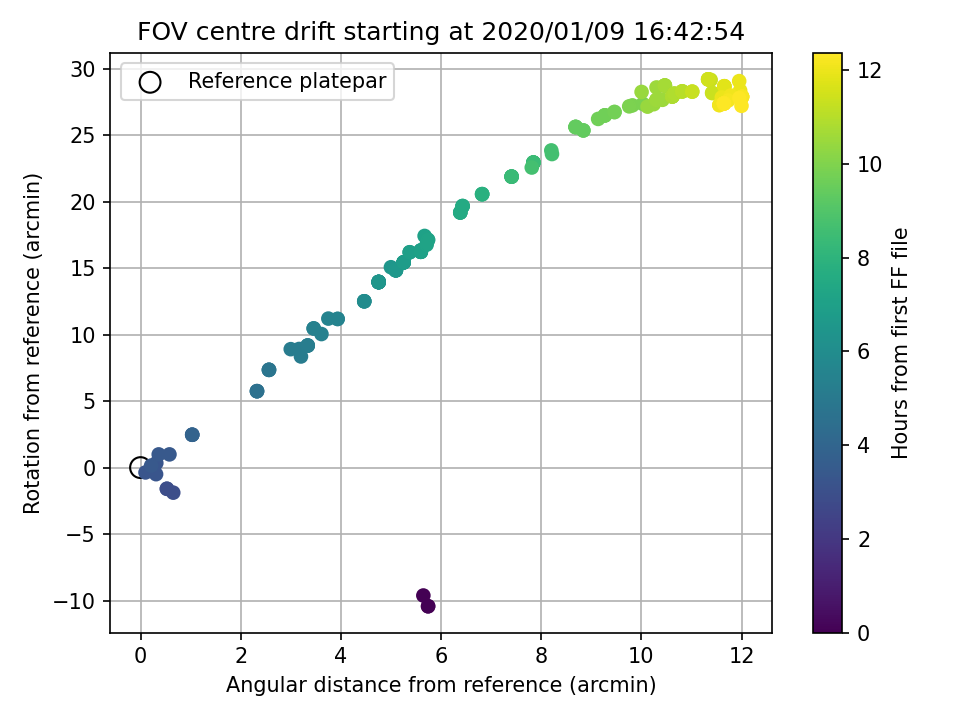}
  \caption{Pointing direction variation color coded by time since the beginning of the night of January 9, 2020 for station HR000K in Croatia.}
  \label{fig:astrometry_variation}
\end{figure}

During the recalibration procedure, the catalog stars and stars detected on the image are paired using progressively smaller match radii, from 10 to 0.5 pixels. The pointing direction is found by minimizing the match cost function for every radius while requiring at least 20 stars to be paired for the match to be considered successful. The cost function, $f$, is defined as:

\begin{equation}
    f = \frac{M^2}{\sqrt{n + 1}} \,,
\end{equation}

\noindent where $M$ is the median distance in pixels between paired stars and $n$ is the number of matched stars. The progressive shrinking of the radius ensures that a rough match is obtained at first, and later any false matches (e.g. binary stars) are excluded from the fit. A maximum median distance of $M = 0.33 \textrm{px}$ is mandated. If the value is higher even after recalibration, the procedure is considered to have failed. Figure \ref{fig:astrometry_calib_report} shows an automated recalibration on one FF file while Figure \ref{fig:astrometry_variation} shows the pointing drift over the course of a night. Shifts of up to 10 arc minutes are often observed. 

If the initial shift is larger than 10 pixels and no stars are matched, the phase correlation image alignment method based on the Fourier transform \citep{reddy1996fft} is used to estimate the translation, rotation, and scale from the initial pointing.

In this approach, a synthetic image is created with a black background and white points placed where detected stars are located. This is then fed into the algorithm, together with a synthetic image generated using the predicted positions of catalog stars. Using a fast Fourier transform and image representation in log-polar form, a cross correlation between the images is performed. This becomes simple multiplication in the Fourier space and can be performed at low computational cost. The method produces values of translation, rotation, and scale that need to be applied for the two input images to match. In practice, this method reliably works for shifts of up to 1/4 of the total image and virtually any rotation.

\begin{figure}
  \includegraphics[width=\linewidth]{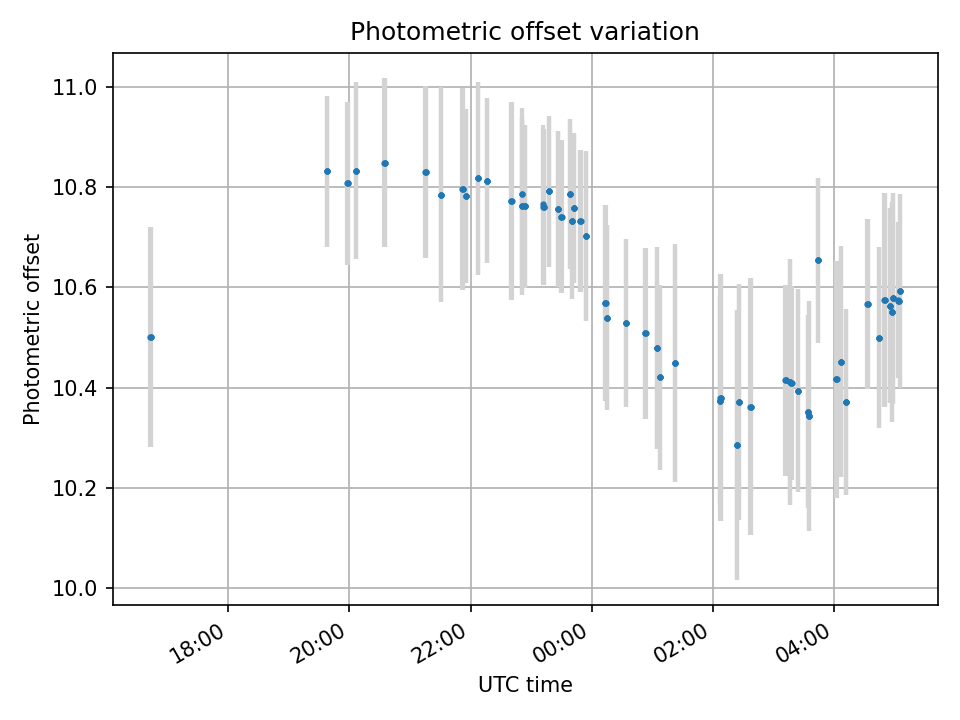}
  \caption{Variation of the photometric offset in the night of January 9, 2020 for the station HR000K in Croatia. The thin lines indicate the one sigma confidence interval of the photometric fits (around $\pm 0.15^M$). The Moon entered the field of view at 00:00 UTC.}
  \label{fig:photometry_variation}
\end{figure}

At the same cadence as the astrometric recalibration (i.e. at the time of every detected meteor), the photometric offset is also recalibrated. It uses the same set of paired stars as used for the astrometric recalibration, while keeping the vignetting coefficient fixed. Frequent photometric recalibration is necessary because the sky transparency may change rapidly due to humidity, clouds, increasing sky glow with the rising Moon, varying aerosol concentration, etc. Figure \ref{fig:photometry_variation} shows an example of how the photometric offset varies during the course of one night. In this case the offset (zero-point) dips as the Moon enters the field of view. Differences throughout the night of more than one magnitude are commonly observed when the full Moon is close to the field of view.

\subsection{Manual data reduction}

The automated RMS meteor detector produces centroids and magnitude measurements for fainter meteors, but the separate fireball detector only stores raw frames for bright events. Due to the complex morphology of fireballs (which often show long wakes, fragmentation, and flares which may saturate a large image area), fireballs are manually reduced using \texttt{SkyFit}.

\begin{figure*}
  \includegraphics[width=\linewidth]{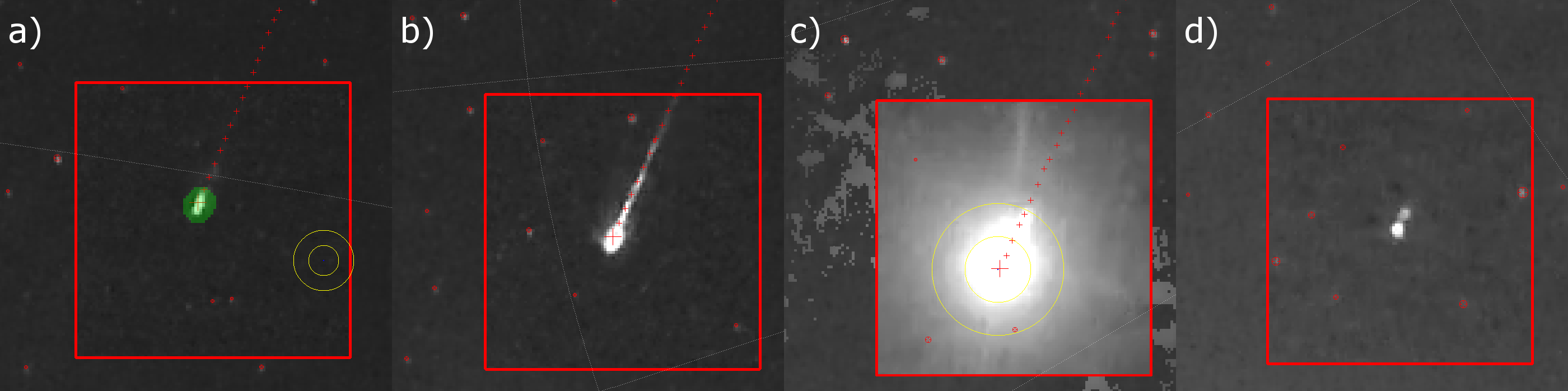}
  \caption{A mosaic of video frames of a fireball observed on January 22, 2020 over Southwestern Ontario. Red squares mark raw frame cutouts produced by the RMS fireball detector, while the background is reconstructed from an FF file. a) The early part of the fireball. Red crosses indicate pick points on the current (large cross) and previous (smaller red crosses) frames, red round markers indicate locations of stars while the green area indicates pixels that will be used for photometry. The large yellow circle with annulus represents the cursor pick location. b) Wake becomes visible. c) A flare occurs necessitating a manual pick. d) Two distinct fragments visible at the end.}
  \label{fig:manual_reduction}
\end{figure*}

When performing fireball reduction in \texttt{SkyFit}, the astrometric plate is manually re-fit using as many as possible stars in the vicinity of the fireball. Figure \ref{fig:manual_reduction} shows an example of how manual astrometric and  photometric picks are done on every frame. For the initial frames, where a fireball often does not saturate, centroiding can be used to make precise picks, and the magnitude can be measured by manually coloring the pixels which the user decides belong to the meteor. 

As the fireball penetrates deeper and develops a wake, the picking becomes more difficult and sometimes has to be manually adjusted so the center of the head is pinpointed. As the fireball becomes brighter and starts producing flares, the picking mostly relies on adjusting the size of the annulus so that it matches the circular leading edge (see the smaller yellow circle in Figure \ref{fig:manual_reduction}), and a pick is forced to be in the middle of the annulus. The reduction requires that care be taken to ensure the pick does not deviate from the line of previous picks, or that it produces a large gap. This process is labour intensive, subjective and requires much experience.  In the latter stages of flight where saturation and wake become less significant again, it is common to resolve several discrete fragments which can be picked separately if desired. In most cases only the brightest fragment is tracked.

After observations from all stations are reduced, the trajectory and orbit is computed using the Monte Carlo method described in \citet{vida2020estimating}. The solution may be inspected and outlier picks improved or removed. We present a full example of a manually reduced fireball trajectory in the Results section \ref{subsec:meteorite_dropper}.

\subsection{Automated trajectory estimation} \label{sec:auto_traj}

All meteor observations are accumulated on the central GMN server and trajectories are computed every 6 hours. For every night and station, a list of meteor detections (time, right ascension, declination, magnitude for every frame) and a calibration file are produced. The calibration files contain the camera location and pointing information. All observations which did not pass the recalibration step described in section \ref{subsec:auto_recalib} are rejected.

The first step is to correlate meteor observations from multiple stations and produce candidate groups of observations of a common meteor. All observations are divided into 24-hour bins, with the bin edges set at 12:00 UTC. Observations within a 10 second window (from the middle of the meteor) and from cameras \SI{5}{\kilo \metre} to \SI{600}{\kilo \metre} distant to one another which have overlapping fields of view are examined as a group to produce candidate trajectories. 

The FOV overlap is checked by confirming that vectors passing through the center of the focal plane of each pair of cameras is visible in the other camera's field of view for heights between \SIrange{50}{130}{\kilo \metre}. Given the location of a reference camera (camera A), its Earth-centered inertial (ECI) coordinates $\boldsymbol{S}_A$ at the time when the meteor was observed \citep[see Appendix D1 of][]{vida2020estimating}, and ECI unit vector $\boldsymbol{\hat{u}}_A$ of the center of the focal plane, the point $\boldsymbol{P}_A$ along that unit vector at height $h$ above ground can be computed as

\begin{align}
    \beta    = {} & a + \arcsin \frac{|\boldsymbol{S}_A| \cos a}{ N + h} \,, \\
    r_A      = {} & ( N + h ) \frac{\cos \beta}{\cos a} \,, \\
    \boldsymbol{P}_A = {} & \boldsymbol{S}_A + r_a \boldsymbol{\hat{u}}_A \,, 
\end{align}

\noindent where $r_A$ is the range from the station to $\boldsymbol{P}_A$, $\beta$ is the angle between $\boldsymbol{S}_A$, the centre of the Earth, and $\boldsymbol{P}_A$, $a$ is the elevation of the centre of the focal plane above horizon, and $N$ is the distance from the center of the Earth to the surface of the WGS84 ellipsoid at the latitude of the camera \citep[see equation D1 in][]{vida2020estimating}. Finally, we can check if this point is visible in the second camera's field of view by testing if the angle between $\boldsymbol{\hat{u}}_B$ and $\boldsymbol{P}_A - \boldsymbol{S}_B$ is smaller than half of the diagonal angle of the field of view. Note that this method assumes a larger circular field of view circumscribed around (drawn outside of) the actual smaller rectangular FOV, but it prevents unphysical trajectory solutions.

If the previous checks are satisfied, an intersecting planes trajectory solution is computed for these two stations using the method described in \cite{vida2020estimating}. Solutions which have meteor end heights higher than begin heights are rejected, as well as all meteors which begin outside the \SIrange{50}{150}{\kilo \metre} range, and end higher than 130 or below \SI{20}{\kilo \metre}. Note that these filters will exclude some deeply penetrating fireballs, but those are rare and are designed to be manually reduced in the GMN workflow. 

Next, the average velocity as seen from both stations separately is computed. The solution is rejected if they do not agree within 25\%, or are outside the \SIrange{3}{73}{\kilo \metre \per \second} range. If all checks are satisfied again, the pair of observations is added as a candidate trajectory. Next, all pairs with common observations are merged into groups of three or more observations if they occurred within a 10 second window and have radiants within \ang{15} from one another. Groups of station detections where the maximum convergence angle between all observations is smaller than \ang{3} are discarded. 

The final complete trajectory solution is done using the Monte Carlo method of \citet{vida2020estimating}, using a minimum of $40\%$ of points from the beginning of the trajectory to compute the initial velocity. This method will automatically estimate timing offsets between observations. Ten Monte Carlo runs are done to estimate the uncertainties, and 20 runs are used if the convergence angle is smaller than \ang{15}. Trajectories are rejected if they do not produce heliocentric orbits or the
trajectory minimization procedure does not succeed. Successful solutions are checked for bad individual measurements from individual stations (those outside the $2\sigma$ confidence interval). If bad observations are found, they are removed and the solution is recomputed. This process continues until stable set of observations is reached. In practice, only 1 or 2 points get trimmed from the ends of the trail which are usually obvious outliers produced by an overzealous detection algorithm. All stations where the root mean square deviation (RMSD) of their observations is larger than 2 arc minutes are rejected. If there are three or more stations, observations that have an RMSD two times larger than the median of all other RMSD's are excluded from the solution, unless the RMSD is less than 0.5 arc minutes. If at any point only one single-station observation satisfying the conditions above remains, the trajectory candidate is rejected. The final average velocity is again checked to ensure it is within the \SIrange{3}{73}{\kilo \metre \per \second} range.

Using the astrometric solution to provide the range to each station per frame, the photometric mass is computed by integrating the \SI{100}{\kilo \metre} range-corrected light curve using the bolometric power of a zero-magnitude meteor $P_{0m} = \SI{1210}{\watt}$ \citep[appropriate for Sony HAD EX-View sensors, which might be different for IMX291; ][]{weryk2013simultaneous} and a dimensionless luminous efficiency $\tau = 0.7\%$ \citep{campbell2013high}. The light curve for integration is computed by combining observations from all stations and only taking the brightest estimate. This method helps reduce effects of saturation if a meteor was observed by a more distant camera that did not saturate. GMN cameras saturate at around magnitude $-1^M$ (depending on the angular velocity of the meteor); thus meteors with brighter peak magnitudes will have underestimated photometric masses. 

Next, a check is made to determine if the meteor began or ended within the field of view of at least one camera. This is done by propagating the meteor trajectory forward/backward by two frames from the observed end/begin point in the image plane and checking if the resulting coordinates are within the image. If only a portion of the meteor was observed, the resulting photometric mass is only the lower limit and is so flagged.

A final set of filters is applied to each event before it is recorded in the GMN database:
\begin{itemize}
    \item The meteor needs to be observed on at least 6 frames by the station which observed it the longest.
    \item A minimum convergence angle of \ang{5} is imposed.
    \item Orbits with eccentricities larger than 1.5 are rejected.
    \item Orbits with geocentric radiant errors $> \ang{2}$ or geocentric velocity errors $> 10\%$ are rejected.
    \item Meteors that start above \SI{160}{\kilo \metre} or end below \SI{20}{\kilo \metre} are rejected.
\end{itemize}

Finally, meteor shower membership is determined using the meteor shower table of \cite{jenniskens2018survey} with a fixed angular radius of association of \ang{3} and a maximum geocentric velocity threshold of $10\%$ \citep[see][for more details about this procedure]{moorhead2020realistic}. All computed trajectories and orbits are accumulated into a table and immediately published on the GMN website. Within a window of seven days, the trajectories are continuously updated with new observations from stations that were late to upload their data.

\section{Results} \label{sec:results}

Since the beginning of semi-regular operations in December 2018, the GMN recorded a total of $\sim 220,000$ meteors for which orbits were computed (up to mid-2021). Every night-time annual meteor shower of significance for the satellite impact hazard \citep{moorhead2019meteor} was observed by the GMN over that period. 

\begin{figure}
  \includegraphics[width=\linewidth]{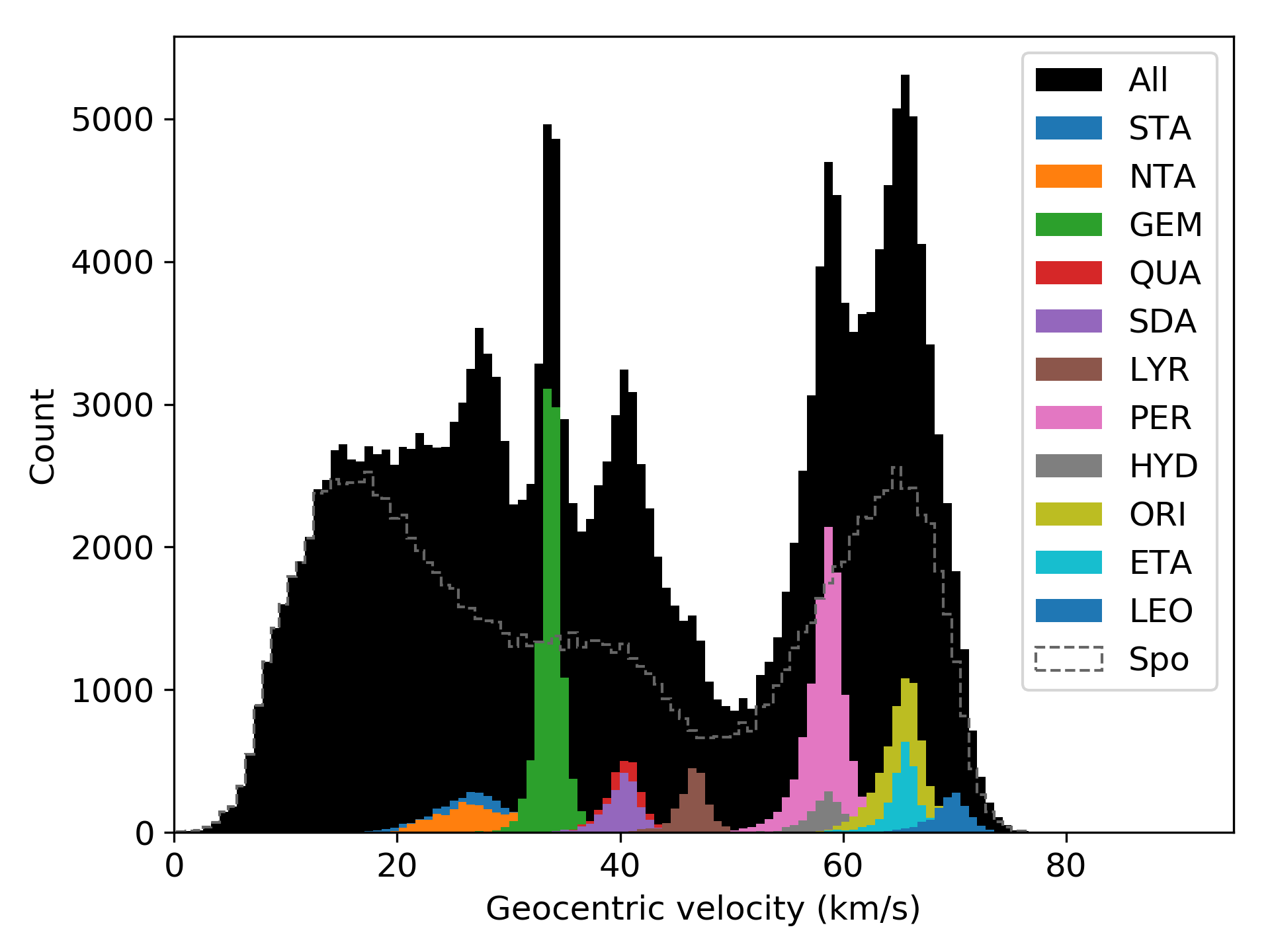}
  \caption{The geocentric velocity distribution of all meteors observed by the GMN from 2018-2021. The meteor showers having the most orbits in the data set are shown as are sporadic meteors.}
  \label{fig:vg_counts}
\end{figure}

Figure \ref{fig:vg_counts} shows the geocentric velocity distribution of all observed meteors, sporadic meteors, and the top eleven showers with the most observed meteors. Meteoroids on asteroidal orbits ($T_J > 3$) make up $28.5\%$ of the data set, Jupiter-family type ($2 < T_J < 3$) orbits $20.5\%$, and Halley-type ($T_J < 2$) orbits $51.0\%$. We note that these raw percentages are subject to a significant observational bias as GMN is a brightness-limited survey; namely the mass limit for fast meteors is much smaller than for slower meteors. At the high velocity extreme, GMN detects meteoroids two to three orders of magnitude smaller in mass than at the low velocity end. As a result the raw number of detections is richer in fast showers. 

We note that at speeds below \SI{20}{\kilo \metre \per \second} there are few known meteor showers and that shower meteors constitute only $5.4\%$ of those orbits. Generally, shower meteors constitute $51.1\%$ and $41.9\%$ of Jupiter-family and Halley-type comet orbits, respectively. A total of 10,400 Perseids, 10,200 Geminids, 6,200 Orionids, 5,400 Taurids (North and South combined), and 2,000 Southern $\delta$-Aquariids have been observed, among the top showers. However, again, these represent raw observed numbers and since meteor showers typically have a shallower size frequency distribution than the sporadic background \citep{koten2019meteors}, they will tend to be over represented relative to sporadics at a given speed. 

\begin{figure}
  \includegraphics[width=\linewidth]{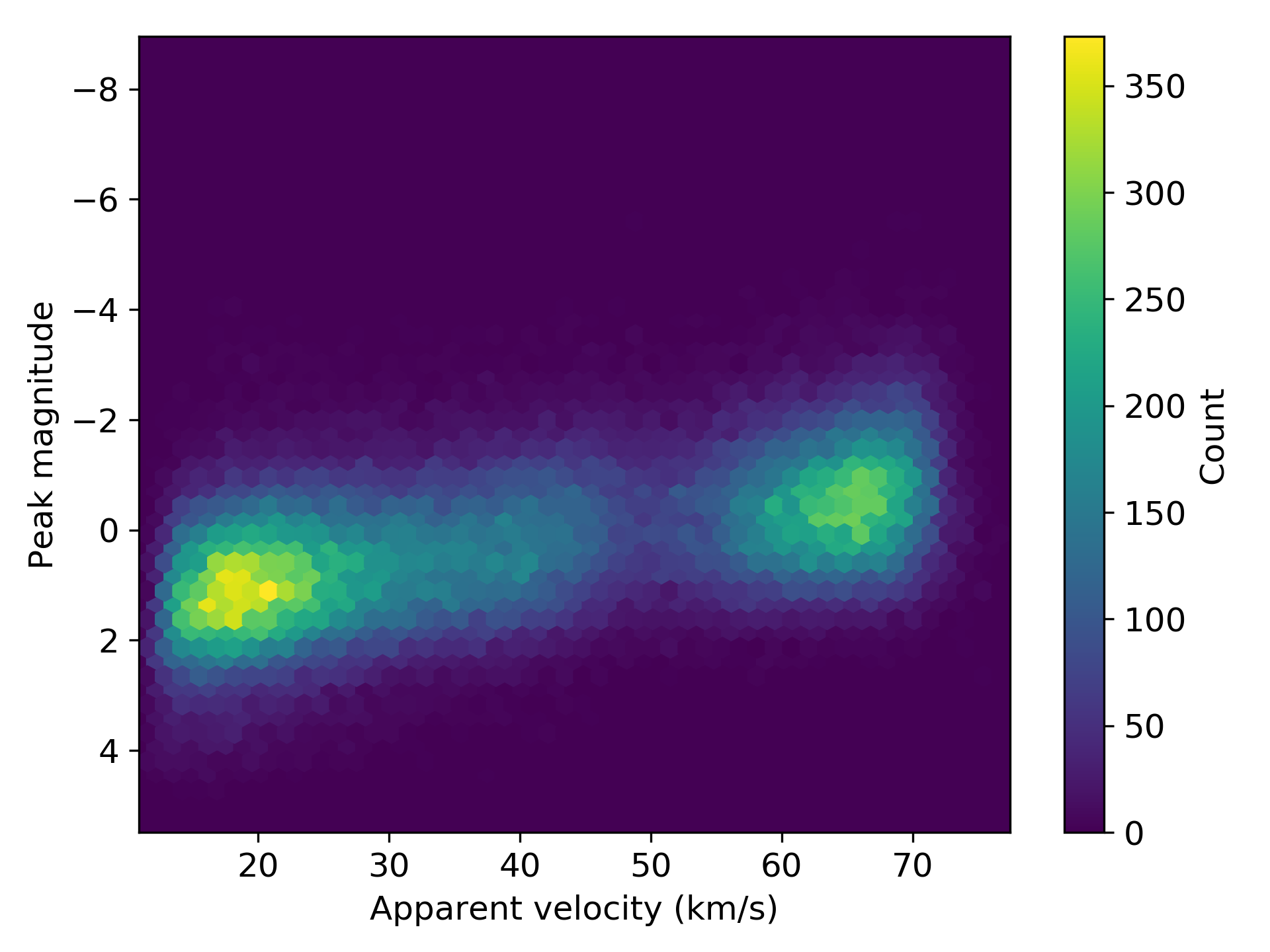}
  \caption{Observed peak meteor magnitude versus observed initial velocity.}
  \label{fig:peak_mag_vs_vel}
\end{figure}

Figure \ref{fig:peak_mag_vs_vel} shows the dependence of the meteor peak magnitude with apparent initial velocity. A clear positive correlation of brighter magnitudes with higher velocities can be seen; at \SI{20}{\kilo \metre \per \second} the average observed peak magnitude is $+1.5^M$, while at \SI{70}{\kilo \metre \per \second} it is $-0.5^M$. This trend is also seen for CAMS meteors \citep{jenniskens2011cams}, though the magnitude difference in that survey between fast and slow meteors is about half the GMN offset. We  also note that compared to the stellar limiting magnitude, peak meteor magnitudes are as much as $4^M$ brighter. This magnitude offset is explained by taking the detection thresholds, a 4-frame duration minimum, noise characteristics and the pixel pitch of the sensor, the PSF of stars, and the average angular velocity of meteors into account.

One factor driving the magnitude trend is the higher angular velocity on average for faster meteors, but another significant difference may be due to the differing spectrum between calibration stars and meteors. The IMX291 sensors are as sensitive in the near infrared (\SIrange{800}{900}{\nano \metre}) as they are in the visible parts of the spectrum (\SIrange{450}{650}{\nano \metre}). Stars of most spectral types used for GMN calibrations have peak emissions blueward of the near-infrared, while meteors emit 30 - 50\% of light at wavelengths \SI{>700}{\nano \metre} \citep{segon2018infrared}, rising with the increasing velocity due to more emissions from atmospheric lines \citep[particularly the OI line at \SI{777.4}{\nano \metre}; ][]{borovivcka2005survey, vojavcek2015catalogue}. Care should be taken when comparing GMN magnitudes with other systems, keeping in mind the very different spectral sensitivity of IMX291 sensors compared to other cameras. The GMN magnitudes will skew observed meteors toward brighter magnitudes, as defined by the GAIA G bandpass used by the GMN.

\subsection{Trajectory precision}

\begin{figure}
  \includegraphics[width=\linewidth]{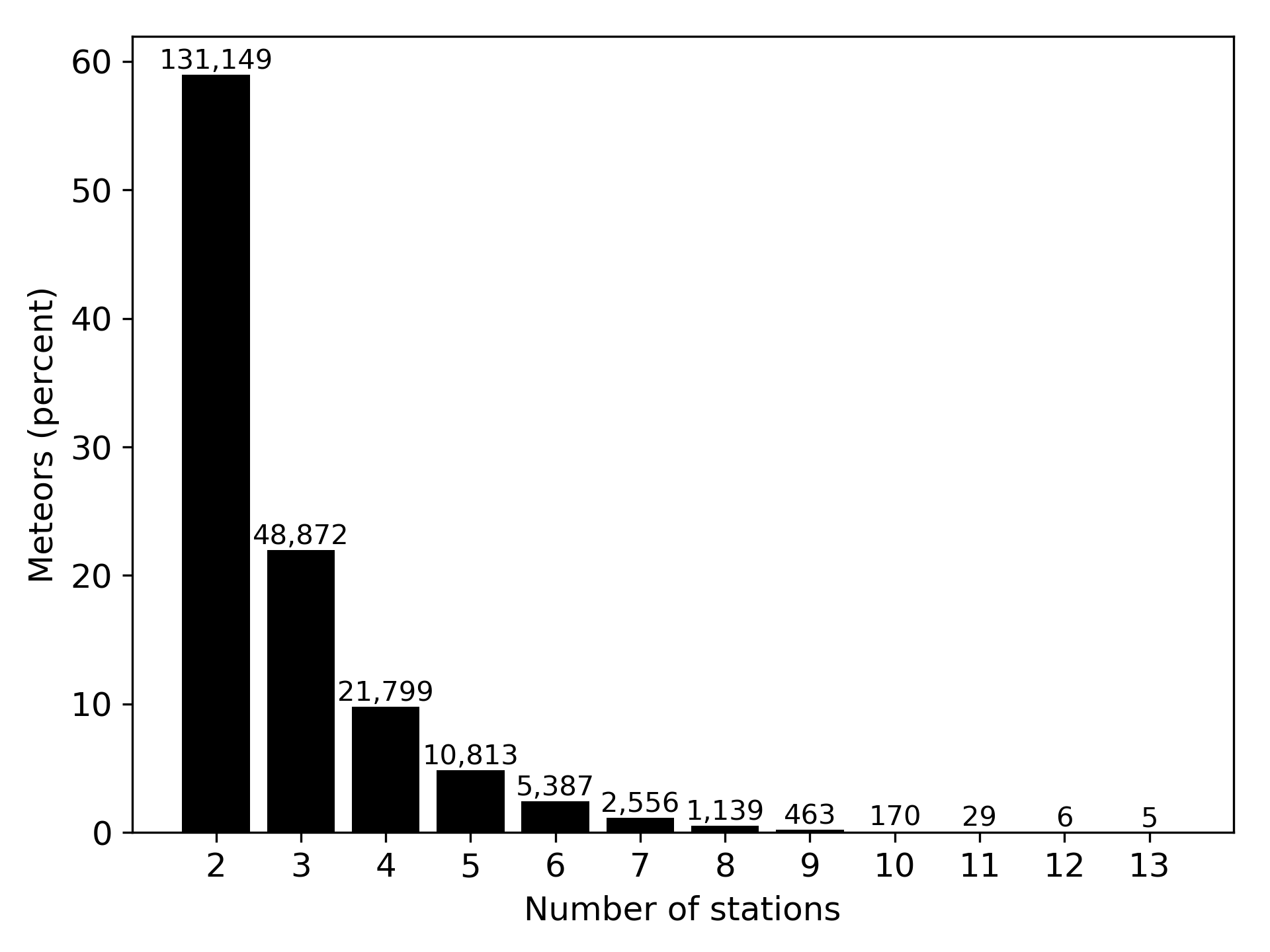}
  \caption{Percentage of all trajectories observed from a given number of stations. The number above each column is the absolute number of trajectories.}
  \label{fig:nstat_freq}
\end{figure}

To accurately measure radiant dispersions of meteor showers and observe radiant sub-structure \citep[e.g.][]{spurny2017discovery}, accurate trajectories and robust estimates of meteor radiant precision are essential. If a meteor is only observed from two stations (and in some cases three stations) with unfavorable geometry, the dynamics is a poor constraint on the trajectory and radiant errors can be as much as \ang{1} \citep{vida2020results}. Figure \ref{fig:nstat_freq} shows the percentage of the GMN trajectories observed from a given number of stations. About 60\% of meteors are observed from only two stations, and about 20\% are higher-quality trajectories observed from four or more stations. Due to the high density of GMN stations in Albuquerque and parts of Central Europe, a handful of meteors are observed from 10+ stations.

\begin{figure}
  \includegraphics[width=\linewidth]{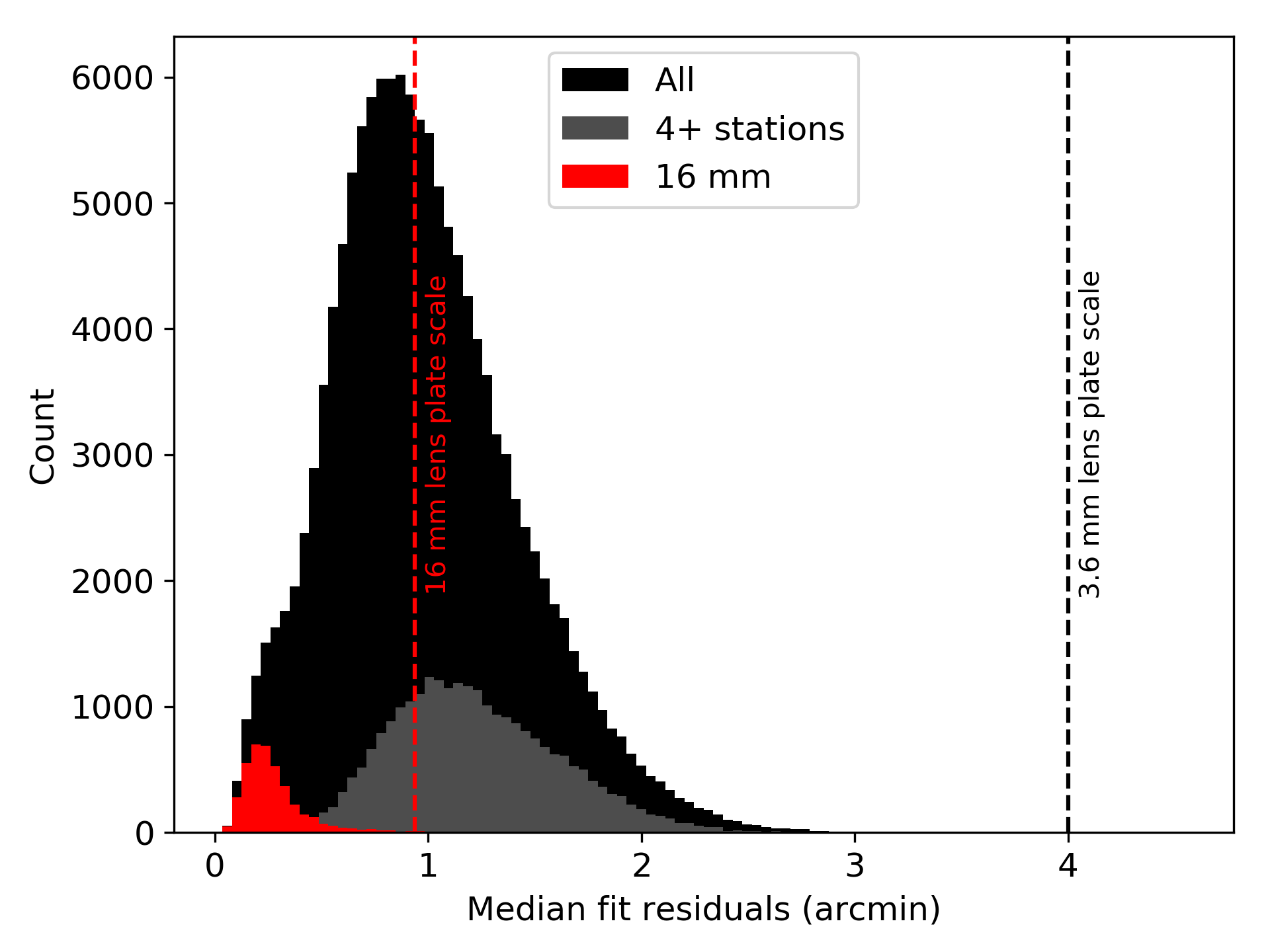}
  \caption{The median fit residuals of sporadic meteors in the GMN data set. This shows the median difference between the apparent angular pick location per frame and the final trajectory solution for each station. Plate scales of \SI{3.6}{\milli \metre} and \SI{16}{\milli \metre} systems are shown as vertical lines. The distributions for cameras using \SI{16}{\milli \metre} lenses and solutions with four or more stations are shown separately.}
  \label{fig:median_fit_residuals}
\end{figure}

Figure \ref{fig:median_fit_residuals} shows the distribution of the median angular deviation of all sightlines from the trajectory solution for GMN data. Only sporadic meteors were used for this analysis to avoid biases introduced by meteor showers. Most trajectories have fit residuals (angular differences between observations and the trajectory) that are around 1 arc minute, about four times smaller than the plate scale of \SI{3.6}{\milli \metre} cameras, indicating the quality of automated picks and astrometry calibration procedure. For meteors observed from 4+ stations, where the residuals include both random and systematic uncertainties, the median fit residuals are about 20\% higher, indicating that the systematic calibration errors are on the order of 20\%. For meteors observed using \SI{16}{\milli \metre} cameras, the median fit residuals are significantly more accurate, at around 10 arc seconds, as expected for the better plate scale.

\begin{figure}
  \includegraphics[width=\linewidth]{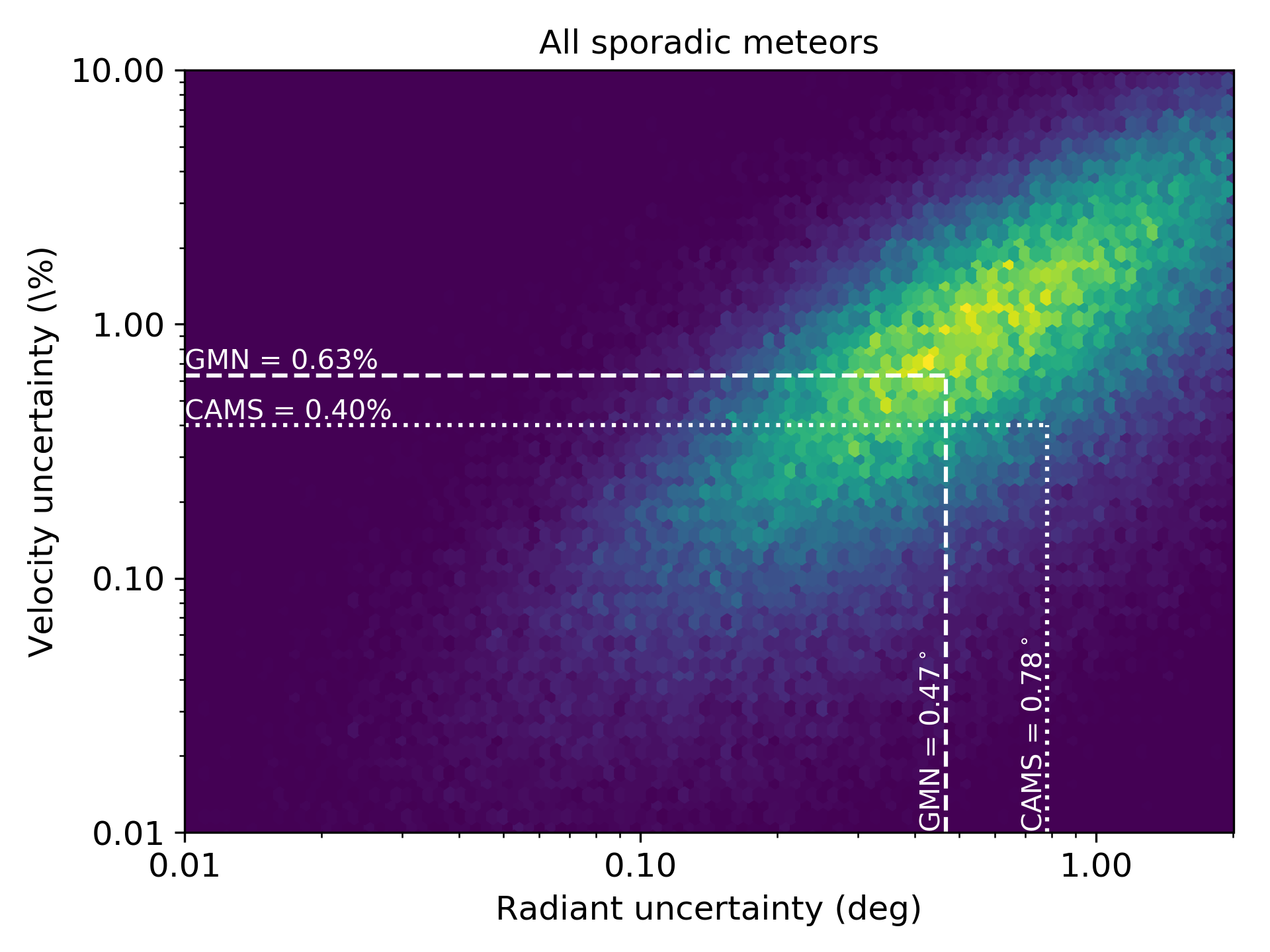} \vfill
  \includegraphics[width=\linewidth]{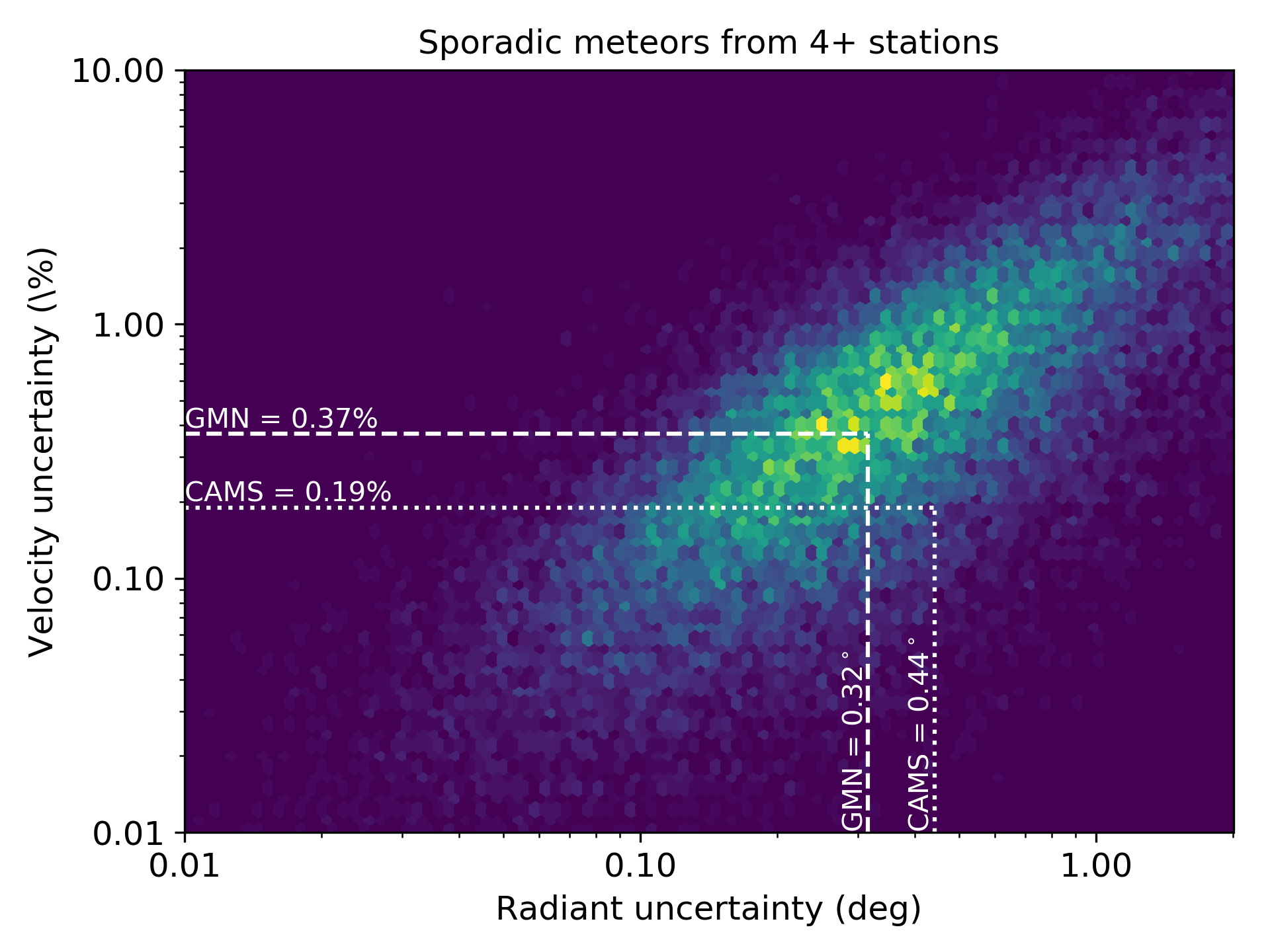} \vfill
  \includegraphics[width=\linewidth]{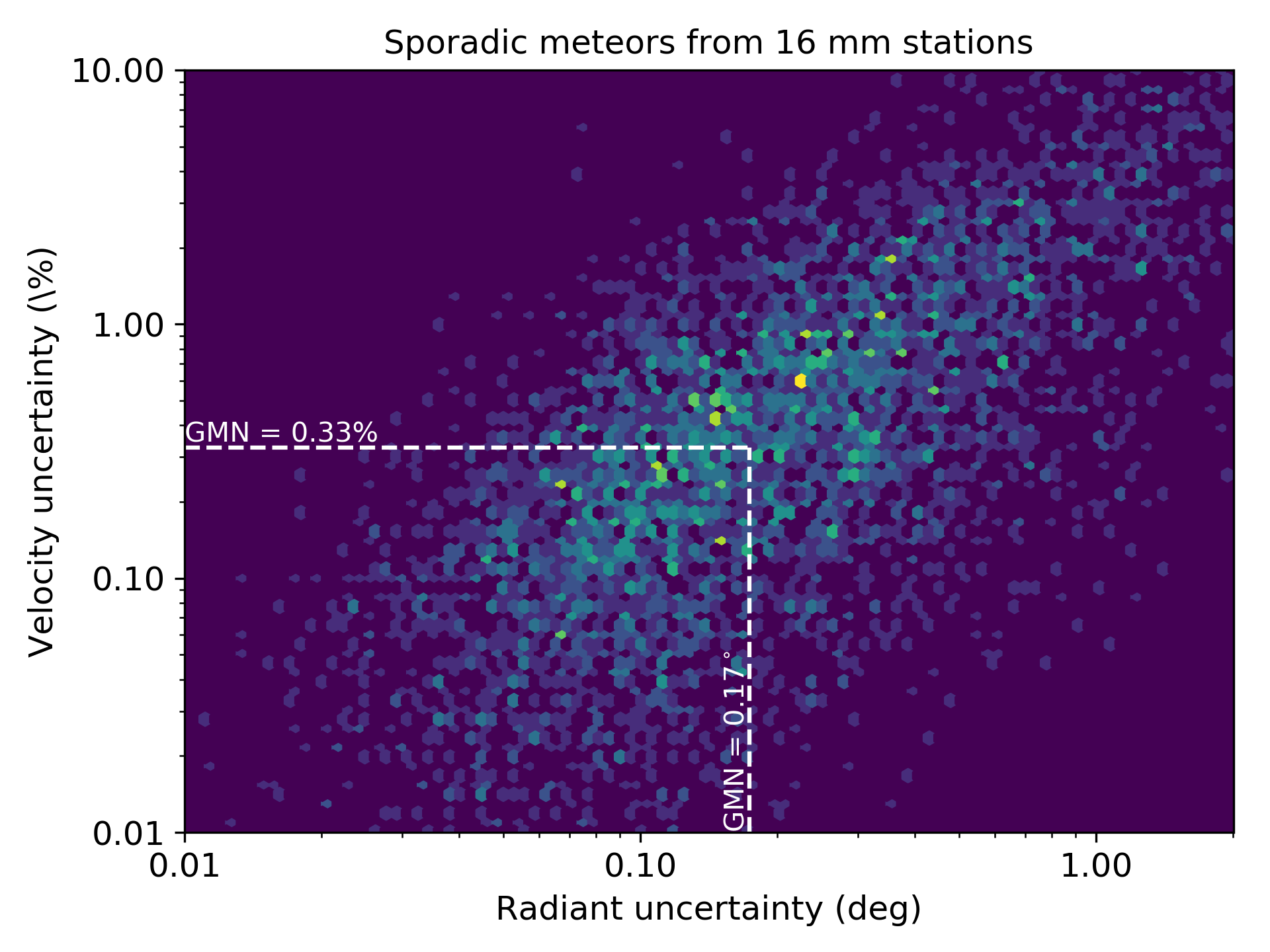}
  \caption{Density plots of radiant and velocity uncertainty estimates for sporadic meteors (brighter bins indicate a higher density). Median uncertainties for GMN are marked by dashed lines, and uncertainties for CAMS are given  by dotted lines. Top: Using all sporadic meteors. Middle: Sporadic meteors observed from 4 or more stations. Bottom: Sporadic meteors observed using \SI{16}{\milli \metre} systems. For GMN all uncertainties are computed using the Monte Carlo approach described in \citet{vida2018modelling}.}
  \label{fig:uncertainty estimates}
\end{figure}

Figure \ref{fig:uncertainty estimates} shows how the trajectory fit residuals translate into estimated radiant precision. The median radiant precision of all sporadic meteors is \ang{0.47}, but this goes down to \ang{0.32} if only 4+ station meteors are selected. Compared to the published CAMS trajectories \citep{jenniskens2016cams}, the estimated GMN radiant precision is 30-40\% better. High-precision \SI{16}{\milli \metre} lenses achieve an average radiant precision of \ang{0.17}. The GMN velocity fit precision is on the order of 0.5\%, which is generally higher than the CAMS values. However, the velocities are most certainly underestimated up to \SI{750}{\metre \per \second} depending on meteoroid speed and mass \citep{vida2018modelling} due to early deceleration before luminous flight so the initial velocity accuracy is worse than this value. We do not apply the deceleration correction proposed in \cite{vida2018modelling} as the range of meteoroid heights of the cometary and asteroidal classes would first have to be calibrated on the GMN data set, which will be done in a future work.  Finally, note that all high velocity errors around 10\% are at speeds below \SI{20}{\kilo \metre \per \second}, where uncertainties of \SI{\sim1}{\kilo \metre \per \second} can appear large.

\subsection{Established meteor showers}

\begin{figure*}
  \includegraphics[width=\linewidth]{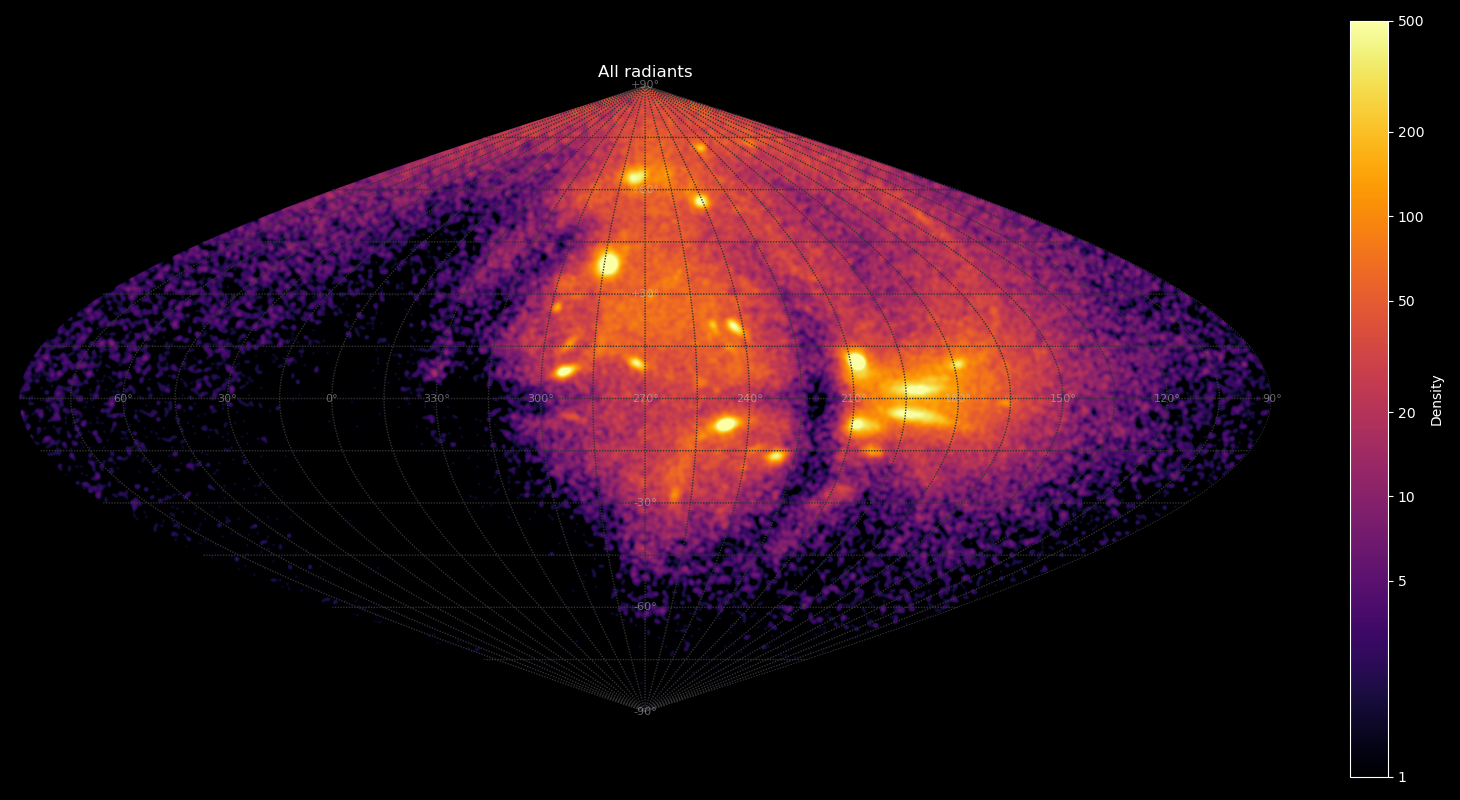}
  \includegraphics[width=\linewidth]{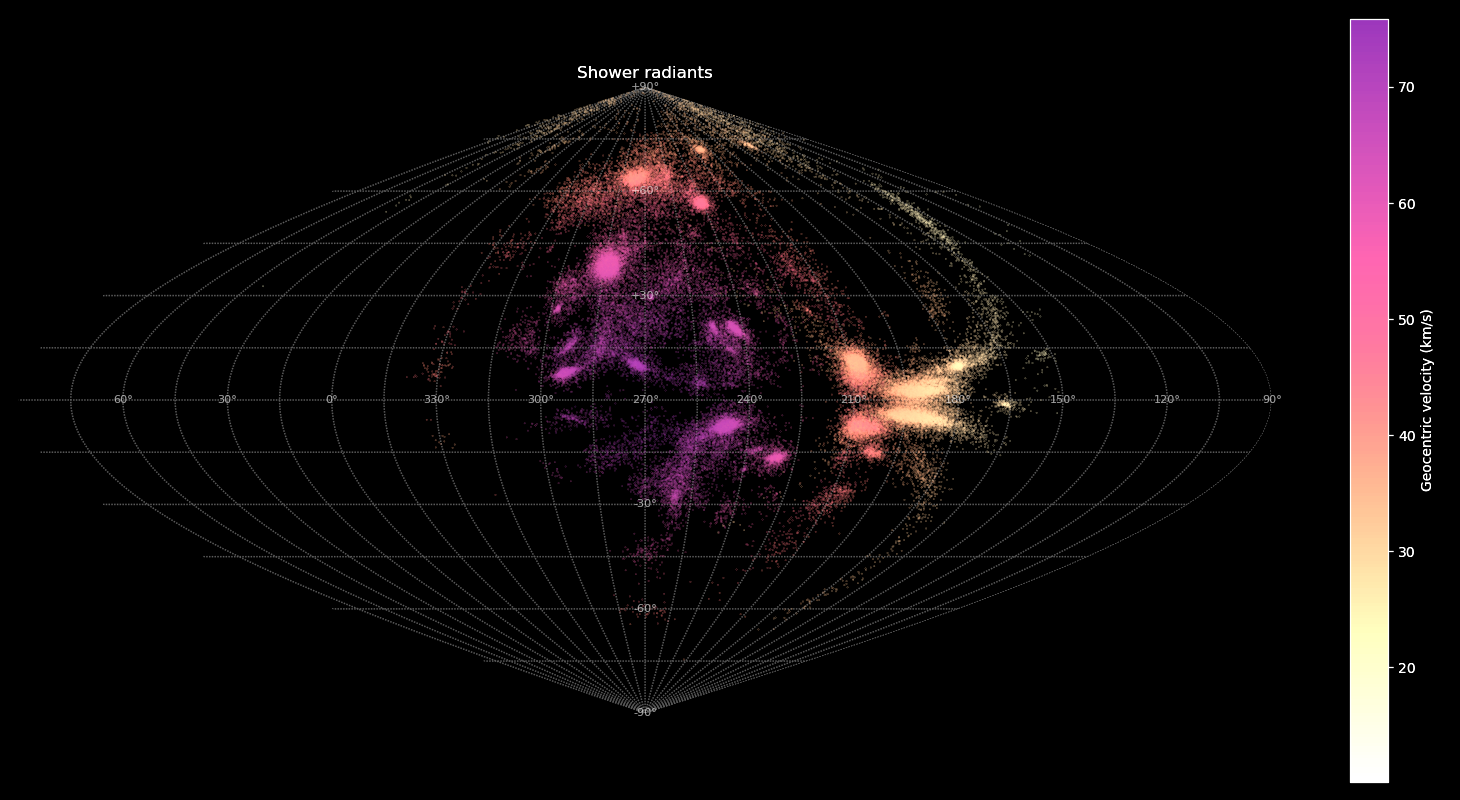}
  \caption{Sun-centered geocentric ecliptic (SCE) plots of GMN radiants. The Sun is at \ang{0} of SCE longitude (horizontal coordinates), and the Earth's apex is in the middle of this plot at \ang{270}. Top: Area density distribution of all GMN radiants (radiants convoluted with a Gaussian where $\sigma = \ang{1}$). Bottom: Individual meteor radiants automatically associated with one of the IAU established meteor showers color coded by geocentric velocity.}
  \label{fig:sce_plots}
\end{figure*}

Figure \ref{fig:sce_plots} is a Sun-centered geocentric ecliptic radiant plot of all observed GMN meteor and meteor shower radiants. The major sporadic radiant sources \citep{Jones1993a} are clearly visible except for the helion source. The majority of established IAU meteor showers\footnote{IAU Meteor Shower List: \url{https://www.ta3.sk/IAUC22DB/MDC2007/}} are also observed. The median radiants and mean orbital elements \citep[computed using the method of][]{jopek2006calculation} of established showers for which at least 50 orbits were observed by GMN between 2018-2021 are given in Table \ref{tab:showers}. 

Established showers that were either not observed or observed in small number are given in Table \ref{tab:showers_not_observed}. The missing showers include daytime showers which are difficult or impossible to observe optically, though we note there are some daytime showers for which GMN captured some orbits. Other showers not detected in significant numbers by GMN are either low-activity showers or those with broad radiants. Some showers which are strongly active on years outside our survey were also not observed in any significant numbers. Note the automated shower association described in Section \ref{sec:auto_traj} was used here and there are likely biases to this automated approach. Namely, the fixed association threshold does not take different dispersions of meteor showers into account, and no attempts to subtract the sporadic background were made. A more detailed analysis will be published in a future paper.

\clearpage
\onecolumn

\LTcapwidth=\textwidth
\begin{longtable}{l l r r r r r r r r r r r r}
\caption{Median radiant locations and mean orbits of IAU established meteor showers as detected by GMN in the December 2018 - June 2021 period. Meteor shower outbursts, observed by the GMN are given below the horizontal line. $\lambda_{\astrosun}$ is the median solar longitude of the shower over the full duration of its activity recorded by GMN, $\alpha_G$  and $\delta_G$ the equatorial geocentric radiant at $\lambda_{\astrosun}$, $\lambda_G - \lambda_{\astrosun}$ and $\beta_G$ the radiant in Sun-centered ecliptic coordinates at $\lambda_{\astrosun}$, $V_G$ the geocentric velocity at $\lambda_{\astrosun}$. N is the total number of GMN observed orbits.}
\label{tab:showers} \\ 
        \hline\hline 
IAU & IAU  &    $\lambda_{\astrosun}$ &  $\alpha_G$  &   $\delta_G$ &   $\lambda_G - \lambda_{\astrosun}$ & $\beta_G$ &   $V_G$ &     q &     e &       i &    $\omega$ &    $\Omega$ &    N \\
No & code & (deg) & (deg) & (deg) & (deg) &  (deg) &  km/s &    AU &       &     (deg) &     (deg) &     (deg) & \\
\hline
\endfirsthead
\caption{continued.} \\
        \hline\hline 
IAU & IAU  &    $\lambda_{\astrosun}$ &  $\alpha_G$  &   $\delta_G$ &   $\lambda_G - \lambda_{\astrosun}$ & $\beta_G$ &   $V_G$ &     q &     e &       i &    $\omega$ &    $\Omega$ &    N \\
No & code & (deg) & (deg) & (deg) & (deg) &  (deg) &  km/s &    AU &       &     (deg) &     (deg) &     (deg) & \\
\hline
\endhead 
\hline
\endfoot
  1 & CAP  & 126.07 & 304.93 &  -9.62 & 179.07 &  +9.70 & 22.36 & 0.584 & 0.764 &   7.297 & 268.564 & 124.708 &   932 \\
  2 & STA  & 209.16 &  41.44 & +11.12 & 192.43 &  -4.76 & 26.99 & 0.364 & 0.786 &   5.362 & 111.405 &  33.693 &  3037 \\
  4 & GEM  & 261.80 & 113.19 & +32.38 & 208.05 & +10.51 & 33.81 & 0.146 & 0.887 &  22.802 & 323.967 & 261.358 & 10190 \\
  5 & SDA  & 126.95 & 340.05 & -16.51 & 208.60 &  -7.52 & 40.24 & 0.079 & 0.963 &  26.868 & 151.131 & 306.977 &  1910 \\
  6 & LYR  &  32.31 & 272.30 & +33.31 & 241.11 & +56.71 & 46.61 & 0.919 & 0.937 &  79.351 & 213.889 &  32.122 &  1823 \\
  7 & PER  & 139.91 &  47.78 & +57.82 & 283.35 & +38.46 & 58.60 & 0.931 & 0.896 & 112.553 & 148.993 & 138.795 & 10424 \\
  8 & ORI  & 209.31 &  96.34 & +15.65 & 246.62 &  -7.57 & 65.50 & 0.550 & 0.903 & 163.828 &  85.264 &  30.025 &  6192 \\
 10 & QUA  & 283.03 & 229.24 & +49.91 & 275.53 & +63.68 & 40.44 & 0.974 & 0.647 &  70.896 & 174.532 & 283.053 &  2639 \\
 11 & EVI  & 357.83 & 185.51 &  +3.48 & 186.03 &  +5.38 & 26.52 & 0.461 & 0.799 &   5.080 & 281.467 & 357.743 &   531 \\
 12 & KCG  & 136.74 & 283.55 & +52.10 & 162.75 & +73.30 & 22.09 & 0.927 & 0.701 &  33.407 & 203.610 & 138.353 &   288 \\
 13 & LEO  & 235.96 & 154.18 & +21.62 & 272.33 & +10.21 & 69.71 & 0.975 & 0.812 & 162.019 & 171.631 & 235.056 &  1338 \\
 15 & URS  & 270.49 & 218.18 & +75.53 & 218.75 & +71.72 & 33.15 & 0.932 & 0.810 &  52.683 & 206.889 & 269.394 &   475 \\
 16 & HYD  & 256.57 & 125.87 &  +2.30 & 230.90 & -16.57 & 58.58 & 0.249 & 0.970 & 128.302 & 119.915 &  78.465 &  1348 \\
 17 & NTA  & 221.92 &  49.60 & +20.75 & 191.07 &  +2.85 & 27.19 & 0.377 & 0.767 &   3.545 & 291.513 & 219.836 &  2316 \\
 18 & AND  & 232.92 &  22.73 & +33.80 & 161.49 & +22.38 & 17.04 & 0.801 & 0.735 &  10.250 & 235.221 & 232.730 &   187 \\
 19 & MON  & 258.53 & 101.03 &  +7.99 & 202.39 & -14.99 & 41.23 & 0.189 & 0.979 &  34.873 & 128.934 &  79.165 &   526 \\
 20 & COM  & 270.82 & 163.78 & +29.31 & 242.71 & +20.79 & 62.73 & 0.538 & 0.928 & 134.257 & 265.219 & 271.354 &  1279 \\
 21 & AVB  &  26.36 & 200.33 &  +4.65 & 170.66 & +12.28 & 19.38 & 0.726 & 0.718 &   7.239 & 248.420 &  29.474 &   365 \\
 22 & LMI  & 209.15 & 160.38 & +36.77 & 298.02 & +26.32 & 61.22 & 0.610 & 0.951 & 123.900 & 102.494 & 210.073 &   243 \\
 23 & EGE  & 205.18 & 100.94 & +28.10 & 254.66 &  +5.07 & 68.34 & 0.773 & 0.871 & 169.992 & 236.569 & 203.666 &   366 \\
 26 & NDA  & 148.11 & 352.74 &  +4.38 & 206.78 &  +6.85 & 37.82 & 0.107 & 0.932 &  19.085 & 326.164 & 149.628 &   890 \\
 27 & KSE  &  25.92 & 247.86 & +16.45 & 216.57 & +37.98 & 46.32 & 0.513 & 0.953 &  74.661 & 269.484 &  25.825 &    65 \\
 31 & ETA  &  47.04 & 338.92 &  -0.43 & 293.29 &  +7.81 & 65.56 & 0.590 & 0.926 & 163.466 &  98.973 &  49.388 &  2401 \\
 33 & NIA  & 163.48 & 359.68 &  +3.91 & 196.56 &  +3.71 & 28.52 & 0.291 & 0.814 &   5.001 & 305.011 & 164.558 &   296 \\
 69 & SSG  &  82.41 & 271.30 & -29.24 & 187.91 &  -5.92 & 25.85 & 0.439 & 0.792 &   5.673 & 106.123 & 264.154 &   124 \\
 96 & NCC  & 290.88 & 120.61 & +22.46 & 189.00 &  +2.37 & 27.52 & 0.411 & 0.801 &   2.519 & 286.386 & 291.294 &   378 \\
 97 & SCC  & 286.98 & 115.50 & +16.33 & 189.29 &  -4.63 & 27.12 & 0.417 & 0.788 &   4.790 & 106.950 & 106.118 &   466 \\
145 & ELY  &  50.12 & 291.06 & +43.64 & 256.92 & +64.44 & 43.70 & 0.998 & 0.914 &  74.093 & 190.356 &  49.778 &   276 \\
151 & EAU  &  87.30 & 302.67 &  +3.90 & 217.43 & +22.52 & 31.93 & 0.174 & 0.782 &  53.325 & 326.296 &  88.096 &    95 \\
164 & NZC  & 106.36 & 313.73 &  -3.34 & 209.16 & +12.63 & 37.63 & 0.120 & 0.891 &  35.603 & 324.815 & 109.698 &   749 \\
165 & SZC  & 107.28 & 321.79 & -27.08 & 208.60 & -11.08 & 38.55 & 0.110 & 0.920 &  33.317 & 146.699 & 287.981 &   140 \\
175 & JPE  & 116.81 & 355.04 & +12.79 & 244.05 & +14.19 & 63.38 & 0.520 & 0.915 & 149.761 & 268.518 & 118.969 &   297 \\
183 & PAU  & 134.70 & 352.78 & -20.73 & 210.49 & -16.46 & 43.84 & 0.132 & 0.955 &  54.660 & 139.807 & 316.258 &    64 \\
184 & GDR  & 125.16 & 279.96 & +50.71 & 166.86 & +73.28 & 27.47 & 0.978 & 0.966 &  40.280 & 202.237 & 125.618 &   150 \\
187 & PCA  & 118.40 &  28.60 & +70.71 & 304.86 & +53.33 & 43.79 & 0.833 & 0.771 &  77.568 & 129.373 & 116.875 &    56 \\
191 & ERI  & 136.90 &  42.90 & -12.77 & 260.11 & -27.57 & 64.00 & 0.933 & 0.911 & 131.998 &  30.655 & 315.691 &   320 \\
197 & AUD  & 147.16 & 270.63 & +59.95 & 127.25 & +81.37 & 21.16 & 0.999 & 0.649 &  33.247 & 184.503 & 147.545 &   640 \\
206 & AUR  & 161.89 &  94.91 & +39.46 & 292.38 & +16.34 & 65.30 & 0.666 & 0.942 & 146.688 & 108.649 & 162.455 &   210 \\
208 & SPE  & 169.50 &  50.51 & +39.71 & 248.99 & +20.33 & 64.12 & 0.683 & 0.918 & 140.299 & 248.370 & 172.953 &   622 \\
242 & XDR  & 222.50 & 206.97 & +74.16 & 270.70 & +68.45 & 38.23 & 0.979 & 0.691 &  64.985 & 180.996 & 221.320 &    93 \\
250 & NOO  & 245.40 &  89.93 & +15.43 & 203.71 &  -7.90 & 42.40 & 0.120 & 0.978 &  22.832 & 139.916 &  66.375 &   886 \\
257 & ORS  & 247.38 &  76.50 & +17.76 & 190.30 &  -4.79 & 27.25 & 0.394 & 0.806 &   5.196 & 108.540 &  69.691 &   668 \\
323 & XCB  & 295.31 & 249.92 & +29.71 & 306.91 & +51.24 & 45.07 & 0.784 & 0.810 &  77.631 & 124.053 & 295.636 &    74 \\
326 & EPG  & 111.65 & 332.51 & +14.08 & 224.94 & +24.93 & 27.40 & 0.167 & 0.761 &  43.289 & 336.581 & 112.227 &    75 \\
331 & AHY  & 283.74 & 125.95 &  -8.17 & 207.49 & -26.38 & 42.96 & 0.286 & 0.958 &  57.136 & 115.683 & 103.097 &   216 \\
333 & OCU  & 202.28 & 145.61 & +64.00 & 279.07 & +46.70 & 55.53 & 0.972 & 0.925 & 100.842 & 163.838 & 202.957 &   123 \\
334 & DAD  & 251.42 & 203.70 & +62.08 & 264.09 & +62.08 & 41.16 & 0.948 & 0.647 &  72.854 & 187.560 & 251.176 &   695 \\
335 & XVI  & 264.74 & 192.12 & -11.06 & 291.09 &  -5.28 & 68.22 & 0.632 & 0.934 & 169.226 & 286.536 &  85.206 &   167 \\
336 & DKD  & 251.29 & 185.94 & +69.43 & 244.04 & +61.19 & 44.10 & 0.924 & 0.895 &  74.337 & 207.961 & 252.226 &   184 \\
337 & NUE  & 175.89 &  74.44 &  +5.08 & 257.35 & -17.05 & 66.11 & 0.809 & 0.833 & 149.653 &  41.839 & 356.232 &  1200 \\
338 & OER  & 229.12 &  56.52 &  -1.53 & 184.95 & -20.64 & 27.68 & 0.512 & 0.855 &  18.935 &  89.156 &  51.420 &   515 \\
339 & PSU  & 252.34 & 169.12 & +43.64 & 258.13 & +35.39 & 61.15 & 0.914 & 0.879 & 117.482 & 211.014 & 252.319 &    82 \\
341 & XUM  & 298.30 & 169.41 & +32.82 & 218.24 & +25.81 & 40.56 & 0.221 & 0.843 &  66.998 & 313.835 & 297.928 &    68 \\
346 & XHE  & 356.07 & 261.33 & +48.09 & 256.97 & +70.48 & 34.75 & 0.981 & 0.656 &  60.523 & 186.683 & 354.317 &   156 \\
348 & ARC  &  37.31 & 320.39 & +46.62 & 309.90 & +57.32 & 40.83 & 0.864 & 0.782 &  69.908 & 136.191 &  37.341 &   219 \\
362 & JMC  &  70.41 &   7.63 & +51.98 & 322.96 & +42.85 & 42.83 & 0.586 & 0.896 &  70.470 &  97.376 &  70.917 &    94 \\
372 & PPS  & 108.92 &  19.94 & +29.27 & 282.19 & +19.31 & 64.93 & 0.867 & 0.808 & 143.930 & 142.242 & 110.635 &   683 \\
388 & CTA  & 214.21 &  57.05 & +25.44 & 206.04 &  +5.33 & 40.80 & 0.097 & 0.954 &  17.041 & 325.767 & 215.655 &   286 \\
390 & THA  & 243.25 &  96.50 & +32.24 & 212.19 &  +9.44 & 31.96 & 0.118 & 0.865 &  23.395 & 329.521 & 246.962 &   160 \\
404 & GUM  & 299.18 & 228.97 & +67.13 & 220.53 & +74.12 & 29.74 & 0.950 & 0.681 &  49.395 & 202.233 & 299.711 &    64 \\
411 & CAN  & 108.35 &  29.94 & +47.32 & 297.83 & +32.77 & 56.76 & 0.674 & 0.891 & 112.455 & 108.594 & 107.311 &   253 \\
428 & DSV  & 276.78 & 213.04 &  +3.34 & 292.59 & +15.25 & 66.00 & 0.616 & 0.922 & 147.880 & 106.948 & 279.723 &   391 \\
445 & KUM  & 223.92 & 146.04 & +45.23 & 267.93 & +29.63 & 64.78 & 0.983 & 0.890 & 128.995 & 187.172 & 224.491 &   111 \\
506 & FEV  & 302.27 & 190.67 & +16.00 & 241.27 & +18.82 & 62.32 & 0.476 & 0.910 & 136.842 & 272.495 & 304.435 &   337 \\
512 & RPU  & 226.08 & 125.45 & -25.19 & 268.82 & -43.14 & 57.41 & 0.981 & 0.862 & 106.816 &   3.500 &  46.237 &    70 \\
529 & EHY  & 263.32 & 138.51 &  +1.07 & 237.48 & -14.37 & 61.95 & 0.359 & 0.949 & 142.925 & 106.086 &  80.646 &   237 \\
530 & ECV  & 300.21 & 190.25 & -17.90 & 255.90 & -11.84 & 67.52 & 0.780 & 0.820 & 157.822 &  52.821 & 121.374 &   134 \\
533 & JXA  & 111.42 &  34.59 &  +8.47 & 284.21 &  -4.63 & 68.12 & 0.827 & 0.903 & 171.343 & 309.562 & 290.863 &    76 \\
549 & FAN  & 117.37 &  26.52 & +46.96 & 286.22 & +33.34 & 59.45 & 0.878 & 0.866 & 119.125 & 136.675 & 116.375 &    80 \\
569 & OHY  & 305.60 & 173.74 & -31.98 & 242.92 & -31.65 & 58.36 & 0.643 & 0.860 & 113.874 &  73.337 & 126.382 &   129 \\
        \hline
246 & AMO  & 239.31 & 117.01 &  +0.69 & 239.63 & -20.09 & 61.18 & 0.463 & 0.937 & 132.797 &  95.455 &  59.311 &    12 \\
343 & HVI  &  38.33 & 202.80 & -10.81 & 166.50 &  -1.20 & 18.54 & 0.753 & 0.737 &   0.601 &  65.244 & 219.105 &   209 \\
1044 & EPU  & 181.48 & 262.98 & +82.42 & 271.57 & +73.79 & 33.71 & 1.002 & 0.680 &  58.305 & 181.423 & 181.475 &    12 \\
1045 & SUT & 179.13 & 66.30 & 24.11 & 249.31 & 2.41 & 67.45 & 0.650 & 0.966 & 175.330 & 253.438 & 179.164 & 22 \\
1046 & PISb  & 232.10 &   6.71 &  -7.86 & 130.83 &  -9.90 & 10.23 & 0.948 & 0.661 &   2.657 &  26.113 &  52.522 &    24 \\
1106 & GAD  & 15.04   &  275.68  & 53.86 & 270.47 & 76.94 & 35.98 & 0.998 & 0.880 & 57.965 & 178.316 & 14.850 & 12 \\
        \hline 
\end{longtable}

\twocolumn

\begin{table}
\caption{List of established meteor showers that were not observed or observed in small number.}
\label{tab:showers_not_observed}
\begin{tabular}{llllll}
Daytime & N  & Low-activity & N  & Not active during survey & N  \\
\hline\hline 
AAN     & 48  & LUM          & 33   & OCT                      & 38 \\
ARI     & 28  & JRC          & 20   & DPC                      & 41 \\
DSX     & 11  & JIP          & 20   & JLE                      & 14  \\
LBO     & 22  & SLD          & 44   & DRA                      & 7   \\
NOC     & 13  & EPR          & 14   & JBO                      & 5   \\
KLE     & 6   & FED          & 13   & TAH                      & 1  \\
SSE     & 5   & ALY          & 7    & BHY                      & 0  \\
OSE     & 2   & BEQ          & 0    & ACE                      & 0  \\
XRI     & 1   &              &      & PHO                      & 0  \\
OCE     & 0   &              &      & COR                      & 0  \\
APS     & 0   &              &      & OCC                      & 0  \\
SMA     & 0   &              &      & PPU                      & 0   \\
ZPE     & 0   &              &      &                          &    \\
XSA     & 0   &              &      &                          &    \\
DLT     & 0   &              &      &                          &    \\
TCB     & 0   &              &      &                          &    \\
ALA     & 0   &              &      &                          &    \\
ZCA     & 0   &              &      &                          &    \\
BTA     & 0   &              &      &                          &    \\
MKA     & 0   &              &      &                          &  
\end{tabular}
\end{table}

\subsection{Meteor shower outbursts}

The period covered by our survey (2018-2021) includes several meteor shower outbursts. Unfortunately, the 2018 Draconid outburst \citep{egal2018draconid, vida2020new} was only observed from a single station as the network was still in development at the time \citep{vida20182018}. Nevertheless, several other outbursts were recorded; the associated mean radiants and orbital elements are given in Table \ref{tab:showers}.

\begin{figure}
  \includegraphics[width=\linewidth]{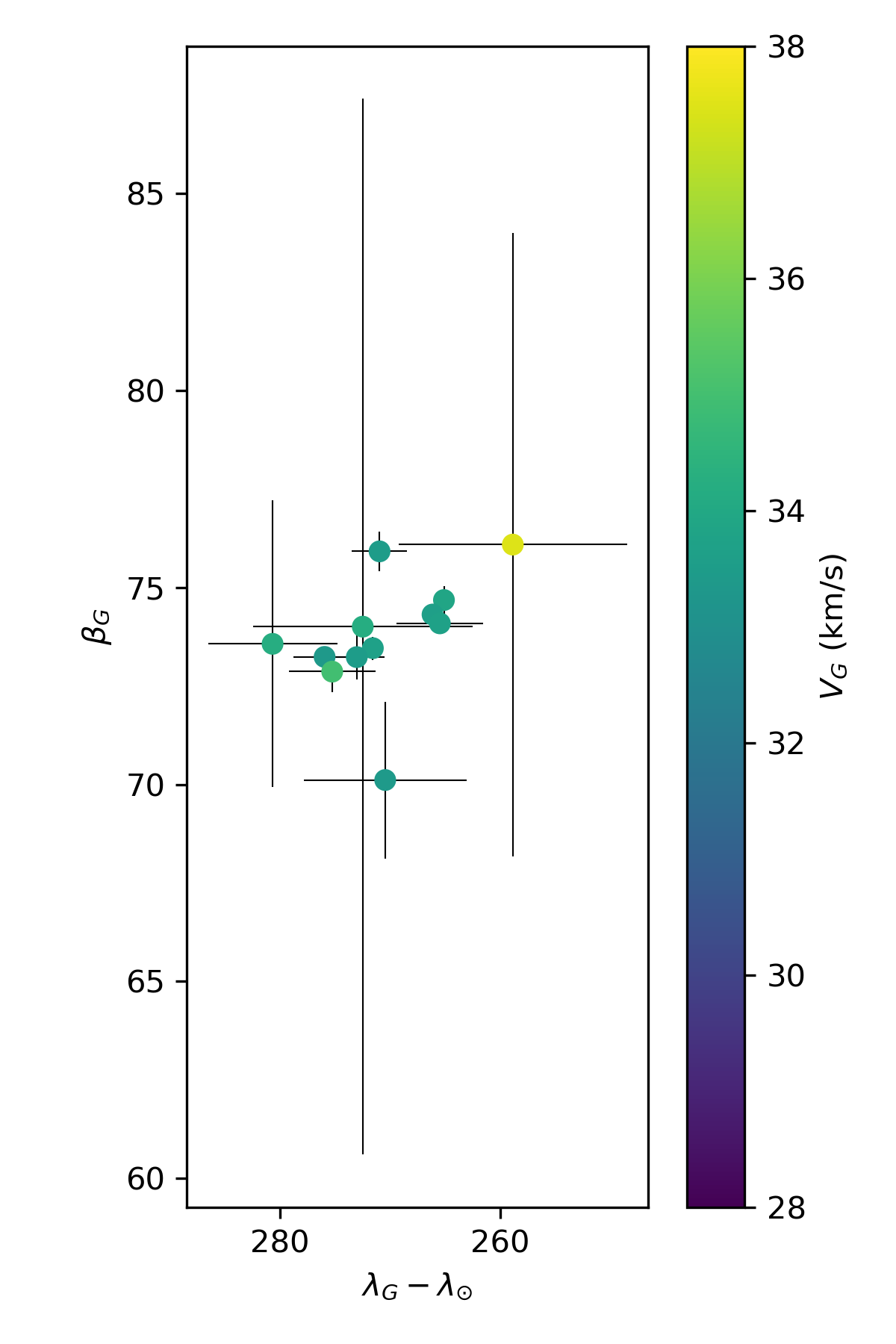}
  \caption{Radiants of the 2019 epsilon Ursae Minorid outburst observed by GMN in Sun-centered geocentric ecliptic coordinates. One sigma uncertainties are shown and the geocentric velocity is color coded. The aspect ratio is adjusted to locally preserve the area.}
  \label{fig:epu_radiants}
\end{figure}

An outburst on September 24, 2019 of the previously unknown high-declination shower, now called the epsilon Ursae Minorids (\#1044 EPU) \citep{sato2020epsilon}, was well recorded in GMN data. Co-authors PK and LP performed manual reduction of the data to ensure the data quality. The derived radiants and the mean orbit match well the values reported by \cite{sato2020epsilon}. The radiant is compact and the measurements with the smallest errors span about \ang{1} (Figure \ref{fig:epu_radiants}). No parent body is known.

\begin{figure}
  \includegraphics[width=\linewidth]{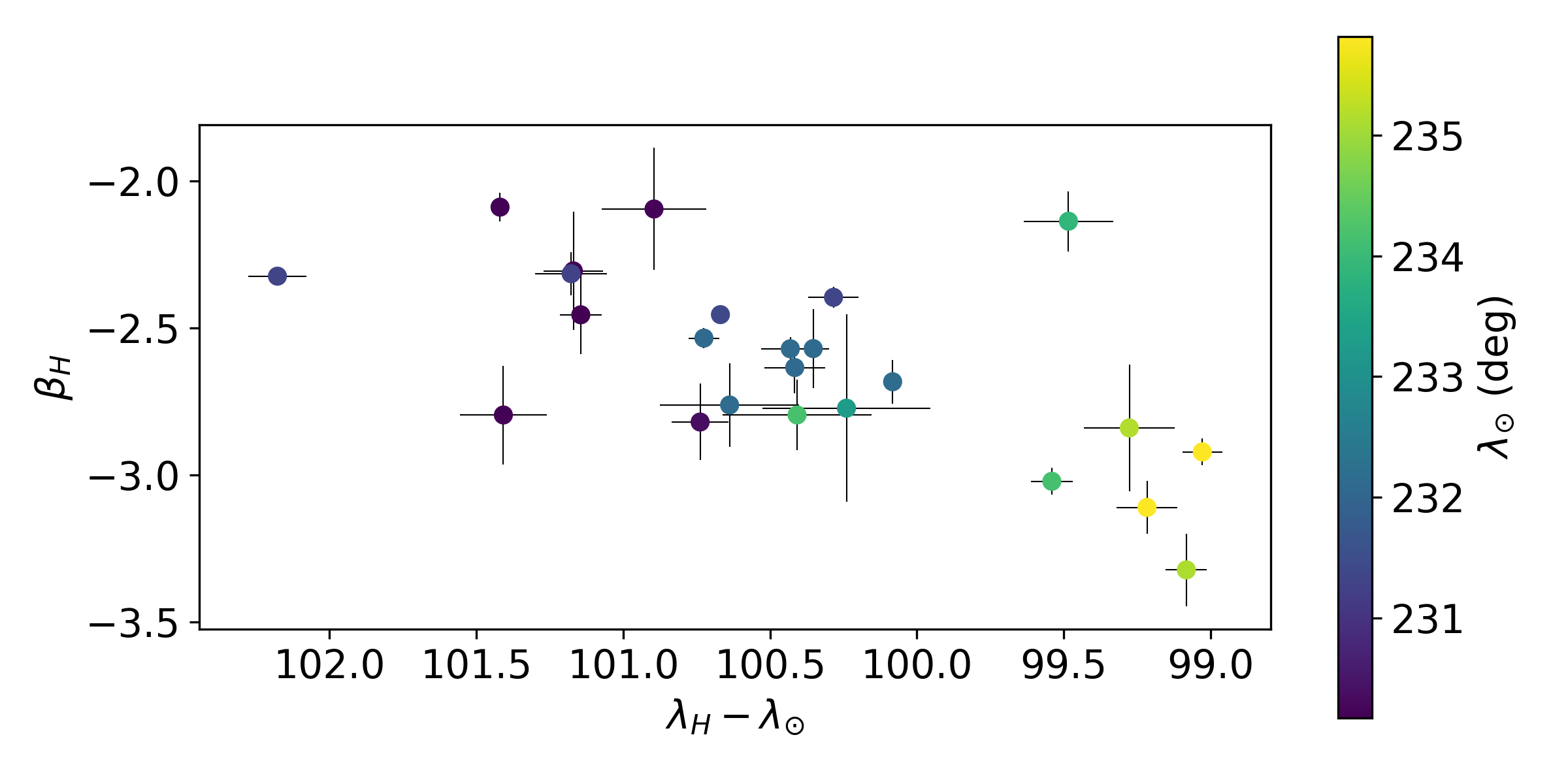}
  \caption{Radiants of the 2019 November outburst of 29 Piscids observed by GMN in Sun-centered heliocentric ecliptic coordinates. The solar longitude is color coded.}
  \label{fig:pis_radiants}
\end{figure}

\cite{jenniskens2020piscid} reported an outburst of a low geocentric velocity (\SI{\sim 10}{\kilo \metre \per \second}) shower in 2019 that had two maxima in activity, the first on October 17-18, and the second one on November 11-18. The shower was later named the 29 Piscids (\#1046 PIS). The GMN only collected three orbits for the first outburst, but it collected 24 orbits associated with the second one. As the shower was at a high elongation from the Earth's apex, it is at the lowest limit of geocentric speeds and as a result the geocentric radiant shows a large apparent motion and dispersion, moving \ang{\sim 7} on the sky in five days and spanning about \ang{2} in width. This effect is caused by the large influence of the Earth's velocity vector on the geocentric radiant \citep{Kresak1970a}. This can be simply mitigated by subtracting Earth's velocity vector from the meteoroid's state vector and thus ``looking" at the radiants in Earth-centered heliocentric ecliptic coordinates \citep{tsuchiya2017correction}, a good strategy for detecting low-velocity showers. Figure \ref{fig:pis_radiants} shows that the radiant only spans \ang{\sim 1} and shows a motion of \ang{3} over five days in heliocentric coordinates, increasing the shower radiant local SNR with respect to the sporadic background and making the shower more easily detectable. Nevertheless, we give the mean radiant for the November 11-18 outburst (PISb in Table \ref{tab:showers}) in geocentric coordinates to adhere to the currently accepted convention.

\begin{figure}
  \includegraphics[width=\linewidth]{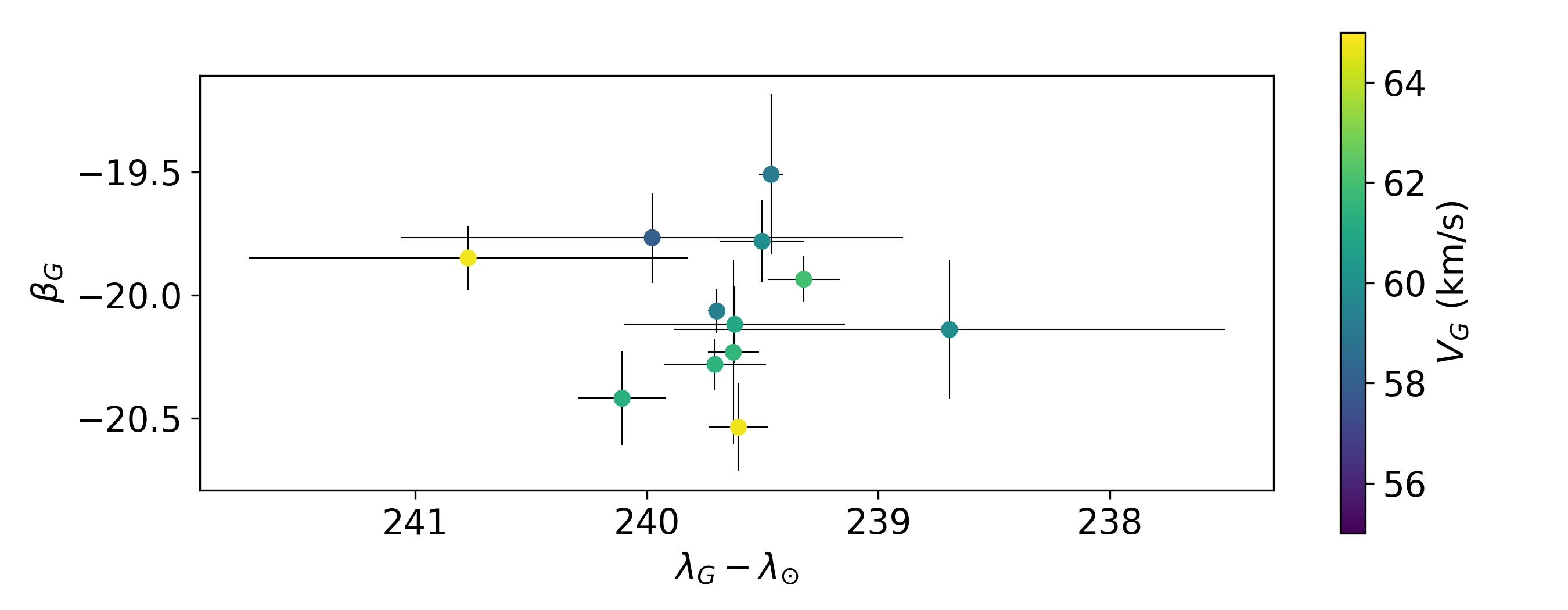}
  \caption{Individual radiants of the Alpha Monocerotid outburst of November 22, 2019 observed by GMN.}
  \label{fig:amo_radiants}
\end{figure}

A total of 12 orbits of the 2019 outburst of the Alpha Monocerotids (\#246 AMO) were observed on November 22, 2019. All meteors were observed within a time window of half an hour, from 04:35 to 05:02 UTC, with the peak of activity after 04:50 UTC when 8 meteors were observed within these 12 minutes \citep{roggemans2020alpha}. Figure \ref{fig:amo_radiants} shows the observed radiants which span about \ang{1}. The observed radiants and the peak time match the predictions of \cite{lyytinen2003meteor} and \cite{lyytinen2020likely} almost exactly.

\begin{figure}
  \includegraphics[width=\linewidth]{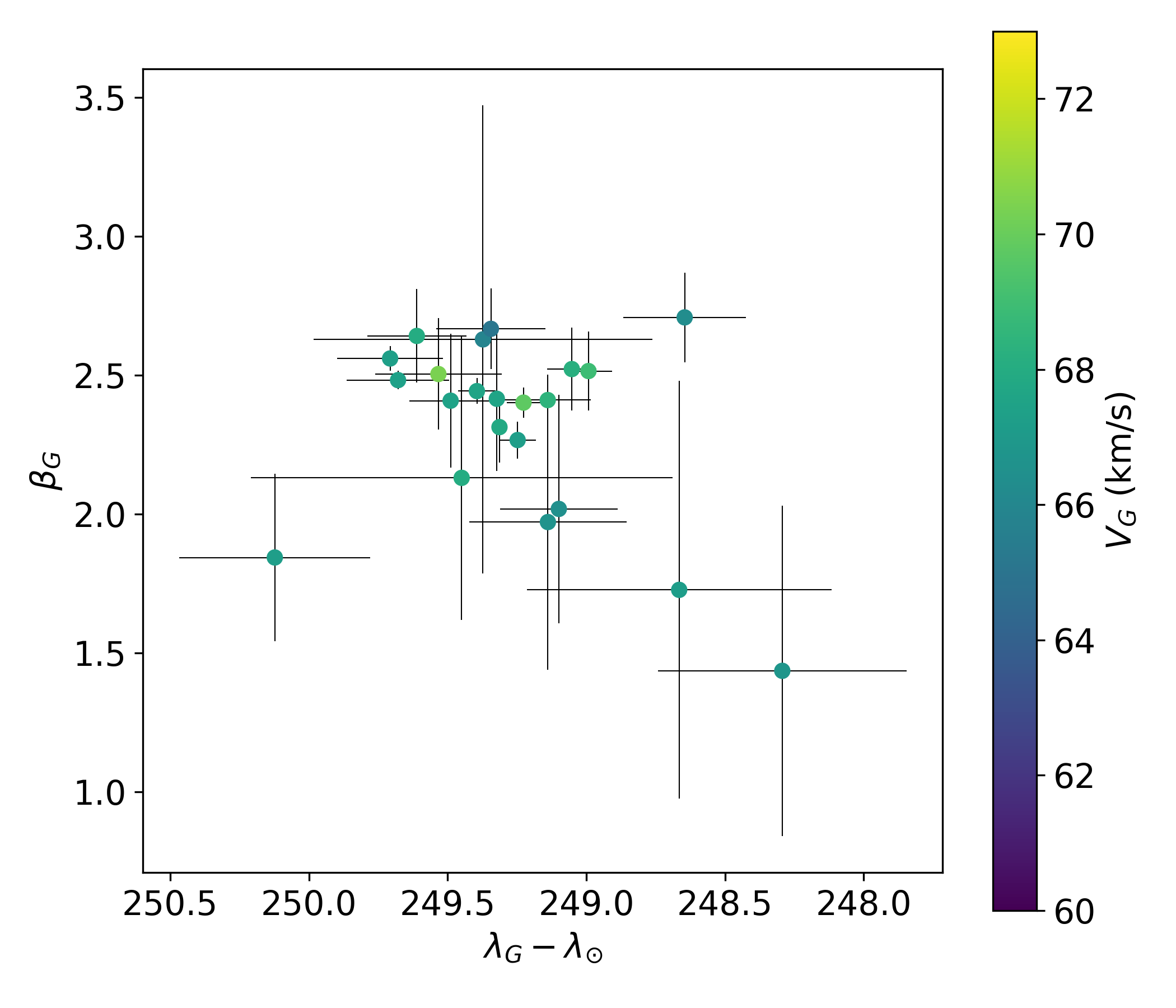}
  \caption{GMN observed radiants of the September upsilon Taurid outburst on September 21, 2020.}
  \label{fig:sut_radiants}
\end{figure}

An outburst of the September upsilon Taurid (\#1045 SUT) \citep{jenniskens2020september} was recorded by the GMN on September 21, 2020 and a total of 22 orbits were collected. Of these, 17 meteors were observed within a narrow time window between solar longitudes \ang{179.1} - \ang{179.2}. A parent body search showed no convincing connections. The radiants were not disperse, as shown in Figure \ref{fig:sut_radiants} - the main bulk of radiants with the smallest uncertainties spanned less than \ang{0.5}.

The GMN also observed the h Virginids during their 2020 outburst. \cite{roggemans2020hvirginids} performed a detailed analysis using GMN data and showed that the results compare well to results obtained from other meteor networks and authors.

Finally, on April 3-5 2021 the GMN observed a outburst of an unknown meteor shower. A total of 12 meteors were observed by cameras in 7 countries. They had tight radiants within a circle of \ang{0.9}. The new shower was named \#1106 Gamma Draconids (GAD) \citep{vida2021gamma} and the details are given in Table \ref{tab:showers}.

\subsection{January 9, 2020 meteorite dropping fireball in Croatia} \label{subsec:meteorite_dropper}

\begin{figure}
  \includegraphics[width=\linewidth]{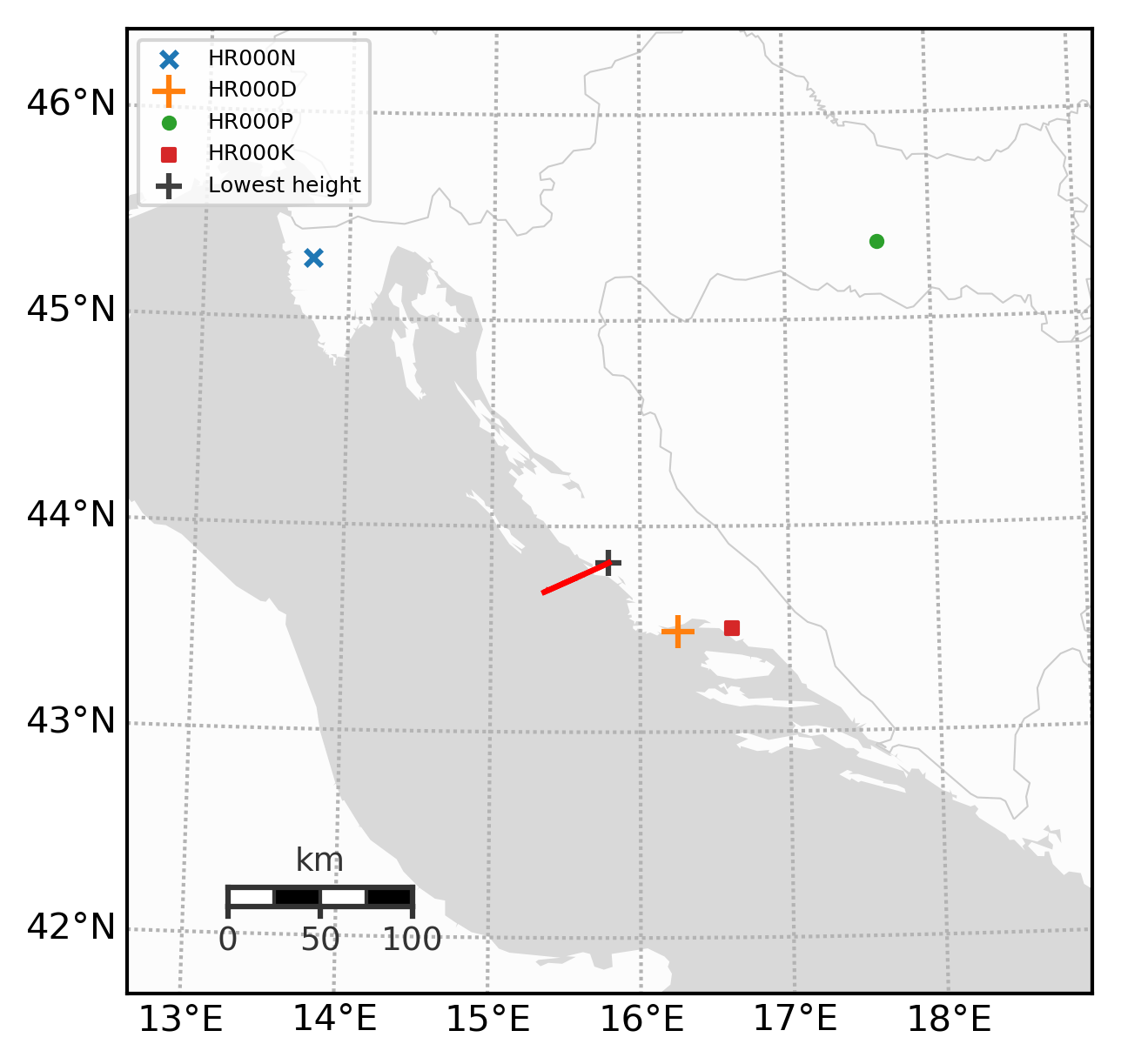}
  \caption{Map of the 2020-01-09 fireball trajectory showing the stations which observed it.}
  \label{fig:20200109_groundmap}
\end{figure}

As a final example of the application of GMN data, we examine a bright fireball lasting \SI{6.2}{\second} observed above southern Croatia on January 9, 2020 at 23:26:40 UTC. It was observed by four GMN stations that were \SIrange{50}{300}{\kilo \metre} away from the fireball, as shown in Figure \ref{fig:20200109_groundmap}. 

\begin{figure}
  \includegraphics[width=\linewidth]{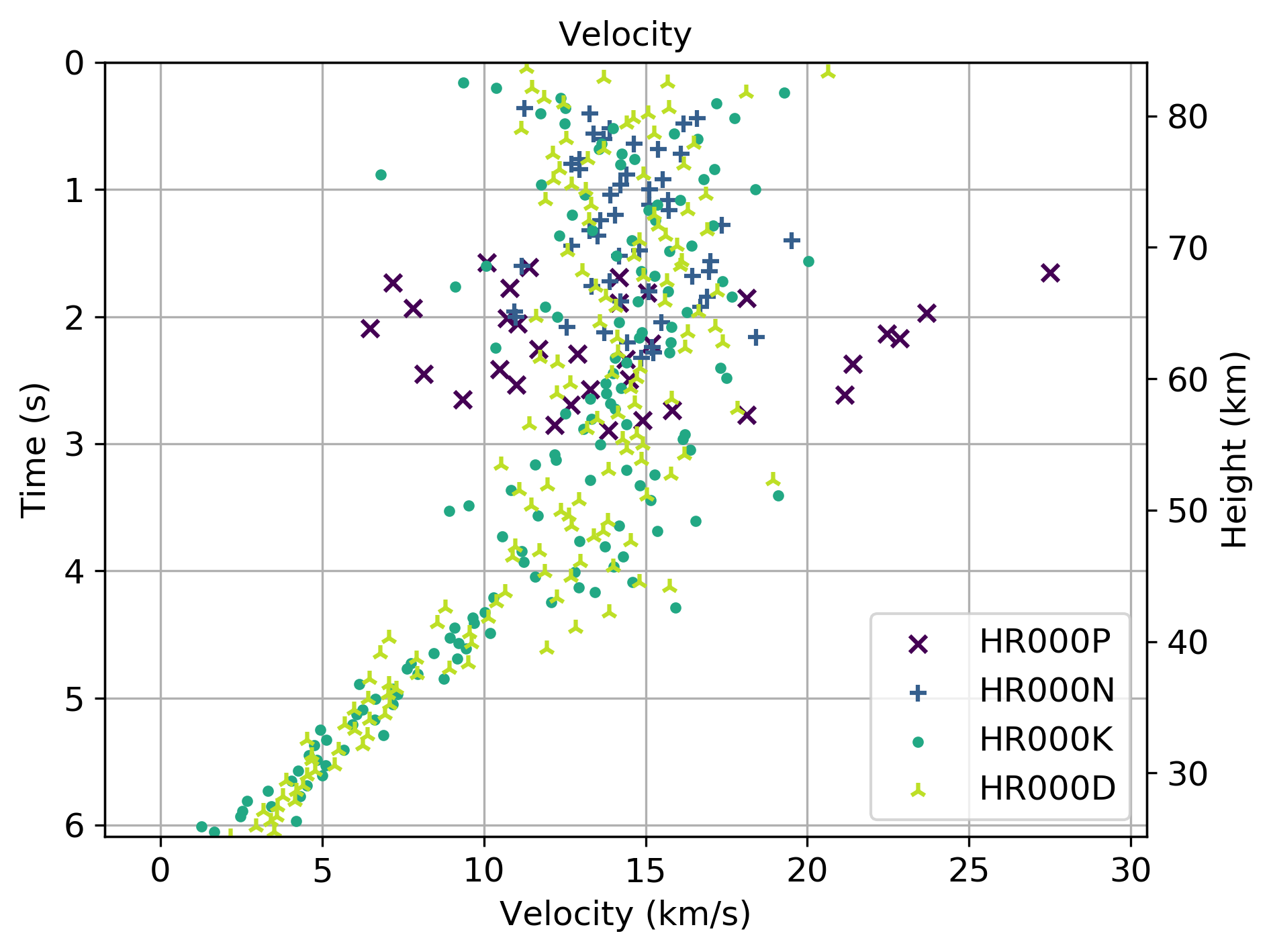}
  \caption{Point-to-point velocities of the January 9 fireball. Observations from more distant stations show larger scatter. The height scale at the right is computed assuming an average velocity.}
  \label{fig:20200109_velocity}
\end{figure}

\begin{figure*}
  \includegraphics[width=\linewidth]{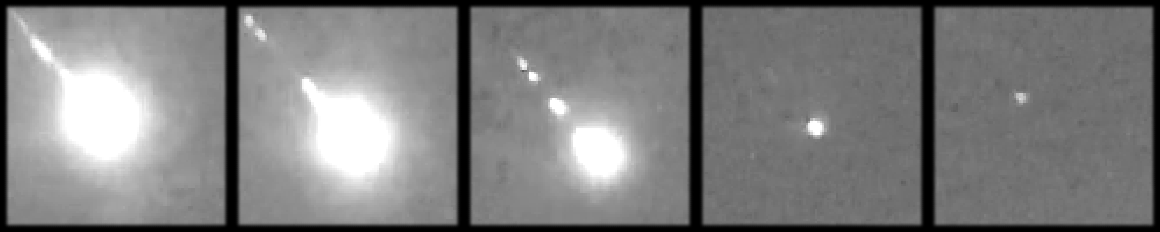}
  \caption{Last \SI{1.4}{\second} of the January 9, 2020 fireball in Croatia as observed from the HR000D station. Every tenth frame is shown, and the last fragmentation occurred on the second frame from the left. A non-fragmenting final fragment with no wake was visible until is became fainter than magnitude $+3^M$.}
  \label{fig:20200109_fireball_frames}
\end{figure*}

It entered the atmosphere at \SI{14.5}{\kilo \metre \per \second} (ground-fixed velocity) with an entry angle of \ang{56.71 \pm 0.01}. It became visible at a height of \SI{85.6}{\kilo \metre}, started developing visible wake at \SI{\sim80}{\kilo \metre}, displayed multiple fragmentations between \SIrange{57}{30}{\kilo \metre}, and was last observed at \SI{25.0}{\kilo \metre} at a velocity of \SI{\sim3}{\kilo \metre \per \second} (Figure \ref{fig:20200109_velocity}). At the height of \SI{30}{\kilo \metre} (at \SI{\sim5.8}{\second} and \SI{\sim7}{\kilo \metre \per \second}), the brightest fragment showed a transverse velocity component of \SI{\sim140}{\metre \per \second}, deviating \SI{200}{\metre} from the initial trajectory over the last \SI{1.4}{\second} as observed from the two closest stations (HR000D at \SI{\sim50}{\kilo \metre}, HR000K at \SI{\sim75}{\kilo \metre}). The transverse velocity component was produced by lateral forces during the last fragmentation as shown in Figure \ref{fig:20200109_fireball_frames}.

We also note a reduction in the magnitude of along-track deceleration around the same time, which also became constant \citep[similar deceleration profile has been observed for e.g. PN 39404, a comparable sized body][]{pecina1983new}. For the discussion about probable physical mechanisms which may cause transverse velocity, see \cite{passey1980effects} and \cite{sansom20193d}. Similar lateral velocities have been observed in the past for other fireballs, e.g. the Mor{\'a}vka and the {\v{Z}}{\v{d}}{\'a}r nad S{\'a}zavou meteorite falls \citep{borovicka2003moravka, spurny2020vzvdar}.

\begin{figure}
  \includegraphics[width=\linewidth]{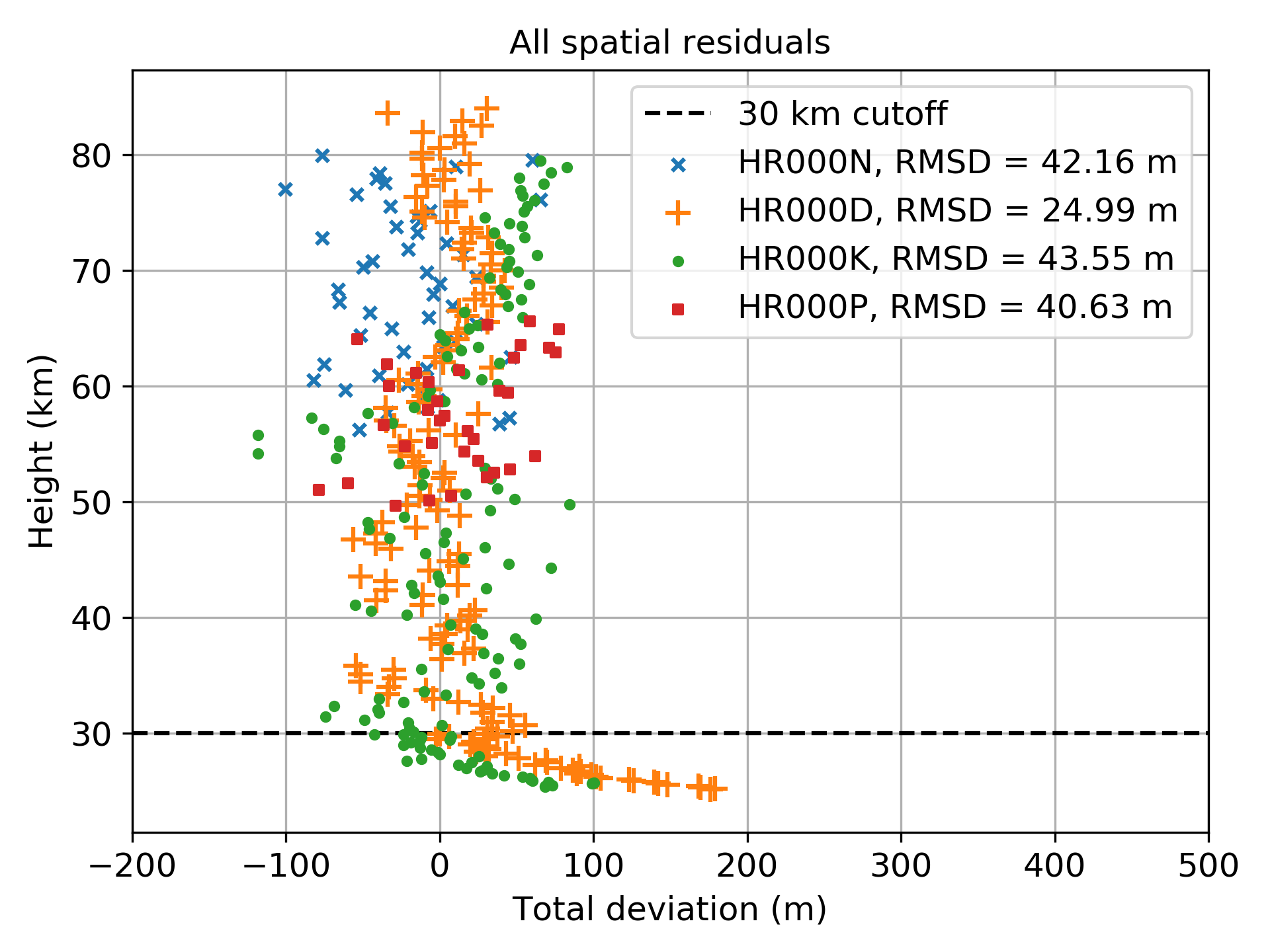}
  \caption{The trajectory fit residuals for the January 9 fireball. Observations below \SI{30}{\kilo \metre} were not included in the trajectory fit used to derive the orbit. Different total residuals below \SI{30}{\kilo \metre} are due to perspective effects.}
  \label{fig:20200109_fit_residuals}
\end{figure}

By fitting a trajectory before and after the force kick which produced the transverse velocity component (see Figure \ref{fig:20200109_fit_residuals}), we determined that the final fragment shifted \ang{1} up and \ang{1.6} towards the south relative to the initial trajectory. The deviation is not caused by bad astrometric calibration in the region of the image where the fireball was observed last, as calibration stars were used in that region, and the deviation was independently observed from two stations. A separate fit on the trajectory portion below \SI{30}{\kilo \metre} yielded fit residuals of \SI{\sim15}{\metre} and enabled an accurate determination of the end point which was about \SI{2}{\kilo \metre} south of the village of Ga\'{c}elezi in Dalmatia. The final dynamic mass, assuming a spherical body and chondritic density, was on the order of \SI{100}{\gram}. Subsequent dark flight analysis resulted in a narrow fall area of $500\times100$ meters, the uncertainty mostly driven by the unknown shape, into a private olive orchard and an area recently cleared of land mines. Further investigation of satellite imagery revealed that the area was burned in a wildfire a few years ago, and medium-sized trees were replaced by knee-high grass, which combined with the local karst geology and a possible danger of leftover land mines made searching for meteorites unfeasible.

Finally, the formal estimate of the radiant and initial velocity errors were only \ang{0.04} and \SI{4}{\metre \per \second}. The estimate was done  using the method of \cite{vida2020estimating}, by adding Gaussian noise on the order of the fit uncertainties and re-fitting the trajectory. Even if the velocity error was underestimated by a factor of 2, the precision is still within the \SI{100}{\metre \per \second} uncertainty limit suggested by \cite{granvik2018identification} that is needed for identifying meteorite source regions. We note that the uncertainly does not include the modelling of pre-luminous flight \citep{pecina1983new}, but that should not have a significant influence on the initial velocity for such large meteoroids. In this case, no significant deceleration was observed above the height of \SI{60}{\kilo \metre}.

\section{Conclusion}

In this paper we have described the hardware, software and data collection methodology used by the Global Meteor Network together with some preliminary observational results. The organizational model adopted by the GMN decentralizes most of the camera deployment and maintenance cost of the network to amateur astronomers, while retaining a high quality standard of data products through a fully automated software pipeline.

Using low-cost security camera boards and Raspberry Pi single-board computers, the hardware used by the GMN is significantly cheaper than what is currently used by other meteor networks. Despite the low cost, the IMX291 CMOS sensors have a factor of 4 larger image resolution and are a factor of two more sensitive than previous comparable meteor camera systems, achieving a limiting stellar magnitude of $+6.0 \pm 0.5^M$ at the resolution of $1280\times720$ px, 25 FPS, and covering a \ang{88}$\times$\ang{48} field of view.

Data formats, meteor detection, astrometric and photometric calibration procedures have been described. Of particular note, we showed that the pointing of meteor cameras needs to be recalibrated on images for every meteor detection to preserve the full astrometric accuracy. Pointing drifts of up to 10 arc minutes over the course of a night are commonly observed for cameras in the network, while the nominal astrometric accuracy is around 1 arc minute. Similarly, photometric recalibration is done for each event as the photometric offset has been found to vary up to one magnitude within a night. We also showed that careful manual reduction of fireball events is necessary to ensure a high level of astrometric accuracy due to the complexity of the morphology caused by fragmentation and disintegration which makes fiducial picks a challenge.

Hundreds of GMN meteor stations upload their observations to a central server where meteor trajectories are computed and published online within 24 hours of observation. Stringent criteria are imposed on automatically computed meteor trajectories, which include limiting heights and speeds of events to the nominal meteor range, and rejecting obvious outliers. We note that peculiar events such as meteors with beginning heights above \SI{160}{\kilo \metre} and meteors on highly hyperbolic orbits might get eliminated, but the raw data is stored so such events may be analyzed in the future.

Between December 2018 to June 2021, over 220,000 meteor orbits have been collected by the GMN, and 70 non-daytime annual IAU established meteor showers were observed, including six meteor shower outbursts (AMO, HVI, EPU, SUT, PIS, and GAD). We provide a table of median radiants and orbital elements of each. The median radiant measurement precision of GMN trajectories is \ang{0.47}, which goes down to \ang{0.32} if only meteors observed from 4+ stations are selected (which include 20\% of all observed meteors). Meteors observed with high-precision \SI{16}{\milli \metre} lenses have a median radiant precision of \ang{0.17}.

Finally, we analyzed one meteorite dropping fireball that was observed by four GMN stations. The mean trajectory fit error was on the order of \SI{40}{\metre} even for \SI{300}{\kilo \metre} distant cameras. Wake and multiple fragmentations were observed between the heights of 80 and \SI{30}{\kilo \metre}, after which a single-body like fragment obtained a transverse velocity on the order of \SI{100}{\metre \per \second}. The dark flight analysis showed that a \SI{\sim100}{\gram} meteorite fell into a potentially dangerous area with high vegetation, thus a search was not done.

\subsection{Note on code availability}

Implementation of all methods used in this work is published as open source on the following GitHub web pages:
\begin{itemize}
    \item RMS software - \url{https://github.com/CroatianMeteorNetwork/RMS},
    \item WesternMeteorPyLib - \url{https://github.com/wmpg/WesternMeteorPyLib}.
\end{itemize}
Readers are encouraged to contact the lead author (DV) in the event they are not able to obtain the code on-line.

\subsection{Data availability}

The orbital and trajectory data underlying this article are freely available online at \url{https://globalmeteornetwork.org/data/}. 
Raw observations of the January 9, 2020 fireball are available on request. A list of stars (their image and celestial coordinates) for every data set in Appendix \ref{appendix:astrometry_methods_comparison} is given in the supplementary materials.

\section{Acknowledgements}

Funding for this work was provided by the NASA Meteoroid Environment Office under cooperative agreement 80NSSC18M0046. We would like to thank the MAGIC Collaboration, the staff of the Museum of Science and the Cosmos San Cristobal de la Laguna in Tenerife, Rijeka Sport Ltd.,  Academical Astronomical Society - Rijeka, the Istria County Association of Technical Culture, and the Croatian Astronomical Union for financial and logistical support.

We would also like to thank the following GMN station operators whose cameras provided the data used in this work and contributors who made important code contributions (in alphabetical order): Victor Acciari, Alexandre Alves, {\v Z}eljko Andrei{\' c}, Georges Attard, Roger Banks, Hamish Barker, Ricky Bassom, Richard Bassom, Jean-Philippe Barrilliot, Dr. Ehud Behar, Josip Belas, Alex Bell, Serge Bergeron, Arie Blumenzweig, Ventsislav Bodakov, Ludger B\"{o}rgerding, Claude Boivin, Bruno Bonicontro, Fabricio Borges, Dorian Bo\v{z}i\v{c}evi\'{c}, Martin Breukers, John Briggs, Laurent Brunetto, Tim Burgess, Peter Campbell-Burns, Pablo Canedo, Seppe Canonaco, Jose Carballada, Gilton Cavallini, Brendan Cooney, Edward Cooper, Dino {\v C}aljku{\v s}i{\' c}, Tim Claydon, Manel Colldecarrera, Christopher Curtis, Ivica {\' C}ikovi{\' c}, J{\" u}rgen D{\"o}rr, Chris Dakin, Alfredo Dal'Ava J{\'u}nior, Steve Dearden, Christophe Demeautis, Paul Dickinson, Ivo Dijan, Pieter D\mbox{ij}kema, Stacey Downton, Zoran Dragi\'{c}, Jean-Paul Dumoulin, Garry Dymond, Robin Earl, Ollie Eisman, Carl Elkins, Rick Fischer, Richard Fleet, Jim Fordice, Mark Gatehouse, Megan Gialluca, Jason Gill, Philip Gladstone, Nikola Gotovac, Neil Graham, Bob Greschke, Sam Green, Daniel J. Grinkevich, Larry Groom, Tioga Gulon, Hugo Gonz\'{a}lez, Uwe Gl\"{a}ssner, Dominique Guiot, Margareta Gumilar, Kees Habraken, Erwin Harkink, Ed Harman, Tim Havens, Richard Hayler, Alex Hodge, Bob Hufnagel, Russell Jackson, Jean-Marie Jacquart, Ron James Jr., Ilya Jankowsky, Klaas Jobse, Dave Jones, Vladimir Jovanovi{\'c}, Milan Kalina, Jonathon Kambulow, Richard Kacerek, Steve Kaufman, Alex Kichev, Jean-Baptiste Kikwaya, Zoran Knez, Dan Klinglesmith, Danko Ko{\v c}i{\v s}, Korado Korlevi{\'c}, Stanislav Korotkiy, Josip Krpan, Zbigniew Krzeminski, Reinhard K\"{u}hn, Ga\'{e}tan Laflamme, David Leurquin, Anton Macan, John Maclean, Igor Macuka, Mirjana Malari{\'c}, Nedeljko Mandi{\'c}, Bob Marshall, Jos\'{e} Luis Martin, Colin Marshall, Andrei Marukhno, Keith Maslin, Bob Massey, Damir Matkovi\'{c}, Sergio Mazzi, Alex McConahay, Robert McCoy, Charlie McCromack, Filip Mezak, Pierre-Michael Micaletti, Matej Mihel{\v c}i{\'c}, Simon Minnican, Wullie Mitchell, Nick Moskovitz, Gene Mroz, Brian Murphy, Carl Mustoe, Przemek Naga{\'n}ski, Jean-Louis Naudin, Damjan Nemarnik, Colin Nichols, Zoran Novak, Michael O'Connell, Gareth Oakey, Washington Oliveira, Thiago Paes, Carl Panter, Filip Parag, Igor Pavleti\'{c}, Richard Payne, Pierre-Yves Pechart, Enrico Pettarin, Alan Pevec, Patrick Poitevin, Pierre de Ponthière, Alex Pratt, Miguel Preciado, Chuck Pullen, Dr. Lev Pustil'nik, Chris Ramsay, Danijel Reponj, David Robinson, Martin Robinson, Heriton Rocha, Herve Roche, Adriana Roggemans, Alex Roig, James Rowe, Dmitrii Rychkov, Michel Saint-Laurent, Jason Sanders, Rob Saunders, William Schauff, Ansgar Schmidt, Jay Shaffer, Jim Seargeant, Ivica Skoki\'{c}, Dave Smith, Tracey Snelus, James Stanley, Peter Stewart, William Stewart, Bela Szomi Kralj, Ian Pass, Rajko Su\v{s}anj, Marko \v{S}egon, Jeremy Taylor, Yakov Tchenak, Eric Toops, Steve Trone, Wenceslao Trujillo, Paraksh Vankawala, Martin Walker, Bill Wallace, Didier Walliang, Jacques Walliang, Christian Wanlin, Tom Warner, Urs Wirthmueller, Steve Welch, Alexander Wiedekind-Klein, John Wildridge, Bill Witte, and Stephane Zanoni. 




\bibliographystyle{mnras}
\bibliography{bibliography} 


\appendix

\section{Astrometric calibration} \label{appendix:astrometric_calibration}

The camera pointing direction is defined by a reference right ascension $\alpha_{C0}$ and declination $\delta_{C0}$ for the centre of projection (optical axis) of the camera/lens combination, at a given reference time (Julian date) $\mathrm{JD}_0$. The reference equatorial coordinates are in the J2000 epoch. This approach allows for a simple way of computing the instantaneous pointing coordinates $\alpha_C$ and $\delta_C$. Assuming that the pointing remains unchanged, the declination of the centre will stay fixed at all times ($\delta_C = \delta_{C0})$, while the right ascension can simply be computed by adding the hour angle difference between the reference time $\mathrm{JD}_0$ and the time of interest. The rotation is defined using a reference position angle $\rho_0$, an angle measured relative to the vertical axis of the focal plane clockwise to the north celestial pole (positive in the positive direction of the right ascension). Finally, the plate scale $F$ is given in pixels per degree. Table \ref{tab:plate_coeffs} gives a summary of astrometric and distortion plate parameters.

Sections \ref{appendix:image_to_eq} and \ref{appendix:eq_to_image} below describe how image coordinates are converted to equatorial coordinates and vice versa. Note that because all modern star catalogs are given in the J2000 epoch, and our procedure does not precess the star positions to the epoch of date for a given recording, the resulting equatorial coordinates will also be in the J2000 epoch. Care must be taken during the trajectory estimation procedure to precess the data to the epoch of date if necessary.

\begin{table*}
    \caption{Description of the astrometric plate parameters.}
    {
    \begin{tabular}{l c l}
    \hline\hline 
    Parameter & Units & Description \\
    \hline 
    \multicolumn{3}{l}{Pointing parameters}\\
    $\mathrm{JD}_0$        & Days         & Reference Julian date at which the $\alpha_{C0}$ and $\delta_{C0}$ are in the centre of projection. \\
    $\alpha_{C0}$ & $^\circ$     & Reference right ascension of the optical centre at the reference time $\mathrm{JD}_0$ in the J2000 epoch. \\
    $\delta_{C0}$ & $^\circ$     & Reference declination of the optical centre at the reference time $\mathrm{JD}_0$ in the J2000 epoch. \\
    $\rho_0$      & $^\circ$     & Reference position angle. \\
    $F$           & px/$^\circ$  & Plate scale. \\
    \hline 
    \multicolumn{3}{l}{Polynomial distortion parameters} \\
    $\boldsymbol{a}$  & None         & Forward mapping X-axis polynomial distortion coefficients. \\
    $\boldsymbol{b}$  & None         & Forward mapping Y-axis polynomial distortion coefficients. \\
    $\boldsymbol{c}$  & None         & Reverse mapping X-axis polynomial distortion coefficients. \\
    $\boldsymbol{d}$  & None         & Reverse mapping Y-axis polynomial distortion coefficients. \\
    \hline 
    \multicolumn{3}{l}{Radial distortion parameters} \\
    $x_{0f}$           & px           & X-axis optical centre offset, forward mapping. \\
    $y_{0f}$           & px           & Y-axis optical centre offset, forward mapping. \\
    $x_{0r}$           & px           & X-axis optical centre offset, reverse mapping. \\
    $y_{0r}$           & px           & Y-axis optical centre offset, reverse mapping. \\
    $s_f$             & None         & Aspect ratio between X and Y axis, forward direction. \\
    $s_r$             & None         & Aspect ratio between X and Y axis, reverse direction. \\
    $\boldsymbol{k}$  & None         & Forward mapping radial distortion coefficients. \\
    $\boldsymbol{m}$  & None         & Reverse mapping radial distortion coefficients. \\
    \hline 
    \end{tabular}
    }
    \label{tab:plate_coeffs}
\end{table*}

\subsection{Astrometric parameter minimization procedure} \label{app:fit_algorithm}

Two different strategies for the minimization of astrometric parameters are adopted, depending on whether the polynomial or the radial distortion is used.

\subsubsection{Polynomial distortion}

Assuming the camera pointing and scale have been roughly estimated and at least 13 stars have been associated between the image and the star catalog, the following astrometric fitting algorithm is applied.

Step 0. All distortion coefficients are initially set to zero, except for $a_0$, $b_0$, $c_0$, and $d_0$ which are set to 0.5 pixels which marks a slight offset from the middle of the image. 

Step 1. The four pointing parameters ($\alpha_{C0}$, $\delta_{C0}$, $\rho_0$, $F$) are fit using a Sequential Least Squares Programming (SLSQP) algorithm \citep{kraft1988software} from the \texttt{scipy} library \citep{virtanen2020scipy}. The cost function is the sum of squared pixel residuals computed using reverse mapping.

Step 2. The reverse distortion coefficients in vectors $\boldsymbol{c}$ and $\boldsymbol{d}$ (from sky to image) are obtained using the Nelder-Mead algorithm \citep{nelder1965simplex}, fitting the X and Y axis independently. For example when the X coefficients are fit (vector $\boldsymbol{c}$), all Y coefficients are set to 0 (except the $c_0$ offset) to ensure they are not used in the fit, and vice versa. When the X or Y coefficients are fit, the cost function is a sum of squared pixel position differences only along the X or Y axis respectively.

Step 3. If this is the first iteration of the fit, the first guess of the forward mapping distortion coefficients are set to be equal to the previously fit reverse coefficients from Step 2 (i.e. $\boldsymbol{a} = \boldsymbol{c}$, $\boldsymbol{b} = \boldsymbol{d}$).

Step 4. The forward mapping coefficients in vectors $\boldsymbol{a}$ and $\boldsymbol{b}$ are fit using the Nelder-Mead algorithm mapping image to sky, solving the X and Y axis independently. The cost function is the sum of squared angular separations in equatorial coordinates.

\begin{figure*}
  \includegraphics[width=\linewidth]{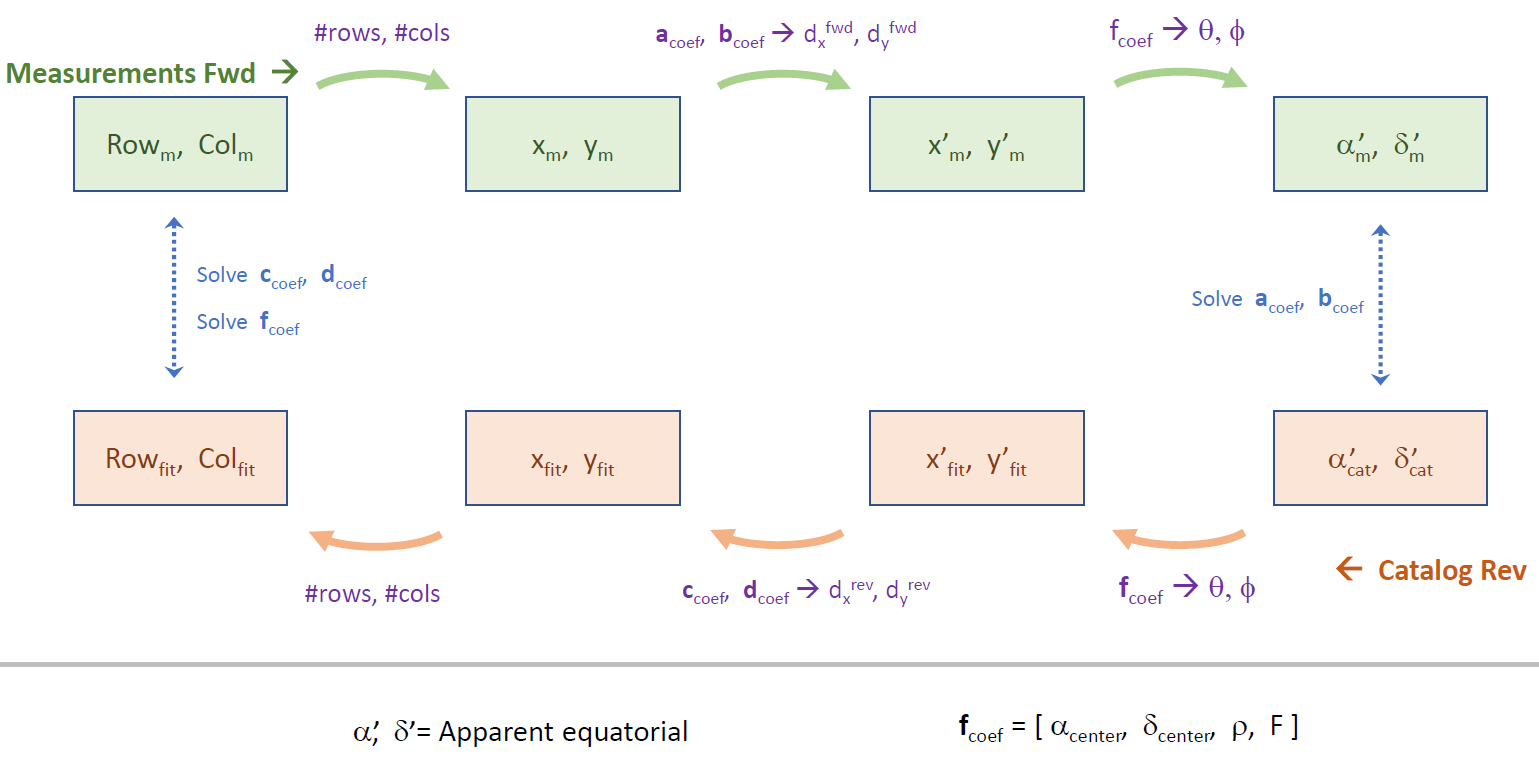}
  \caption{Diagram illustrating the astrometry calibration minimization procedure for the polynomial distortion. Values with the index ``m" indicate image positional measurements taken off of the focal plane propagated forward through the transformations to equatorial coordinates, and values with the index ``fit" indicate the reverse mapping from celestial coordinates to image coordinates.}
  \label{fig:astrometry_fitting_diagram}
\end{figure*}

The pointing parameters, the reverse and the forward mapping distortion coefficients, usually converge in 2-3 iterations. After the first iteration, Step 3 is skipped. Note that during the minimization procedure, $\alpha_{C0}$ and $\rho_0$ need to be wrapped to the \ang{0} - \ang{360} range, $\delta_{C0}$ to the \ang{-90} - \ang{+90} range, and the plate scale $F$ should be forced to be positive. Figure \ref{fig:astrometry_fitting_diagram} illustrates the fitting procedure.

Please note that the $\alpha_{C0}$ and $\delta_{C0}$ do not represent the equatorial coordinates of the centre of the focal plane, but of the centre of projection. The focal plane center coordinates would need to be calculated from the final astrometry solution using the middle position row and column.

\subsubsection{Radial distortion}

As determining the exact centre of projection is essential for the radial projection, the pointing and distortion parameters are estimated together. Steps 0 and 0.5 below are only run during the first initialization of the astrometric plate so a rough pointing direction is obtained.

Step 0. All distortion coefficients are initially set to zero, except for $x_{0f}$, $y_{0f}$, $x_{0r}$, and $y_{0r}$ which are set to 0.5 pixels which marks a slight offset from the middle of the image, and an equal aspect ratio is assumed ($s_f = 1$).

Step 0.5. The four pointing parameters ($\alpha_{C0}$, $\delta_{C0}$, $\rho_0$, $F$) are fit using the SLSQP algorithm. The cost function is the sum of squared pixel residuals computed using reverse mapping.

Step 1. The four pointing parameters and the forward mapping radial coefficients ($x_{0f}$, $y_{0f}$, $s_f$, and vector $\boldsymbol{k}$) are fit together using the Nelder-Mead algorithm. The cost function is the sum of squared angular separations in equatorial coordinates.

Step 2. If this is the first iteration of the fit, the first guess of the reverse mapping distortion coefficients are set to be equal to the previously fit forward coefficients from Step 1 ($\boldsymbol{m} = \boldsymbol{k}$, $s_r = s_f$, $x_{0r} = x_{0f}$, $y_{0r} = y_{0f}$).

Step 3. Only the reverse mapping coefficients ($x_{0r}$, $y_{0r}$, $s_r$, and vector $\boldsymbol{m}$) are fit using the Nelder-Mead algorithm. The cost function is a sum of squared pixel position differences for both the X and Y axis together ($\sqrt{ \Delta X^2 + \Delta Y^2 }$).

The approach usually converges within 1-2 iterations. Note that users should decide whether the aspect ratio and the asymmetry correction (see Section \ref{subsec:radial_distortion_fwd}) should be used or not, depending on the optical setup. For the IMX291 which has square pixels the aspect ratio is kept at unity. The asymmetry correction is used by default and usually improves the fit errors by $10\% - 25\%$.

\subsection{Transforming image coordinates to equatorial coordinates} \label{appendix:image_to_eq}

Given a pair of image column and row coordinates ($X, Y$), and the Julian date/time of the video frame or image (JD), the corresponding right ascension $\alpha$ and declination $\delta$ in the J2000 epoch can be computed as follows.

First, the column X and row Y image coordinates are normalized to centre-offset coordinates ($x, y$) :

\begin{align}
    x = {} & X - \frac{X_{\mathrm{res}}}{2} \,, \\
    y = {} & Y - \frac{Y_{\mathrm{res}}}{2} \,,
\end{align}

\noindent where $X_{\mathrm{res}}$ is the column and $Y_{\mathrm{res}}$ is the row image resolution.

Next, the undistorted coordinates $x'$ and $y'$ are computed by applying the distortion correction. Note that there are two sets of distortion coefficients, one mapping image to undistorted coordinates (which we define as forward mapping), and another set mapping undistorted coordinates to image coordinates (reverse). In this case, we use the forward mapping coefficients.

\subsubsection{Polynomial distortion correction}
If the distortion is modelled as a third order polynomial with added radial terms, the following equations are used to compute the transformation to undistorted coordinates:

\begin{equation} \label{eq:distortion_forward}
\begin{aligned}
    r =  {} & \sqrt{(x - a_0)^2 + (y - b_0)^2} \,, \\
    dx_\mathrm{FWD} = {} & a_0 + a_1 x + a_2 y + a_3 x^2 \\
            & + a_4 x y + a_5 y^2 + a_6 x^3 + a_7 x^2 y \\
            & + a_8 x y^2 + a_9 y^3 + a_{10} x r + a_{11} y r \,, \\
    dy_\mathrm{FWD} = {} & b_0 + b_1 x + b_2 y + b_3 x^2 \\
            & + b_4 x y + b_5 y^2 + b_6 x^3 + b_7 x^2 y \\
            & + b_8 x y^2 + b_9 y^3 + b_{10} x r + b_{11} y r \,,
\end{aligned}    
\end{equation}

\noindent where $r$ is the radius from the centre of projection, $dx_\mathrm{FWD}$ the offset in X and $dy_\mathrm{FWD}$ the offset in Y image directions. $\boldsymbol{a}$ is the forward X-axis distortion coefficient vector, and $\boldsymbol{b}$ is the forward Y-axis distortion coefficient vector.

\subsubsection{Radial distortion} \label{subsec:radial_distortion_fwd}
Following a modified approach of \cite{jeanne2019calibration}, a radius corrected for distortion is computed using odd terms up to the ninth order, $r_\text{corr} = r + k_3 r^3 + k_4 r^5 + k_5 r^7 + k_6 r^9$. It can be limited up to the $3^{\mathrm{rd}}$, $5^{\mathrm{th}}$, $7^{\mathrm{th}}$, and $9^{\mathrm{th}}$ order in the software, and we have found that in most cases no further improvement is achieved after the $7^{\mathrm{th}}$ order is included. To apply the radius distortion correction to image coordinates, the scale of the radius change can computed as $r_s = \frac{r_\text{corr}}{r} - 1$. Substituting $r_\text{corr}$ into $r_s$, the final form of the equations is the following:

\begin{equation} \label{eq:radial_distortion_forward}
\begin{aligned}
    r' =  {} & \sqrt{(x - x_{0f})^2 + s_f^2(y - y_{0f})^2}  \,, \\
    r  =  {} & \frac{1}{X_{\mathrm{res}}/2} \left( r' + k_1 s_f(y - y_{0f})\cos k_2 \right. \\
    {} & \left. \, - k_1 (x - x_{0f}) \sin k_2 \right) \,, \\
    r_s = {} & k_3 r^2 + k_4 r^4 + k_5 r^6 + k_6 r^8 \,, \\
    dx_\mathrm{FWD} = {} & (x - x_{0f}) r_s - x_{0f} \,, \\
    dy_\mathrm{FWD} = {} & s_f(y - y_{0f}) r_s - s_f y_{0f} + (s_f - 1) y \,, \\
\end{aligned}    
\end{equation}

\noindent where $r$ is the radius from the centre of projection normalized such that $r = 1$ at the middle of the left/right side of the image and corrected for asymmetry, $s_f$ is the aspect ratio between X and Y axes \citep[similar to the approach by][but usually kept at unity for square pixels]{palotai2019analysis}, $X_{\mathrm{res}}$ is the X-axis image resolution, $\boldsymbol{k}$ is the radial distortion coefficient vector (forward mapping), $x_{0f}$ and $y_{0f}$ offsets from the optical centre, $dx_\mathrm{FWD}$ is the offset in the X and $ dy_\mathrm{FWD}$ is the offset in the Y image direction. The asymmetry correction is applied through the $k_1 s_f (y - y_{0f}) \cos k_2 - k_1 (x - x_{0f}) \sin k_2$ terms \citep{borovicka1995new, jeanne2019calibration}, which can be disabled in the software by the user.

\subsubsection{Computing equatorial coordinates}

With the pixel offsets in hand, depending on which type of distortion was used, the distortion-corrected image coordinates are computed as:

\begin{align}
    x' = {} & x + dx_\mathrm{FWD} \,, \\
    y' = {} & y + dy_\mathrm{FWD} \,.
\end{align}

Next, the undistorted coordinates are represented in a polar form normalized by the plate scale $F$ (in px/$^\circ$):

\begin{equation} \label{eq:projection_fwd}
\begin{aligned}
    \phi = {} & \sqrt{ \frac{x'^2 + y'^2}{F} } \,, \\
    \theta = {} & \left ( \ang{90} -  \rho_0 + \atantwo ( y', x' ) \right ) \mod \ang{360} \,,
\end{aligned}
\end{equation}

\noindent where $\phi$ is the angular separation from the object to the centre of projection at right ascension $\alpha_C$ and declination $\delta_C$, $\theta$ is the angle between the object and the north celestial pole, and $\rho_0$ is the reference position angle.

The instantaneous declination of the centre of projection $\delta_C$ remains the same as the reference declination $\delta_{C0}$. The instantaneous right ascension $\alpha_C$ is computed by adding the difference between Greenwich hour angles at the reference time $\mathrm{JD}_0$ and at the time of interest JD to the reference right ascension $\alpha_{C0}$. See equation E3 in Appendix E of \cite{vida2020estimating} for computing the Greenwich hour angle for both the reference time ($\mathrm{GHA}_0$) and the time of interest ($\mathrm{GHA}$):

\begin{align} \label{eq:ra_ref_now}
    \alpha_C = {} & ( \alpha_{C0} + \mathrm{GHA} - \mathrm{GHA}_0 + \ang{360} )  \mod \ang{360} \,.
\end{align}

Next, refraction needs to be applied to the reference centre of projection coordinates $(\alpha_C, \delta_C)$ following the steps in Appendix \ref{appendix:refraction_true_to_apparent}, producing apparent equatorial coordinates of the centre of projection $(\alpha'_C, \delta'_C)$. Note that correcting for refraction is not absolutely necessary if the chosen distortion model can account for the distortion induced by refraction and the pointing of the camera is absolutely fixed. However, eliminating most of the effects of refraction with the simple formulae of Appendix \ref{appendix:refraction_true_to_apparent} and \ref{appendix:refraction_apparent_to_true} leaves more degrees of freedom available to fit the lens distortions.

The apparent equatorial coordinates $(\alpha', \delta')$ of the object in the J2000 epoch can then be computed as:

\begin{equation}
\begin{aligned}
    \zeta = {} & \sin \delta'_C \cos \phi + \cos \delta'_C \sin \phi \cos \theta \,, \\
    \delta' = {} & \atantwo \left ( \zeta, \sqrt{(1 - \zeta^2} \right ) \,,  \\
    \alpha' = {} & \Bigg( \alpha'_C - \atantwo \left (  \frac{\sin \theta \sin \phi  }{\cos \delta'}, \frac{\cos \phi - \sin \delta' \sin \delta'_C}{ \cos \delta' \cos \delta'_C}  \right ) \\
    {} & + \ang{360} \Bigg) \mod \ang{360} \,.
\end{aligned}
\end{equation}

Finally, true equatorial coordinates $(\alpha, \delta)$ are obtained by correcting the apparent coordinates for refraction using equations in Appendix \ref{appendix:refraction_apparent_to_true}.

We note that many additional steps might be avoided if the reference centre of projection is defined in apparent horizontal coordinates (which are by definition in the epoch of date), which would make the procedure more straightforward. Nevertheless, the same result can be obtained using the equations above if care is taken to apply them properly.

\subsection{Transforming equatorial coordinates to image coordinates} \label{appendix:eq_to_image}

Given equatorial coordinates $\alpha, \delta$ of an object or a star in the J2000 epoch (e.g. as provided by the star catalog) and the time of observation JD, apparent equatorial coordinates $\alpha', \delta'$ with the refraction taken into account can be computed using the equations in Appendix \ref{appendix:refraction_true_to_apparent}. The apparent equatorial coordinates of the centre of projection $\alpha'_C, \delta'_C$ are computed as described in Appendix \ref{appendix:image_to_eq}.

The undistorted coordinates in the polar form are computed as follows:

\begin{equation}
\begin{aligned}
    \phi = {} & \arccos \left ( \sin \delta'_C \sin \delta' + \cos \delta'_C \cos \delta' \cos(\alpha' - \alpha'_C) \right ) \,,  \\
    \theta = {} & -\atantwo \left (  \frac{ \cos \delta' \sin (\alpha' - \alpha'_C) }{ \sin \phi }, \right. \\
    {} & \left. \frac{ \sin \delta' - \sin \delta'_C \cos \phi }{\cos \delta'_C \sin \phi} \right ) + \rho_0 - \ang{90} \,,
\end{aligned}
\end{equation}

\noindent where $\phi$ is the angular separation between the centre of projection and the object, $\theta$ is the angle between the object and the north celestial pole, and $\rho_0$ is the reference position angle.

The undistorted coordinates are then simply:

\begin{align}
    x' = {} & F \phi \cos \theta \,, \\
    y' = {} & F \phi \sin \theta \,,
\end{align}

\noindent where F is the plate scale in px/$^\circ$. Note that $\phi$ also needs to be given in degrees.

\subsubsection{Polynomial distortion}
The reverse polynomial distortion is applied using the following equations (same form as equation \ref{eq:distortion_forward}):

\begin{equation}
\begin{aligned}
    r = {} & \sqrt{(x' - c_0)^2 + (y' - d_0)^2} \,, \\
    dx_\mathrm{REV} = {} & c_0 + c_1 x' + c_2 y' + c_3 x'^2 \\
            & + c_4 x' y' + c_5 y'^2 + c_6 x'^3 + c_7 x'^2 y' \\
            & + c_8 x' y'^2 + c_9 y'^3 + c_{10} x' r + c_{11} y' r \,, \\
    dy_\mathrm{REV} = {} & d_0 + d_1 x' + d_2 y' + d_3 x'^2 \\
            & + d_4 x' y' + d_5 y'^2 + d_6 x'^3 + d_7 x'^2 y' \\
            & + d_8 x' y'^2 + d_9 y'^3 + d_{10} x' r + d_{11} y' r \,,
\end{aligned}    
\end{equation}

\noindent where $r$ is the radius from the centre of projection, $dx_\mathrm{REV}$ the offset in X and $dy_\mathrm{REV}$ the offset in Y image direction. $\boldsymbol{c}$ is the reverse X-axis distortion coefficient vector, and $\boldsymbol{d}$ is the reverse Y-axis distortion coefficient vector.

\subsubsection{Radial distortion}
If the radial distortion is used, the offsets are computed using the following equations:

\begin{equation}
\begin{aligned}
    r' =  {} & \sqrt{x'^2 + y'^2} \,, \\
    r =   {} & \frac{1}{X_{\mathrm{res}}/2} \left ( r' + m_1 y' \cos m_2 - m_1 x' \sin m_2 \right ) \,, \\
    r_s = {} & m_3 r^2 + m_4 r^4 + m_5 r^6 + m_6 r^8 \,, \\
    dx_\mathrm{REV} = {} & x' r_s - x_{0r} \,, \\
    dy_\mathrm{REV} = {} & \frac{y' r_s}{s_r} - y_{0r} + (1 - 1/s_r) y' \,, \\
\end{aligned}    
\end{equation}

\noindent where $r$ is the radius from the centre of projection normalized such that $r = 1$ at the middle of the left/right side of the image, $s_r$ is the aspect ratio between X and Y axes, $X_{\mathrm{res}}$ is the X-axis image resolution, $\boldsymbol{m}$ is the radial distortion coefficient vector (reverse mapping), $x_{0r}$ and $y_{0r}$ offsets from the optical centre, $dx_\mathrm{REV}$ is the offset in X and $dy_\mathrm{REV}$ is the offset in the Y image direction.

\subsubsection{Obtaining image coordinates}

The image coordinates $(X, Y)$ are then simply computed as:

\begin{align}
    X = {} & x' - dx_\mathrm{REV} + \frac{X_{\mathrm{res}}}{2} \,, \\
    Y = {} & y' - dy_\mathrm{REV} + \frac{Y_{\mathrm{res}}}{2} \,,
\end{align}

\noindent where $X_{\mathrm{res}}$ is the column and $Y_{\mathrm{res}}$ is the row image resolution.

\subsection{Atmospheric refraction correction - apparent to true} \label{appendix:refraction_apparent_to_true}

As the refraction is dependent on the local horizontal altitude, the equatorial coordinates (which are in the J2000 epoch as defined by this procedure) need to be precessed to the epoch of date using equations given in Appendix H of \cite{vida2020estimating}, and converted into azimuth and altitude using equations in Appendix G from the same paper.

Given the apparent altitude $a'$ in degrees, the true altitude $a$ can be computed following \cite{bennett1982calculation}:

\begin{equation}
    a = a' - \frac{1}{60} \cot \left ( a' + \frac{7.31}{a' + 4.4} \right) \,.
\end{equation}

Note that we assume a fixed atmospheric pressure of \SI{101}{\kilo \pascal} and a temperature of \SI{10}{\celsius}. Furthermore, for apparent altitudes $a' < \ang{-0.5}$ we assume no further increase of the refraction correction, as otherwise the equation gives large errors \citep{wittmann1997astronomical}.

The horizontal coordinates (azimuth remains unchanged) are then converted back into equatorial coordinates in the epoch of date using equations in Appendix F, and precessed back to J2000 using equations in Appendix H of \cite{vida2020estimating}.

\subsection{Atmospheric refraction correction - true to apparent}
\label{appendix:refraction_true_to_apparent}

Computing the apparent equatorial coordinates from true coordinates is done in the same way as for the inverse approach in \ref{appendix:refraction_apparent_to_true}, only a different equation is used to transform the true horizontal altitude $a$ to apparent altitude $a'$ in degrees \citep{saemundsson1986astronomical}:

\begin{equation}
    a' = a + \frac{1.02}{60} \cot \left ( a + \frac{10.3}{a + 5.11} \right) \,.
\end{equation}

\subsection{Performance comparison to other astrometry calibration methods} \label{appendix:astrometry_methods_comparison}

This section compares various published methods of astrometric calibration and distortion compensation as applied to three different data sets: the all-sky GFO image taken with a Samyang 8mm f/3.5 lens (see Figure \ref{fig:dfn_calib}), a Canon 700D DSLR image taken with a Samyang \SI{16}{\milli \metre} f/2.0 lens, and a GMN image taken with a \SI{3.6}{\milli \metre} lens (see Figure \ref{fig:es0002_calib}). Table \ref{tab:astrometry_calib_info} summarizes the properties of every data set.

The comparison is given in Table \ref{tab:astrometry_calib_comparison}. Note the $r^2$ term referred to in the table in the GMN case as well as the CAMS-based cubic and the quadratic polynomial models, is an ``implicit" term in that one has $x^2$ and $y^2$ components in the distortion model \citep{steyaert1990photographic}. The same goes for the azimuthal distortion terms, as some models do not use explicitly a sinusoidal term, but instead the x and y polynomial mixture model provides some degree of azimuthal distortion.

The examples were chosen to represent typical wide-field data sets one may encounter when reducing meteor data. Narrow field of view imagery (e.g. GMN data collected with \SI{16}{\milli \metre} lenses) was not considered, as the lens distortion is simpler in those cases and can be easily modelled using cubic polynomials \citep[e.g.][]{jenniskens2011cams, vida2020high}. Wide-field lenses are more challenging to model due to the ``moustache" distortion which requires higher order radial terms to compensate. On the other hand, radial models may be more difficult to fit to data for simpler distortions due to being highly non-linear, while polynomial models behave linearly and are invariant to the distortion center \citep{tang2017precision} due to the appearance of translational terms in the formulation. Parameters for the investigated Center of Projection (COP) methods were estimated using the Particle Swarm Optimization algorithm \citep{kennedy1995particle}. Note that the Jeanne entry lines in the table are for variants of the original paper with radial terms included up to, but not higher than the indicated radial power.

Images from a Global Fireball Observatory camera near Tavistock, Ontario were used as an example of an all-sky data set. The 16-bit color images are $7380 \times 4928$ pixels large and have a pixel scale of 1.9 arcmin/px. The main limiting factor on the accuracy of the fit is the point spread function of stars near the edges of the field of view (\ang{\sim20} above the horizon and below), which shows large deformations and thus determining exact positions of stars becomes uncertain. 66 stars were picked across the whole field of view, to reflect how a typical all-sky astrometric plate might be done. Our radial distortion model with terms up to the $7^{\textrm{th}}$ order performed best, while additional higher-order terms did not improve the fit. The astrometric fit can be seen in Figure \ref{fig:dfn_calib}.

\begin{figure*}
  \includegraphics[width=\linewidth]{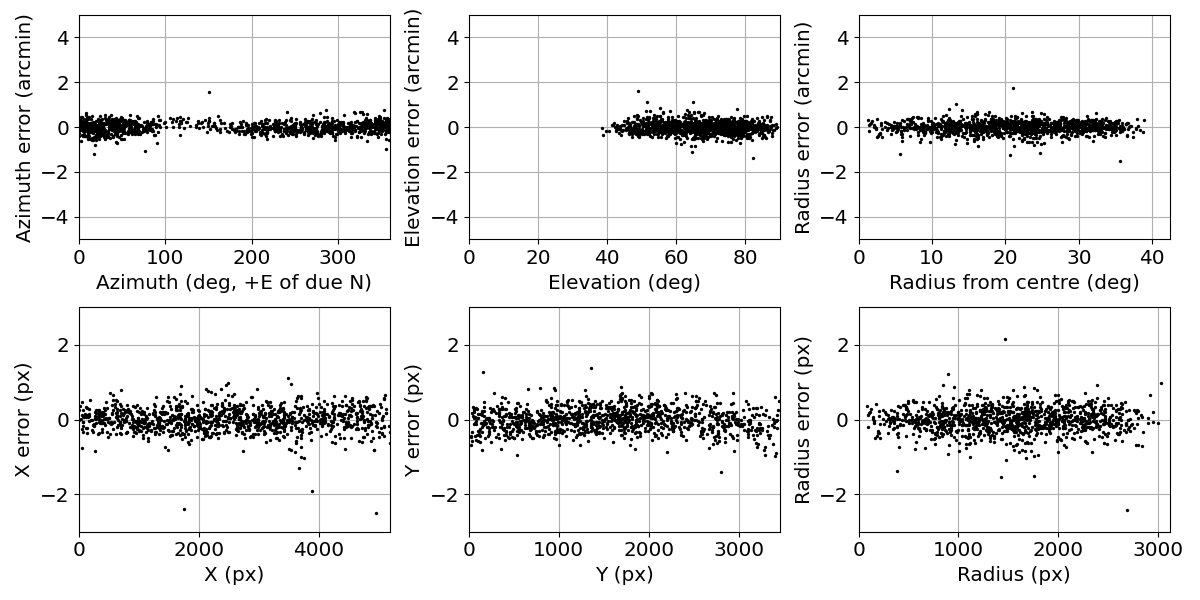}\hfill
  \caption{Astrometric calibration residuals for the DSLR \SI{16}{\milli \metre} data set.  The fit was done using the radial method with odd terms up to the $7^{\textrm{th}}$ order. The asymmetry correction was used, but the aspect ratio was kept at unity.}
  \label{fig:dslr_calib_radial}
\end{figure*}

Next, an image taken by a consumer-grade DSLR camera was used, which is equipment that is often used for casual recordings of fireballs. Many modern DSLR lenses have a large radial ``moustache" distortion which may be difficult to fit using conventional polynomial methods. For that reason, over 1100 stars were manually picked on the image to accurately model the distortion. Again, our radial model with terms up to the $7^{\textrm{th}}$ order performed best. The fit residuals can be seen in Figure \ref{fig:dslr_calib_radial}.

Finally, an image taken by a typical GMN system with the \SI{3.6}{\milli \metre} lens was used. These lenses also have a significant distortion near the edges of the field of view. 100 stars were picked to ensure that a good fit to the distortion coefficients was achieved and we were not undersampled relative to the number of free parameters. In practice, a typical number of picked stars would be around 40-50. Again, our radial model with terms up to the $7^{\textrm{th}}$ order performed the best, achieving a fit residual of 0.78 arcmin. The plot of fit residuals is given in Figure \ref{fig:es0002_calib}. This is significantly better than other methods, including our polynomial method (second best) with fit residuals of 1.27 arcmin.

In conclusion, our novel radial method performs best for all types of tested lenses in this work. In supplementary materials a list of stars (their image and celestial coordinates) for every data set used herein is given, together with fit parameters for the $r^7$ method.

\begin{table}
    \caption{Description of example data sets used for evaluating the performance of the astrometric calibration methods.}
    {
    \begin{tabular}{l c r r}
    \hline\hline 
    Name & FOV   & Pixel scale & Num stars \\
         & (deg) & (arcmin/px) &           \\
    \hline 
    GFO                         & all-sky & 1.9 & 66 \\
    DSLR \SI{16}{\milli \metre} & \ang{69} $\times$ \ang{49} & 0.86 & 1118 \\
    GMN \SI{3.6}{\milli \metre} & \ang{88} $\times$ \ang{48} & 4.1 & 100 \\
    \hline 
\end{tabular}
    }
    \label{tab:astrometry_calib_info}
\end{table}

\begin{table*}
    \caption{Comparison of fit residuals achieved with different methods of astrometry calibration. All fits include refraction correction. COP means Center of Projection (5 optical-axis parameters). $\cent$ indicates that the center of FOV is in equatorial coordinates. NC means ``non-convergent". Poly means polynomial in x and y. The entries are sorted by increasing values of RMS residuals for the GFO all-sky data set. Some entries in this table represent all-sky algorithms applied to narrower fields of view that were not pointing at the zenith. To use the all-sky algorithms in those cases, the normal azimuth and zenith angle were reprojected into a pseudo-azimuth and pseudo-zenith angle whose center of projection was the middle of the focal plane, similar to the projection used in \ref{eq:projection_fwd}. }
    {
    \begin{tabular}{l r c c c S[table-format=3.2] S[table-format=3.2] S[table-format=3.2]}
    \hline\hline 
    Model & \# Fit & Wrap Model & Distortion & Distortion & \multicolumn{3}{c}{Root Mean Square residual (arcmin)} \\
          & Coeffs & General    & Radial     & Azimuthal  & {GFO all-sky} & {DSLR \SI{16}{\milli \metre}} & {GMN \SI{3.6}{\milli \metre}} \\
    \hline 
GMN $r^7$, 1* & 13 & Roll, F, Aspect Ratio, \cent & r, $r^3$, $r^5$, $r^7$ & x, y scale & 0.46 & 0.35 & 0.78 \\
Jeanne $r^9$, 2* & 12 & COP & r, $r^3$, $r^5$, $r^7$, $r^9$ & x, y scale & 0.51 & 0.43 & 1.60 \\
Jeanne $r^7$, 2* & 11 & COP & r, $r^3$, $r^5$, $r^7$ & x, y scale & 0.52 & 0.43 & 1.60 \\
Barghini, 3* & 12 & COP ( xz, yz ) & r, exp(r), exp($r^2$) & x, y scale & 0.65 & 1.33 & 2.46 \\
Borovicka 1995, 4* & 12 & COP & r, exp(r), exp($r^2$) & x, y scale & 0.66 & 0.55 & 1.67 \\
GMN poly, 1* & 26 & Cubic, Roll, F, \cent & xr,  yr,  $r^2$ & x,y poly & 0.67 & 0.72 & 1.27 \\
Jeanne $r^5$, 2* & 10 & COP & r, $r^3$, $r^5$ & x, y scale & 0.72 & 0.43 & 1.63 \\
GMN $r^5$, 1* & 12 & Roll, F, Aspect Ratio, \cent & r, $r^3$, $r^5$ & x, y scale & 0.79 & 0.35 & 0.81 \\
Howell, 5* & 10 & COP, Aspect ratio & r, exp(r) & x, y scale & 1.63 & 1.32 & 1.64 \\
Borovicka 1992, 6* & 8 & COP & r, exp(r) & none & 1.66 & {NC} & 1.87 \\
Cubic Poly, 7* & 22 & Cubic, Roll, Scale, \cent & $r^2$ & x,y poly & 1.96 & 3.11 & 1.86 \\
GMN $r^3$, 1* & 11 & Roll, F, Aspect Ratio, \cent & r, $r^3$ & x, y scale & 2.05 & 1.61 & 2.0 \\
Jeanne $r^3$, 2* & 9 & COP & r, $r^3$ & x, y scale & 2.23 & 1.67 & 2.42 \\
Bannister, 8* & 8 & COP & r, $r^2$ & none & 6.58 & 3.30 & 5.21 \\
Wray, 9* & 14 & Quad, Roll, Scale, \cent & xr,  yr,  $r^2$ & x,y poly & 12.0 & 11.0 & 5.92 \\
Quadratic Poly, 7* & 14 & Quad, Roll, Scale, \cent & $r^2$ & x,y poly & 61.0 & 11.0 & 28.0 \\
    \hline 
\end{tabular}
\\References: 1* - this work, 2* - \cite{jeanne2019calibration}, 3* - \cite{barghini2019astrometric}, 4* - \cite{borovicka1995new}, 5* - \cite{palotai2019analysis}, 6* - \cite{borovivcka1992astrometry}, 7* -  \cite{jenniskens2011cams}, 8* - \cite{bannister2013numerical}, 9* - \cite{wray1967computation}.
    }
    \label{tab:astrometry_calib_comparison}
\end{table*}

\section{Photometric calibration} \label{appendix:photometric_calibration}

Photometry is a process of computing stellar or meteor magnitudes from observational data. This is a transformation from the sum of image pixel intensities into magnitudes, where we take care to compensate the sensitivity losses from optical vignetting and atmospheric extinction. Photometric calibration is done by fitting sums of image pixel intensities of stars to their known catalog magnitudes using a model. Table \ref{tab:photometry_coeffs} gives a summary of parameters used for the RMS photometric calibration.

\begin{table*}
    \caption{Description of photometric parameters.}
    {
    \begin{tabular}{l c l}
    \hline\hline 
    Parameter & Units & Description \\
    \hline 
    $\gamma$    & None      & Camera gamma correction. \\
    $L$         & $^\circ$/px    & Vignetting coefficient. \\
    $p_0$       & magnitude & Photometric offset. \\
    $s_e$       & None      & Extinction correction scale. \\
    \hline 
    \end{tabular}
    }
    \label{tab:photometry_coeffs}
\end{table*}

As discussed by \cite{jenniskens2011cams}, an important aspect of the calibration is the spectral band-pass of the sensor, which needs to roughly match the spectral band of stars in the catalog. RMS supports several star catalogs: the BSC5 catalog \citep{hoffleit1987bright} with V-band magnitudes, the SKY2000 catalog \citep{myers2015vizier} with BVRI filters \citep[a synthetic spectral band may be constructed by taking ratios of each available band, see][]{jenniskens2011cams}, and the GAIA DR2 catalog \citep{gaia2018dr2} with the GAIA G band magnitudes (a broad spectral sensitivity band, not to be confused with the green filter). The GAIA DR2 catalog is used by RMS by default as the GAIA G band matches the spectral sensitivity of the deployed GMN camera's CMOS sensors. The BSC5 catalog is used for calibrating all-sky cameras as the GAIA DR2 is not complete for stars brighter than $+3^M$.

The photometric calibration is used to compute magnitudes of meteors from background-subtracted sum of pixel intensities belonging to a meteor on a given video frame. Determining which pixels belong to the meteor can be done either automatically (e.g. by image thresholding) or by manual data reduction. The pixel sum is traditionally computed as \citep{roelandts2012beware}:

\begin{align} \label{eq:photometry_sum}
    S_{px} = \sum_{i=0}^n \left( I_{i}^{1/\gamma} - B_{i}^{1/\gamma} \right) \,,
\end{align}

\noindent where $S_{px}$ is the background-subtracted sum of pixel intensities, $I_{i}$ is the intensity of a given pixel, and $B_{i}$ the background intensity (i.e. the brightness of the sky without the meteor). Most professional-grade cameras produce data where the pixel intensities are linear ($\gamma = 1$), but some commercial-grade cameras raise the pixel intensities to the power of $\gamma = 0.45$ and they need to be corrected prior to performing any linear operations. If the correction is not done, linear operations cannot be performed using pixel intensities, making even the subtraction in equation \ref{eq:photometry_sum} invalid.

Due to vignetting, the sensitivity across the field of view drops off radially from the optical centre. Traditionally, there are two ways of compensating for vignetting: applying a flat field correction \citep[see][]{berry2005handbook} which also takes care of other optical imperfections caused by dust or dirt on the lens, or modelling the vignetting as a $\cos^4$ sensitivity drop off from the optical centre \citep{jenniskens2011cams}. RMS software supports both methods, although in practice flat field calibration requires careful work and the construction of specialized calibration equipment (light boxes). It is possible to make a flat field image by computing a per-pixel median using images taken throughout the night to remove stars and only get an estimate of the background sky glow. Nevertheless, such calibration images also often contain the gradient in brightness of the observed night sky, and not just the loss of sensitivity due to vignetting. Thus, RMS operationally corrects the vignetting using the following relation:

\begin{align}
    S'_{px} = \frac{S_{px}}{\cos^4 ( L r )} \,,
\end{align}

\noindent where $S'_{px}$ is the vignetting-corrected pixel intensity sum, $L$ is the vignetting coefficient (in degrees per pixel), and $r$ is the radial distance from the optical centre in pixels. Note that this approach assumes that the measured meteor segment on a given frame is a point source and that the vignetting is locally constant across the apparent size of the segment, which is a good approximation for this application. A typical vignetting coefficient for GMN cameras with \SI{3.6}{\milli \metre} lenses is 0.057 deg/px at the video resolution of $1280\times720$ - the sensitivity in the corner of the image plane is only $\sim 40\%$ relative to the center.

Finally, from the definition of the magnitude \citep{ehlert2017improving}, it follows that the apparent meteor magnitude $M'$ at a given time can be computed as:

\begin{align} \label{eq:magnitude}
    M' = -2.5 \log_{10} (S'_{px}) + p_0 \,,
\end{align}

\noindent where $p_0$ is the photometric offset.

\subsection{Atmospheric extinction} \label{appendix:exitinction}

Atmospheric extinction, the stellar or meteor brightness loss at low elevations due to atmospheric scattering, is modelled following \cite{green1992magnitude}. First, the approximation of the total air mass along the observer's line of sight is computed:

\begin{equation}
    X(z) = \frac{1}{\cos z + 0.025 \exp (-11 \cos z)} \,,
\end{equation}

\noindent where $X$ is the air mass and $z$ is the zenith angle (\ang{90} - altitude) of the line of sight.

According to \cite{hayes1975rediscussion}, there are three sources of atmospheric extinction - Rayleigh scattering $A_\mathrm{Ray}$, aerosol scattering $A_\mathrm{aer}$, and ozone scattering $A_\mathrm{oz}$:

\begin{align}
    A_\mathrm{Ray}(h) = {} & 0.1451 \exp(-h/7.996) \,, \\
    A_\mathrm{aer}(h) = {} & 0.12 \exp(-h/1.5) \,, \\
    A_\mathrm{oz} = {} & 0.016 \,,
\end{align}

\noindent where $h$ is the elevation of the observer above sea level in kilometers. As outlined by \cite{green1992magnitude}, there is a significant variability in the aerosol contribution which mostly depends on humidity and amount of volcanic aerosols, but we use an average value that the author suggested. The total magnitude difference is then simply:

\begin{equation}
    \Delta M (z, h) = \left ( A_\mathrm{Ray}(h) +  A_\mathrm{aer}(h) + A_\mathrm{oz}\right ) X(z) \,.
\end{equation}

Finally, given an apparent magnitude $M'$, the no-atmosphere magnitude $M$ is computed as:

\begin{equation}
    M = M' - s_e \left( \Delta M(z, h) - \Delta M (\ang{0}, h) \right) \,,
\end{equation}

\noindent where $s_e$ is the extinction correction scaling which is kept at unity by default, but may be adjusted manually as the equations above are given for the V spectral band. Also, note that we only perform a differential magnitude correction between the altitude of the line of sight and the zenith, while the photometric calibration compensates for any differences between the no-atmosphere magnitude and the apparent magnitude at the zenith.


\bsp	
\label{lastpage}
\end{document}